\documentclass[aps,twocolumn,prd,superscriptaddress,showpacs,nofootinbib]{revtex4-1}

\usepackage[utf8]{inputenc}

\usepackage{diagbox}

\usepackage{mathtools}
\usepackage{amsfonts}
\usepackage{mathrsfs}
\usepackage{bbm}
\usepackage{slashed}

\usepackage{graphicx}
\usepackage{color}
\usepackage{array}

\usepackage{placeins}
\usepackage{booktabs}
\usepackage{makecell}
\usepackage{epstopdf}
\usepackage[caption=false]{subfig}

\usepackage{xspace}
\usepackage{siunitx}
\usepackage{hyperref}
\usepackage[nameinlink]{cleveref}
\usepackage{appendix}

\usepackage{xifthen}
\usepackage{xcolor}
\hypersetup{
	colorlinks,
	linkcolor={red!75!black},
	citecolor={blue!75!black},
	urlcolor={blue!75!black}
}

\setkeys{Gin}{width=0.48\textwidth}

\graphicspath{
	{./figures/}
	{../figures/}
}

\def\pslash{p\llap{/}}
\def\qslash{q\llap{/}}
\def\kslash{k\llap{/}}
\def\dr{{D\!\llap{/}}\,}

\def\s0#1#2{\mbox{\small{$ \frac{#1}{#2} $}}}
\def\0#1#2{\frac{#1}{#2}}

\newcommand{\ltpre}[1][0pt]{\mathrel{\raisebox{#1}{\scriptsize$\parallel$}}}
\newcommand{\lt}{{\ltpre[1.2pt]}}
\newcommand{\trans}{ \bot}

\def\eq#1{\eqref{#1}}
\def\Eq#1{Eq.~\eqref{#1}}

\def\fig#1{Figure \ref{#1}}
\def\Fig#1{Figure~\ref{#1}}

\def\Tab#1{Table~\ref{#1}}

\def\Sec#1{Section~\ref{#1}}
\def\App#1{Appendix~\ref{#1}}

\newcommand{\tr}{{\text{tr}}}

\newcommand{\sumint}{\int\hspace{-4.8mm}\sum}
\newcommand{\imag}{\text{i}}

\newcommand{\gettitle}{QCD phase structure from functional methods}

\hypersetup{
	pdftitle={\gettitle},
	pdfauthor={Gao, Pawlowski},
	pdfkeywords={functional methods}
		{correlations functions} {phase structure}
		{functional renormalisation group} {Dyson-Schwinger equation},
	bookmarksopen=true,
	bookmarksopenlevel=2,
	bookmarksnumbered=true
}

\begin{document}

\title{\gettitle}

\author{Fei Gao}
\affiliation{Institut f{\"u}r Theoretische Physik,
	Universit{\"a}t Heidelberg, Philosophenweg 16,
	69120 Heidelberg, Germany
}

\author{Jan M. Pawlowski}
\affiliation{Institut f{\"u}r Theoretische Physik,
	Universit{\"a}t Heidelberg, Philosophenweg 16,
	69120 Heidelberg, Germany
}
\affiliation{ExtreMe Matter Institute EMMI,
	GSI, Planckstr. 1,
	64291 Darmstadt, Germany
}

                             % Classification Scheme.

\begin{abstract}
  We discuss the QCD phase structure at finite temperature and chemical potential for $2$-flavour and $2+1$-flavour QCD. The results are achieved by computing QCD correlation functions within a generalised functional approach that combines Dyson-Schwinger equations (DSE) and the functional renormalisation group (fRG). In this setup fRG precision data from  \cite{Cyrol:2017ewj} for the vacuum quark-gluon vertex and gluon propagator of $2$-flavour QCD are used as input, and the respective DSEs are expanded about this input. While the vacuum results for other correlation functions serve as a self-consistency check for functional approaches, the results at finite temperature and density are computed, for the first time, without the need of phenomenological infrared parameters.
\end{abstract}

\maketitle

\section{Introduction}\label{sec:Introduction}

A detailed understanding of the QCD phase structure at finite temperature and density are essential for our understanding of the formation of matter and the evolution of the universe. It is also necessary for interpreting data collected at running and planned heavy-ion experiments as well as further predictions, for reviews see
\cite{Luo:2017faz, Adamczyk:2017iwn, Andronic:2017pug, Stephanov:2007fk, Andersen:2014xxa,  Shuryak:2014zxa, Pawlowski:2014aha, Roberts:2000aa, Fischer:2018sdj, Yin:2018ejt}.

In the present work we access the QCD phase structure within a generalised functional approach, that combines Dyson-Schwinger equations (DSE) and the functional renormalisation group (fRG). In \cite{Cyrol:2017ewj} the fRG flows of vertices and propagators in unquenched $2$-flavour QCD have been solved within a quantitative approximation, and in particular the quark-gluon vertex has been fully resolved. The solution of the gap equation of the quark propagator solely requires the knowledge of the quark-gluon vertex and the gluon propagator. These correlation functions carry the full information about confinement and chiral symmetry breaking, and their quantitative computation within functional methods has been the subject of many works in the past two decades, for fRG works see e.g.\ \cite{Braun:2008pi, Braun:2009gm, Fister:2011uw, Mitter:2014wpa, Braun:2014ata, Rennecke:2015eba, Fu:2016tey, Rennecke:2016tkm, Cyrol:2016tym, Cyrol:2017ewj, Cyrol:2017qkl, Fu:2018qsk, Fu:2019hdw, Leonhardt:2019fua, Braun:2019aow, Braun:2020ada}, for DSE works see e.g.\ \cite{Roberts:2000aa, Qin:2010nq, Fischer:2011mz, Fischer:2013eca, Fischer:2014ata, Eichmann:2015kfa, Gao:2015kea, Gao:2016qkh, Tang:2019zbk, Fischer:2018sdj, Gunkel:2019xnh, Isserstedt:2019pgx, Reinosa:2015oua, Reinosa:2016iml, Maelger:2017amh, Maelger:2018vow, Maelger:2019cbk, Aguilar:2016lbe, Aguilar:2017dco, Aguilar:2018epe}. For related lattice studies see e.g.\ \cite{Bazavov:2012vg, Borsanyi:2013hza, Borsanyi:2014ewa, Bonati:2015bha, Bellwied:2015rza, Bazavov:2017dus, Bazavov:2017tot, Bonati:2018nut, Borsanyi:2018grb, Bazavov:2018mes, Guenther:2018flo, Ding:2019prx,Borsanyi:2020fev}.
	
Functional investigations of the gluon sector have revealed a very interesting structure. While a quantitative access to the gluon propagator in Yang-Mills theory without any phenomenological input requires very elaborate approximations, see \cite{Cyrol:2016tym}, its modification in the presence of dynamical quarks converge already in rough  approximations to the matter fluctuations: the  dominating effect of the matter fluctuations by far is the change in the momentum scale running in the ultraviolet which is already captured very well by perturbation theory. This property is also reflected in the fact that the gluon propagator does not change significantly if changing the pion mass from the chiral limit to masses of about 400\,MeV, see \cite{Cyrol:2017ewj}. Moreover, the light flavour quark-gluon vertex does not change significantly in the presence of a heavy strange quark: without flavour changing parts the presence of the strange quark in first order only changes the momentum running of the gluon propagator. Note that this approximate decoupling or rather separation of fluctuations changes in the chiral limit with a vanishing strange quark mass.	

In combination this provides us with an optimised expansion scheme of DSEs for QCD, based on the quantitative fRG results in \cite{Cyrol:2017ewj}: All correlation functions except the quark propagator are expanded about the $2$-flavour case. This allows us to access correlation functions and the phase structure of QCD at baryon-chemical potential up to $\mu_B/T\lesssim 3$ with quantitative reliability, but the results extend beyond this regime. In the vacuum, at finite temperature, and at not too high baryon-chemical potential quantitative functional and lattice results serve as benchmark tests for our method.

\section{fRG-assisted Dyson-Schwinger equations}\label{sec:FRG-DSE}

Functional relations for QCD such as functional renormalisation group equations or Dyson-Schwinger equation are formulated as master equations for the effective action $\Gamma[\Phi]$ with the superfield $\Phi=(A_\mu,c,\bar c, q, \bar q)$. The derivatives of $\Gamma$ with respect to the fields,
\begin{align}
\label{eq:Gn}
\Gamma^{(n)}_{\Phi_{i_1}\cdots \Phi_{i_n}}(p_1,...,p_n)=\frac{\delta\Gamma[\Phi]}{\delta\Phi_{i_1}(p_1)\cdots \delta\Phi_{i_n}(p_n)}\,,
\end{align}
 are the one-particle-irreducible (1PI) correlation functions and are derived by taking field derivatives of the fRG or DSE for the effective action. This leads to one-loop (fRG) or two-loop (DSE) exact relations between full 1PI correlation functions of QCD.

It is worth emphasising that the notion one- or two-loop has nothing to do with perturbation theory. The latter actually can be rederived by a recursive solution of the relations about the classical correlation functions. While being relations for the same set of correlation functions, in non-perturbative approximations they differ due to the different non-perturbative loop structure. Accordingly, an important respective self-consistency check is provided by agreeing results obtained in given approximations to different functional hierarchies. Moreover, their structure of closed one- or two loop relations between correlation functions allows to use correlation functions from external input such as other functional computations or lattice simulations, see e.g.\ \cite{Braun:2007bx, Fischer:2011mz, Fister:2013bh, Fischer:2013eca, Fischer:2014ata, Eichmann:2015kfa, Fischer:2018sdj, Gunkel:2019xnh, Isserstedt:2019pgx, Reinosa:2015oua, Reinosa:2016iml, Maelger:2017amh, Maelger:2018vow, Maelger:2019cbk, Aguilar:2016lbe, Aguilar:2017dco, Aguilar:2018epe}. By now, the lattice approach to Landau gauge QCD has produced quantitatively reliable vacuum ghost, gluon and quark propagators with small statistical errors as well as gluon propagators at finite temperature. Results for full dynamical QCD at finite temperature and physical quark masses are still lacking. Moreover, while impressive progress has been made in the computation of vertices, so far these computations lack the statistical precision required for a quantitative input.

In turn, functional computations in vacuum QCD have by now reached a quantitative level of precision not only for the propagators but also for vertices such as the quark-gluon vertex. In the present work we use the quark-gluon vertex and gluon propagator fRG data of vacuum $2$-flavour QCD at physical quark masses from \cite{Cyrol:2017ewj} as an external input in the DSEs for propagators and vertices in $2$- and $2+1$-flavour QCD at finite temperature and density.

\begin{figure}[t] %[hdbp]
	\includegraphics[width=0.9\columnwidth]{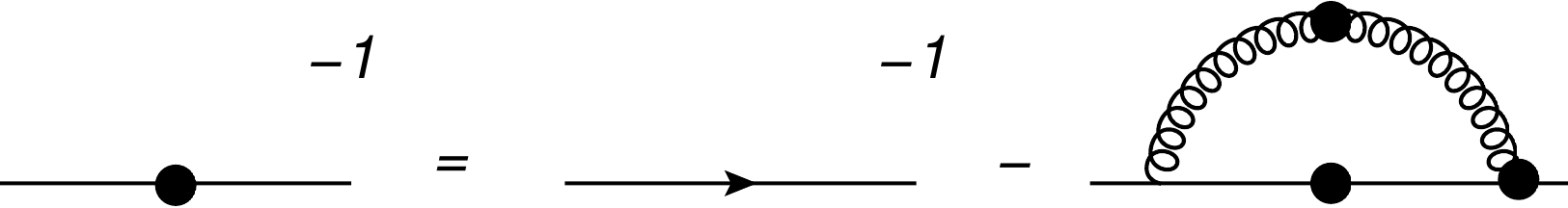}
	\caption{Quark gap equation. The lines with a blob are the full propagators, that without is the classical one. Vertices with blobs are full vertices, those without are  classical ones. }\label{fig:QuarkDSE}
\end{figure}

We pursue slightly different strategies for the correlation functions involved. The quark propagator is computed from its DSE, the gap equation, see \fig{fig:QuarkDSE} for all temperatures and densities. For the gluon propagator and the quark-gluon vertex we expand the respective DSEs about the fRG results of vacuum $2$-flavour QCD from \cite{Cyrol:2017ewj}: the DSE for the inverse gluon propagator $\Gamma_{AA}^{(2)}(p)$ and the quark-gluon vertex $\Gamma^{(3)}_{q \bar q A}(p_1, p_2)$ are reformulated as one for the difference between the vacuum two-flavour gluon propagator and quark-gluon vertex and the full gluon propagator and quark-gluon vertex. For the gluon propagator this DSE is solved, while we employ an ansatz for the strange, thermal and density fluctuations for the quark-gluon vertex. The latter is built on regularity relations between the different tensor structures in the vertex as well as the Slavnov-Taylor identities (STI).

\subsection{Quark gap equation and chiral phase transition}\label{sec:GapEquation}

The quark gap equation relates the inverse quark propagator $\Gamma^{(2)}_{q\bar q}$ to its classical counter part $S^{(2)}_{q\bar q}$, to the quark and gluon propagators, and the classical and full quark-gluon vertex, see \fig{fig:QuarkDSE}. In the vacuum we write
\begin{align}\label{eq:quark2point}
\Gamma^{(2)}_{q\bar q}(p) = Z_q(p^2)\left[\imag\,\slash{\hspace{-.193cm}p} + M_q(p)\right]\,,
\end{align}
using the notation of \cite{Cyrol:2017ewj}. In \eq{eq:quark2point} the Dirac tensor structure is proportional to the quark wave function renormalisation $Z_q(p)$. Its RG-scale or momentum derivative is the quark anomalous dimension. In turn, the scalar dressing $M_q(p)$ is the quark mass function and carries the information of chiral symmetry breaking.

 Finite temperature and density singles out a rest frame and we parameterise,
\begin{align}\label{eq:quark2pointTmu}
\Gamma^{(2)}_{q\bar q}(\tilde p) = Z_q(\tilde p) \left[  \frac{Z^{\lt}_q(\tilde p)}{ Z_q(\tilde p) }\Bigl[  \gamma_0 \,\imag \tilde p_0 +  M_q(\tilde p)\Bigr]+
\vec \gamma \,\imag \vec p \right]\,,
\end{align}
with the wave function renormalisation for the mode parallel to the rest frame, $Z^{\lt}_q$. The overall factor $Z_q=Z^{\bot}_q$ is the wave function renormalisation for the transverse modes perpendicular to the rest frame. We have pulled out the transverse wave function renormalisation since the transverse modes carry more weight in the DSE loop integrals: within our construction of the thermal and density fluctuations of the quark-gluon vertex used in the DSE, see \Sec{sec:DSEQuarkGluon}, we shall use a uniform wave function renormalisation $Z_q$. This implies the approximation $ Z^{\lt}_q/Z_q\approx 1$ in the vertices.

The quark momentum $\tilde p$ includes the chemical potential via $\tilde p_0 = p_0- i/3 \mu_B$. \Eq{eq:quark2pointTmu} straightforwardly extends to general correlation functions \eq{eq:Gn} with $\tilde p_{i_j}$ with $j=1,...,n$ that carry the chemical potential of the respective field $\Phi_{i_j}$,  for example,
\begin{align}
\label{eq:tildep}
\tilde p= \left(p_0 - i \,s_{\Phi_i}\mu_B\,, \,{\bf p}\right)\,.
\end{align}
The factor $s_{\Phi_i}$ is the baryon number of the component field $\Phi_i$ of the superfield $\Phi=(A_\mu,c,\bar c, q, \bar q)$. We have $s_q=1/3$, $s_{\bar q}=-1/3$, and $s_{\Phi_i}=0$ for $\Phi_i\neq q,\bar q$. Note also that the correlation functions at finite density and/or finite temperature also carry a genuine dependence on the baryon-chemical potential.

 \Eq{eq:quark2pointTmu} signifies the mass function $M_q(p)$ as the 'pole mass' function of the quark as the full mass function is $Z_q^\lt\,M_q$. Alternatively, we could have defined the full mass function $Z_q\,M_q$. Then $M_q$ would be the 'screening mass' function since $Z_q=Z^\bot_q$.  The quark DSE reads

\begin{eqnarray}\nonumber
\hspace{-.2cm}\Gamma^{(2)}_{q\bar q}(\tilde p)-S^{(2)}_{q\bar q}(\tilde p)
&=& \sumint\frac{dq_0}{2\pi} \! \int\frac{d^3{q}}{(2\pi)^3}\;   G_{AA,\mu\nu} (q+p) \quad  \\[1ex]
& & \hspace{-.3cm}\times \frac{\lambda^a}{2} {(-i\gamma_{\mu})} G_{q\bar q}(\tilde{q})
\Gamma^{(3)}_{q\bar qA}{}^a_\nu (\tilde q,-\tilde p)\, .
\label{eq:DSEq}\end{eqnarray}
In \eq{eq:DSEq} the quark and gluon propagators are given by $G_{q\bar q} = (1/\Gamma^{(2)})_{q\bar q}$ and $G_{AA}= (1/\Gamma^{(2)})_{AA}$ respectively, and all  momenta are counted as incoming. At finite temperature the frequency integrals turn into Matsubara sums with
\begin{align}
\!\! A_\mu, c,\bar c: \ \omega_n= 2 \pi T n\,,\quad q, \bar q:\  \omega_n=&\, 2 \pi T\left(  n+\frac12\right)\,.
\label{eq:Matsubara}
\end{align}
\Eq{eq:DSEq} can be solved with the knowledge of the gluon propagator $G_A$ and the quark-gluon vertex $\Gamma_{q\bar q A}^{(3)}$. In \Sec{sec:DSEGluon} and \Sec{sec:DSEQuarkGluon} we detail how these correlation functions at finite temperature and density are computed on the basis of the vacuum two flavour fRG-data from \cite{Cyrol:2017ewj}.

\subsection{DSE for the strange quark, temperature, and density corrections to the gluon propagator}\label{sec:DSEGluon}

\begin{figure}[t] %[hdbp]
	\vspace{.3cm}
	
	\includegraphics[width=1\columnwidth]{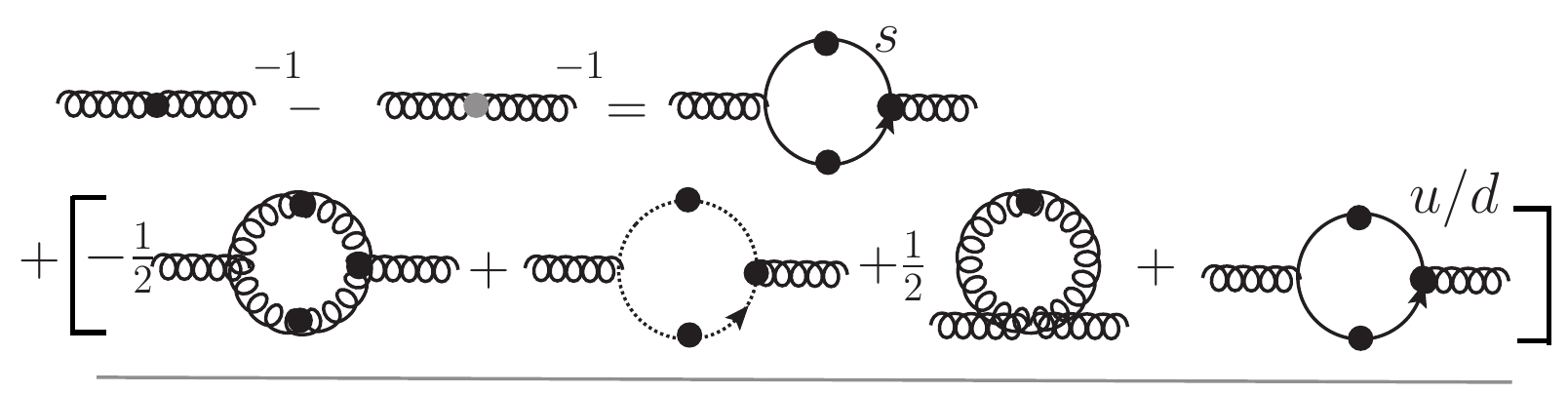}
	\vspace{-.2cm}
	
	\includegraphics[width=0.85\columnwidth]{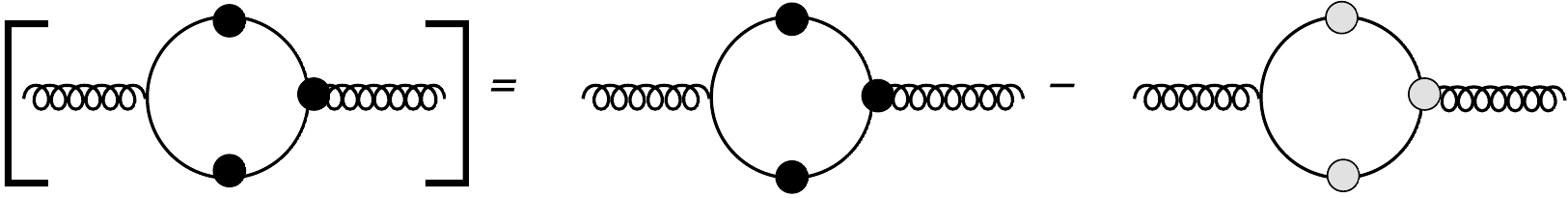}
	\caption{Gluon DSE for the difference $\Delta\Gamma_{AA}^{(2)}$ between the full gluon propagator and the vacuum $2$-flavour gluon propagator. Lines and vertices with black blobs are the full propagators and vertices at finite temperature and density. Lines and vertices with grey blobs are full vacuum propagators and vertices for $N_f=2$. The square bracket contains the temperature and density fluctuations and does not require renormalisation. We have dropped the part of the classical gluon propagator that carries the renormalisation of the strange loop. }\label{fig:DSEDeltaGluon}
\end{figure}

The DSE for the inverse gluon propagator for general flavours $N_f$ at finite temperature and density is expanded about that in the vacuum for two-flavour QCD. This reads schematically
\begin{align} \label{eq:GA}
\left. \Gamma^{(2)}_{AA}(p) \right|_{T,\mu_B, N_f}  =
\left. \Gamma^{(2)}_{AA}(p)\right|_{0,0,2} +
\Delta\Gamma^{(2)}_{AA}(p)\,,
\end{align}
and leaves us with a DSE for the difference $\Delta\Gamma^{(2)}_{AA}$ between the full inverse gluon propagator and that of vacuum $2$-flavour QCD, depicted in \Fig{fig:DSEDeltaGluon}.
At finite temperature and density the gluon two-point function has color-electric and color-magnetic components. Then, the two-point function $ \Gamma^{(2)}_{AA}(p)$ reads
\begin{align} \label{eq:GA1}
\Gamma^{(2)}_{AA}(p) =p^2\left[Z^M_A(p)\Pi^M_{\mu\nu}(p)+Z^E_A(p)\Pi^E_{\mu\nu}(p)
\right]+\frac{p_\mu p_\nu }{\xi}\,,
\end{align}
where we have separated the gauge-fixing part $p_\mu p_\nu /\xi$. In \eq{eq:GA1} we introduced the color-magnetic wave function renormalisation $Z^M_A$ and the color-electric one, $Z^E_A$. The projection operators onto the color-electric and color-magnetic directions are given by
\begin{align} \label{eq:GA2}
&\Pi^M_{\mu\nu}(p)=(1-\delta_{0\mu})(1-\delta_{0\nu})\left(\delta_{\mu\nu}-\frac{p_\mu p_\nu}{\vec{p}^2}\right)\,,\notag\\
&\Pi^E_{\mu\nu}(p)=\delta_{\mu\nu}-\frac{p_\mu p_\nu}{{p}^2}-\Pi^M_{\mu\nu}(p)\,.
\end{align}
 In the present work we only consider the Landau gauge, $\xi=0$. The loop integrals in the DSE are dominated by the color-magnetic components and in the following we will use the $O(4)$-symmetric approximation $Z_A^E(p) = Z_A^M(p)$. In the quark gap equation this approximation is reflected in the identification of the color-magnetic and electric running couplings $\alpha_s^E(p)= \alpha_s^M(p)$.

 Now we discuss the advantages of the split \eq{eq:GA} as well as generic strategies for its numerical solution. Since the propagator shows a subleading dependence on both the additional quark flavour (for $N_f=2+1$) as well as temperature and density, the DSE for $\Delta\Gamma_{AA}^{(2)}$ can e.g.\ be iterated starting with $\Delta\Gamma^{(2)}_{AA}=0$. The subdominant behaviour of changes in the quark look has been evaluated in \cite{Cyrol:2017ewj} with a study of the missing current-quark mass dependence in two-flavour QCD: the strange quark loop is quantitatively captured with its induced perturbative momentum and RG-scale running of the gluon propagator. The latter behaviour is already captured in simple approximations. This argument also carries over to the mild temperature and in particular very mild chemical potential dependence, see e.g.~\cite{Fu:2019hdw, Fischer:2018sdj, Gunkel:2019xnh}. These studies also provide further numerical justification for the approximation $Z_A^E(p) = Z_A^M(p)$.

Note also that the expansion in \eq{eq:GA} reduces the renormalisation subtleties in the vacuum. Moreover, if iterating about the respective vacuum result for $N_f=2$ and $N_f=2+1$, the temperature- and density independence of the renormalisation procedure is apparent in the finiteness of $\Delta\Gamma^{(2)}_{AA}$. A further numerical stabilisation is
achieved by separating the thermal sums and integrals in $\Delta\Gamma_{AA}^{(2)}$ as follows
\begin{align}\nonumber
&\hspace{-1cm} T \sum_{\omega_n} \textrm{loop}_{T,\mu_N}(\tilde q,p)-\int \frac{d \omega}{2 \pi}\textrm{loop}_{\textrm{vac}}(q,p) \\[1ex] \nonumber
=&
 \left[T \sum_{\omega_n}\textrm{loop}_{\textrm{vac}}(\tilde q,p) -\int \frac{d \omega}{2 \pi}\textrm{loop}_{\textrm{vac}}(q,p)\right]\\[1ex]
& + T \sum_{\omega_n}\Biggl[ \textrm{loop}_{T,\mu_B}(\tilde q,p)-\textrm{loop}_{\textrm{vac}}(\tilde q,p)\Biggr]  \,,
\label{eq:NumStab} \end{align}
with $q_0 =\omega_n$ at $T\neq 0$ and $q_0=\omega$ at $T=0$, and
$\textrm{loop}(q,p)$ stands for the loops in the second line of \fig{fig:DSEDeltaGluon}. Then, the second line
in \eq{eq:NumStab} encodes the direct effect of thermal and density fluctuations, namely substituting the $q_0$-integral by the Matsubara sum as well as introducing the chemical potential in the quark loop while keeping the vacuum correlation functions. Evidently this is finite and numerically stable. Moreover, it can be computed from the knowledge of the vacuum correlation functions only. The third line vanishes for $\Delta\Gamma^{(2)}_{AA}=0$, and the difference of loops in the square bracket decays faster than $\textrm{loop}(q,p)$ itself.
This results in a numerically stable second line which is only triggered by the direct effects of thermal and density fluctuations in the third line. Moreover, it is subleading for sufficiently small $T,\mu_B$.

 In the present work we utilise the quantitative smallness of the correction $\Delta\Gamma^{(2)}_{AA}$, and resolve it in a simplified approximation: First of all   the temperature dependence of ghost-gluon correlation function has been shown to be subdominant for the temperatures considered, both with functional methods and on the lattice , e.g.\  \cite{Cucchieri:2007ta,Ilgenfritz:2006he}. The dependence on the chemical potential is even more suppressed as it enters the DSEs for ghost-gluon correlations only via the chemical-potential dependence of the quark loops in gluonic correlations. Consequently, we approximate ghost propagator and the ghost-gluon vertex at finite temperature and density on the right hand side of \eq{eq:NumStab} with their vacuum counterparts in \cite{Cyrol:2017ewj}.

Moreover, we have approximated the vacuum three-gluon vertex with the classical one. In this approximation ghost-gluon contributions to the third line in \eq{eq:NumStab} vanish. This approximation is trustworthy as long as $\Delta\Gamma^{(2)}_{AA}$ remains a small perturbation. We have monitored this property in our explicit computation. Due to the finiteness of the momentum integrals this approximation is easily lifted which will be considered elsewhere.

\subsection{Quark-gluon vertex at finite $T$ and $\mu_B$}\label{sec:DSEQuarkGluon}
In this Section we derive an Ansatz for the strange quark, thermal and density corrections $\Delta\Gamma^{(3)}_{q\bar q A}$ of the quark-gluon vertex for $2$- and $2+1$-flavour QCD based on the $N_f=2$ flavour quark-gluon vertex in the vacuum computed in \cite{Cyrol:2017ewj}, see also \cite{Williams:2014iea, Williams:2015cvx, Aguilar:2016lbe, Aguilar:2018epe}. The strange-quark vertex in the vacuum is identified with the two-flavour vertex: for perturbative momentum scales with $p^2/m_s^2\ll 1$ the mass-dependence is sub-leading and for infrared momenta the symmetry-breaking part of the mass function dominates. Hence, similarly to the parameterisation of the gluon two-point function \eq{eq:DSEq} we expand the vertex about the two-flavour vertex in the vacuum,
\begin{align}
\left.\Gamma^{(3)}_{q\bar q A}(p_1,p_2)\right|_{ T,\mu_B, N_f}\!\! =\left.\Gamma^{(3)}_{q\bar q A}(p_1,p_2)\right|_{0, 0, 2}\!\! +\Delta\Gamma^{(3)}_{q\bar q A}(p_1,p_2)\,,
\label{eq:GqbarqA}\end{align}
for both the light quarks, $q=l$, and the strange quark, $q=s$. Note that in the present work we assume isospin symmetry with $m_l=m_u=m_d$, see \eq{eq:mpi-mq2} for two-flavours and \eq{eq:mpi-mq2+1} for 2+1 flavours. At finite temperature and density the vertex has color-electric and magnetic components. We identify them within our $O(4)$-symmetric approximation, which is consistent with the $O(4)$-symmetric approximation of the gluon two-point function.

\begin{figure}[t] %[hdbp]
	\vspace{.3cm}
	\includegraphics[width=0.95\columnwidth]{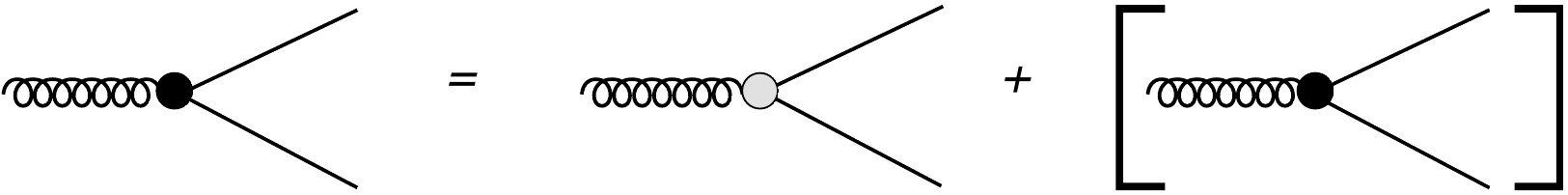}
	\caption{Quark-gluon DSE for the difference $\Delta\Gamma_{q\bar q A}^{(3)}$ between the full quark-gluon vertex (with black blob) and the vacuum $2$-flavour quark-gluon vertex (with grey blob). As in \Fig{fig:DSEDeltaGluon} the temperature and density fluctuations in the square bracket do not require renormalisation.}\label{fig:DSEDeltaQuarkGluon}
\end{figure}

\Eq{eq:GqbarqA} leaves us with the task of solving the Dyson-Schwinger equation for the difference of flavours and temperature/chemical potential, depicted in \Fig{fig:DSEDeltaQuarkGluon}. Evidently, for small corrections $\Delta \Gamma^{(3)}_{q\bar q A}$ this could be easily solved within an iteration about $\Delta \Gamma^{(3)}_{q\bar q A}=0$. For the gluon DSE this is described explicitly in \Sec{sec:DSEGluon}. Moreover, if neglecting subleading flavour-changing processes the change of the quark-gluon vertex for $N_f=2+1$ only arises from the change in the momentum running of the vertices and propagators, no new diagram is present.

In the present work we refrain from computing the corrections $\Delta\Gamma_{q\bar q A}^{(3)}$, their computation will be presented elsewhere. Here we follow a strategy that reduces the computational costs significantly: We construct vertices by utilising firstly non-trivial relations between the dressings of the different tensor structures of the quark-gluon vertex, secondly its regularity at momenta $p^2\gtrsim 1$\,GeV, and thirdly the STIs for the dressings of the different tensor structures. Apart from the reduced numerical costs this construction guarantees the quantitative compatibility of the running of the different 'avatars' of the strong couplings derived from the purely gluonic, ghost-gluon and quark-gluon vertices, see \cite{Cyrol:2017ewj}: small deviations in the perturbative and semi-perturbative momentum runnings of the respective couplings have a huge effect on the chiral symmetry breaking scale.

Within this construction the vertex dressings of all tensor structures of the quark-gluon vertex can be expressed in terms of quark, gluon and ghost dressings and the dressing of the ghost-gluon vertex. This allows us to make use of the hierarchy for the latter dressings in terms of their $T,\mu_B$ and $s$-quark dependence: While ghost-gluon correlations are effectively independent of $T,\mu_B$ and $s$-quark fluctuations, pure gluon correlations show a mild $T$ dependence. The dependence of pure gluon correlations on the strange current-quark mass $m_s^0$ is negligible. This argument extends to the $\mu_B$-dependence of pure gluon correlation, which is also subleading. Consequently, only the quark dressings show a significant $T,\mu_B$-dependence, that cannot be neglected in the computation of the quark-gluon vertex. The above hierarchy also entails, that to leading order differences between the light-quark--gluon vertex, $\Gamma_{l\bar l A}^{(3)}$,  and the strange-quark--gluon vertex $\Gamma_{s\bar s A}^{(3)}$ are solely induced by differences in the quark dressings. This supports the self-consistent and quantitative nature of the present approximation scheme.

The idea behind this construction can be nicely illustrated by discussing the quark-gluon coupling $\alpha_{q\bar q A}(\bar p)$ and the ghost-gluon coupling $\alpha_{c\bar cA}(\bar p)$
given by
\begin{align}
\label{eq:alphas}
\alpha_{q\bar q A}(\bar p)= \frac{1}{4 \pi} \frac{\left(\lambda_{\bar q q A}^{(1)}(\bar p)\right)^2}{Z_A(\bar p)Z_q^2(\bar p)}\,,\
\alpha_{c\bar c A}(\bar p)= \frac{1}{4 \pi} \frac{\left(\lambda_{\bar c c A}^{(1)}(\bar p)\right)^2}{Z_A(\bar p)Z_c^2(\bar p)}\,,
\end{align}
Here, $\lambda_{\bar q q A}^{(1)}\,,\,\lambda_{\bar c c A}^{(1)}$ are the dressings of the classical tensor structures of the respective vertices. For the classical action we have $\lambda^{(1)}_{q\bar q A}=g=\lambda^{(1)}_{c\bar c A}$. Furthermore, we identify the wave function renormalisations of all quark flavours, $Z_q=Z_l=Z_s$ due to their very mild dependence on the quark mass. This implies that in the present approximation differences in the vertex dressing for $\Gamma^{(3)}_{l\bar l A}$ and $\Gamma^{(3)}_{s\bar s A }$ are only triggered by the quark mass functions consistent with our discussion above. In \eq{eq:alphas} the vertices are evaluated at the symmetric point
\begin{align}\label{eq:SymPoint}
\bar p^2=\frac{p^2+q^2+k_+^2}{3}\,, \quad \textrm{with} \quad k_\pm =  p\pm q\,,
\end{align}
where $p$ is the incoming quark (ghost) momentum, $q$ is the incoming anti-quark (anti-ghost) momentum and $k_+$ is the incoming gluon momentum. Moreover, $Z_A(\bar p)$ and $Z_c(\bar p)$ are gluon and ghost dressing function, for more details on the notation see
\cite{Cyrol:2017ewj}.

The two couplings in \eq{eq:alphas} are related with STIs and two-loop universality. They agree to a quantitative degree due to rather trivial quark-gluon scattering kernels for perturbative and semi-perturbative momenta $\bar p^2 \gtrsim 3$\, GeV. There we have
\begin{align} \label{eq:alphaID}
\lambda_{\bar q q A}^{(1)}(\bar p) \approx Z_q(\bar p) \,\frac{\lambda_{\bar c c A}^{(1)}(\bar p)}{Z_c(\bar p)}\,,
\end{align}
see \Fig{fig:STI-Regular} in \App{app:STI}.
\Eq{eq:alphaID} is a rather interesting relation: the ghost-gluon dressing $\lambda_{\bar c c A}$ and the ghost wave function renormalisation $Z_c(\bar p)$ are effectively independent of the strange quark, temperature and in particular of density fluctuations. For $T$ and $\mu$ this reads
\begin{align} \label{eq:ghost0}
\left(\frac{\partial}{\partial\mu_B} \frac{\lambda_{\bar c c A}^{(1)}(\bar p)}{Z_c(\bar p)} \ , \  \frac{\partial}{\partial T}  \frac{\lambda_{\bar c c A}^{(1)}(\bar p)}{Z_c(\bar p)} \ , \frac{\partial}{\partial m_s^0}  \frac{\lambda_{\bar c c A}^{(1)}(\bar p)}{Z_c(\bar p)}\right) \approx 0\,,
\end{align}
with the strange current-quark mass $m^0_s$.
The very mild temperature and current-quark mass dependence has been checked explicitly with functional methods and lattice simulations. In turn, the chemical potential dependence of ghost correlation functions is only triggered very indirectly: The explicit chemical potential dependence of the quark propagator triggers a very mild one in gluon correlations which then feeds into the ghost correlations.

In summary, for the temperatures and chemical potentials relevant for the phase structure we can rely on vacuum ghost-gluon correlations. As an important consequence the whole strange-quark, temperature and chemical potential dependence of the quark-gluon dressing $\lambda^{(1)}_{q\bar q A}$ of the classical tensor structure in the perturbative and semi-perturbative regime stems from the factor $Z_q(\bar p)$ in \eq{eq:alphaID}. Using \eq{eq:GqbarqA} for the dressing $\lambda^{(1)}_{q\bar q A}$ leads us to the general parametrisation
\begin{align} \label{eq:Deltalambda1}
\lambda^{(1)}_{q\bar q A}=(\lambda^{(1)}_{q\bar q A})^{(\textrm{in})}+\Delta\lambda^{(1)}_{q\bar q A}\,,
\end{align}
where the superscript $\lambda^{(\textrm{in})}$ indicates the input data, here the 2-flavour QCD dressing in the vacuum.
With \eq{eq:alphaID} the correction $\Delta\lambda^{(1)}_{q\bar q A}$ is given by
\begin{align}\label{eq:Delta1Sym}
\Delta\lambda^{(1)}(\bar p) = \tilde \lambda^{(1)}_{q\bar q A}(\bar p)\left[Z_q - Z^{(\textrm{in})}_q\right](\bar p)\,,
\end{align}
with
\begin{align}
 \tilde \lambda^{(1)}_{q\bar q A}(\bar p)=
\left(\frac{\lambda^{(1)}_{q\bar q A}(\bar p)}{Z_q(\bar p)}\right)^{(\textrm{in})}\,.
\end{align}
Hence, the strange quark, thermal and density dependence of $\lambda^{(1)}_{q\bar q A}$ is entirely carried by $Z_q$. Note that $\tilde \lambda^{(1)}_{q\bar q A}$ carries a non-trivial momentum dependence also for $\bar p\gtrsim 1$\,GeV as we refrained from dividing out the ratio $\lambda^{(1)}_{c\bar c A}/Z_c$. As discussed above, this ratio has a negligible $s,T,\mu_B$-dependence, see \eq{eq:ghost0}, and in the present approximation we simply use the vacuum values. Then it is a global prefactor in the difference and can be
absorbed in $\tilde \lambda^{(1)}_{q\bar q A}$ as done in \eq{eq:Delta1Sym}.

The above derivation sets the stage for the discussion of the full quark-gluon vertex. We discuss regularity and STI-constraints for the different dressings, which finally provide us with a vertex ansatz. To that end we expand the full quark-gluon vertex in a full tensor basis which contains twelve elements. In \cite{Cyrol:2017ewj} the full transversely projected quark-gluon vertex has been written as follows,
\begin{align}\nonumber
\Pi^\bot_{\mu\nu}(k_+)\Gamma^{(3)}_{q\bar q A,\nu} (p, q)=& \sum_{i=1}^8 \lambda^{(i)}_{q\bar q A}(p,q) \\[1ex]
 & \hspace{.2cm}\times\Pi^\bot_{\mu\nu}(k_+)\left[{\cal T}^{(i)}_{q\bar q A}\right]_\nu (p,q)\,,
\label{eq:FullGqbarqA}\end{align}
with $k_+=p+q$, see \eq{eq:SymPoint}, and the transverse and longitudinal projection operators
\begin{align}\label{eq:Projections}
\Pi^\trans_{\mu\nu}(k)=&\, \delta_{\mu\nu}-\0{k_\mu k_\nu}{k^2}\,,
\qquad \Pi^{\lt}_{\mu\nu}(k) = \0{k_\mu k_\nu}{k^2}\,.
\end{align}
The tensor basis ${\cal T}_{q\bar q A}^{i}$ in \eq{eq:FullGqbarqA} is given by
\begin{align}
\begin{array}{lcl}
\left[{\cal T}^{(1)}_{q\bar q A}\right]^\mu(p,q) =-i \gamma^{\mu}\,, &\qquad&
\left[{\cal T}^{(5)}_{q\bar q A}\right]^\mu(p,q) =i{\kslash}_+ k_-^\mu\,,\\[2ex]
\left[{\cal T}^{(2)}_{q\bar q A}\right]^\mu(p,q) =k_-^\mu\,, &\qquad&
\left[{\cal T}^{(6)}_{q\bar q A}\right]^\mu(p,q)  = i\kslash_- k_-^\mu\,,\\[2ex]
\left[{\cal T}^{(3)}_{q\bar q A}\right]^\mu(p,q) = {\kslash_-}\gamma^\mu\,, &\qquad&
\left[{\cal T}^{(7)}_{q\bar q A}\right]^\mu(p,q) =\frac{i}{2} [\pslash,\qslash]\gamma^\mu\,, \\[2ex]
\left[{\cal T}^{(4)}_{q\bar q A}\right]^\mu(p,q) =\kslash_+\gamma^\mu\,, &\qquad&
\left[{\cal T}^{(8)}_{q\bar q A}\right]^\mu(p,q) = -\frac12[\pslash,\qslash] k_-^\mu\,.
\end{array}
\label{eq:TensorsQuarkGluon}\end{align}
$\lambda^{(1)}_{q\bar q A}$ is the dressing of the classical tensor structure.
The basis \eq{eq:TensorsQuarkGluon} can be derived from quark--anti-quark and gluon derivatives of the gauge invariant operators
$\bar q \dr q\to{\cal T}^{(1)}_{q\bar q A}$, $\bar q \dr^2 q\to{\cal T}^{(2-4)}_{q\bar q A}$, $\bar q \dr^3 q\to{\cal T}^{(5-7)}_{q\bar q A}$ and $\bar q \dr^4 q\to{\cal T}^{(8)}_{q\bar q A}$, see \cite{Mitter:2014wpa, Cyrol:2017ewj}. The tensors ${\cal T}^{(1,5-7)}_{q\bar q A}$ are chirally symmetric, while ${\cal T}^{(2-4,8)}_{q\bar q A}$ break chiral symmetry.

\begin{figure}[t] %[hdbp]
	\includegraphics[width=0.9\columnwidth]{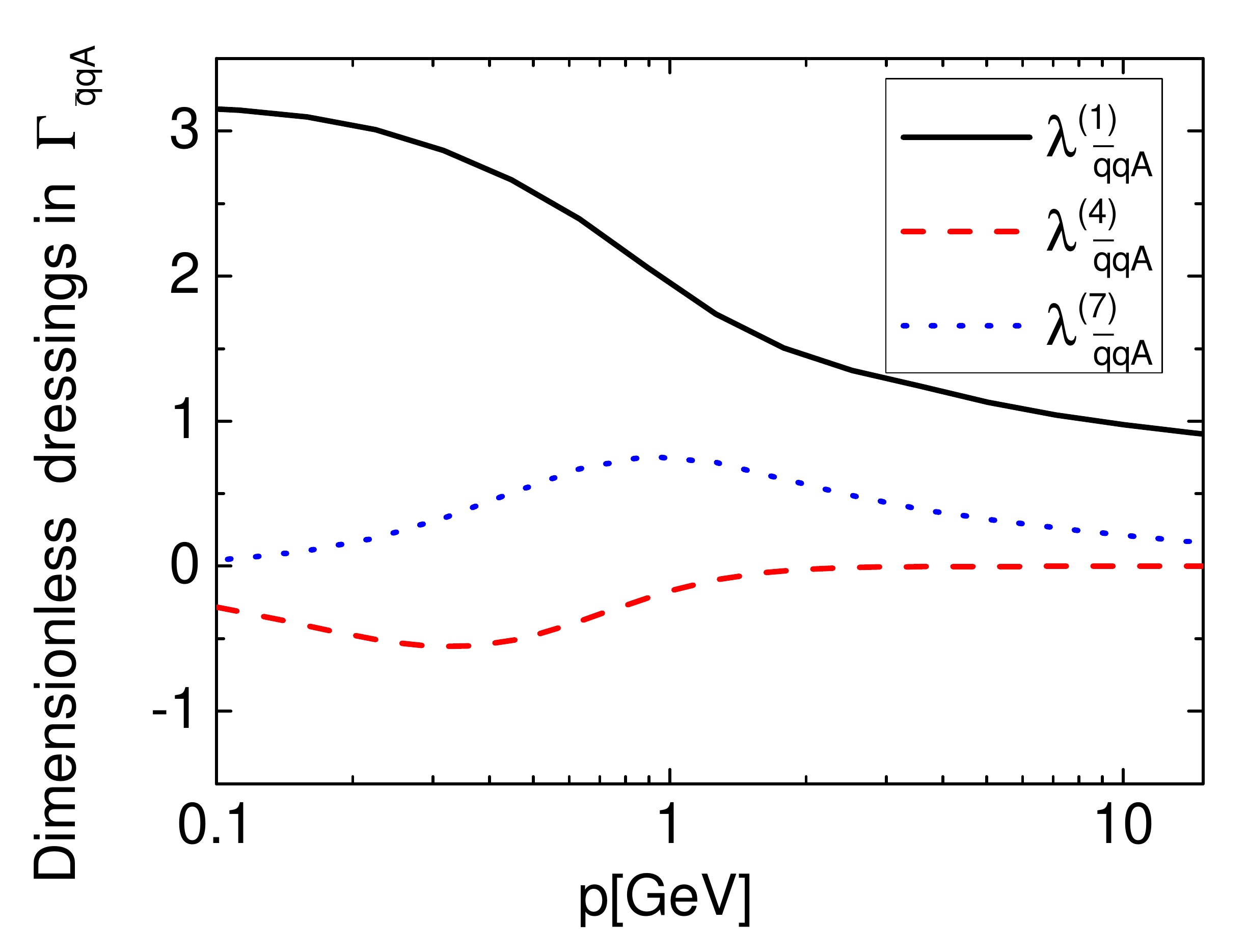}
	\vspace{-.1cm}
	
	\caption{The angular averages, defined in \eq{eq:thetaAv}, of the dimensionless dressings for the three dominant dressings  $\lambda^{1,4,7}_{q\bar q A}(p)$ of the tensors ${\cal T}^{(1,4,7)}_{q\bar q A}$ defined in  \eq{eq:TensorsQuarkGluon}.}\label{fig:Dressing}
\end{figure}
In \cite{Mitter:2014wpa, Cyrol:2017ewj} is has been also  observed that for perturbative and semi-perturbative momenta the different dressings $\lambda^{(i)}_{q\bar q A}$ of those tensor structures obtained from the same operator are indeed related at the symmetric point. Moreover, in \cite{Mitter:2014wpa, Cyrol:2017ewj} it has been found that the dressings $\lambda^{(i)}_{q\bar qA}$ only show a mild angular dependence. For similar considerations in a basis with defined $C$-parity see \cite{Maris:1999nt, Williams:2014iea, Williams:2015cvx, Eichmann:2016yit}. This allows us to employ angular averages
of the input dressings,
\begin{align}
\label{eq:thetaAv}
\lambda^{(i)}_{q\bar q A}(p^2,q^2,\theta)\to \bar{\lambda}^{(i)}_{q\bar q A}(p^2,q^2)=\frac{1}{\pi}\int_\theta\lambda^{(i)}_{q\bar q A}(p^2,q^2,\theta)\,,
\end{align}
for the transverse dressings $i=1,...,8$. The longitudinal dressings do not contribute to the gap equation. The average \eq{eq:thetaAv} simplifies the numerical computations, and also allows for a more direct access to the physics mechanisms at work. The averaged dressings for three dominant components $\lambda_{q\bar q A}^{(1,4,7)}$  are shown in \Fig{fig:Dressing}.

The derivations and discussions are deferred to \App{app:STI}. Here we emphasise the important results. First of all the dominant dressings are given by $\lambda^{(1,4,7)}_{q\bar q A}$, while $\lambda^{(3,8)}_{q\bar q A}\approx 0$. Moreover, for momenta $\bar p\gtrsim 1$\,GeV we have
\begin{align} \label{eq:256}
\lambda^{(2)}_{q\bar q A} \approx \frac{\lambda^{(4)}_{q\bar q A}}{2}\,,\quad \lambda^{(5)}_{q\bar q A} \approx \frac{\lambda^{(7)}_{q\bar q A}}{2}\,,
	\quad 	\lambda^{(6)}_{q\bar q A} \approx \frac{\lambda^{(7)}_{q\bar q A}}{12}\,,
\end{align}
which leaves us with the task of writing $\Delta\lambda^{(1,3,4,7,8)}_{q\bar q A}$ in terms of  quark, gluon and ghost dressings and the dressing of the ghost-gluon vertex with the help of the STIs. The validity of the latter procedure is checked in the vacuum, see \App{app:STI}. We are led to
\begin{align}\nonumber
\Delta\lambda^{(1,5,6,7)}_{q\bar q A}(p,q)= &\,\tilde\lambda^{(1,5,6,7)}_{q\bar q A}\left[ \Sigma_{Z_q}-\Sigma^{\textrm{(in)}}_{Z_q}\right](p,q) \,, \\[1ex]\nonumber
\Delta\lambda^{(2,3,4)}_{q\bar q A}(p,q)= &\,\tilde\lambda^{(2,3,4)}_{q\bar q A}\Bigl[ Z_A^{1/2}\Delta_{Z_qM_q}\\[1ex] \nonumber
& \hspace{1.2cm}-( Z_A^{1/2} \Delta_{Z_qM_q})^{\textrm{(in)}}\Bigr](p,q) \,,\\[1ex]
\Delta\lambda^{(8)}_{q\bar q A}(p,q)= &\,\tilde\lambda^{(8)}_{q\bar q A}\left[ \Delta_{Z_q M_q}-\Delta^{\textrm{(in)}}_{Z_q M_q}\right](p,q) \,,
\label{eq:Delta1-8}\end{align}
with
\begin{align}\nonumber
 \tilde\lambda^{(1,5,6,7)}_{q\bar q A}(p,q)= &\,\left(\frac{\lambda^{(1,5,6,7)}_{q\bar q A}(p,q)}{\Sigma_{Z_q}(p,q)} \right)^{(\textrm{in})}\,, \\[1ex]\nonumber
\tilde\lambda^{(2,3,4)}_{q\bar q A}(p,q)= &\, \left(\frac{\lambda^{(2,3,4)}_{q\bar q A}(p,q)}{Z_A^{1/2}(p+q)\Delta_{Z_q M_q}(p,q)}\right)^{(\textrm{in})}\,, \\[1ex]
\tilde\lambda^{(8)}_{q\bar q A}(p,q)= &\, \left(\frac{\lambda^{(8)}_{q\bar q A}(p,q)}{
	\Delta_{Z_q M_q}(p,q)}\right)^{(\textrm{in})}\,.
\label{eq:tildel1-8}\end{align}
In \eq{eq:Delta1-8} and \eq{eq:tildel1-8} we have used the abbreviations
\begin{align}\label{eq:SigDel}
\Sigma_X=&\,X(p^2)+X(q^2)\,,\quad
\Delta_X=&\,\frac{X(p^2)-X(q^2)}{p^2-q^2}\,,
\end{align}
for dressings factors that originate from the quark legs of the vertex.  The dressings $\tilde\lambda_{q\bar q A}^{(i)}$ with $i=1,...,8$, partially supplemented with trivial momentum dependences and $\lambda^{(1)}_{c\bar c A}$, are depicted in \Fig{fig:Vertexdressings} and \Fig{fig:AllVertexdressings} in \App{app:STI} at the symmetric point. Within our approximation with identical dressings for all quarks, $Z_q=Z_l=Z_s$, the quark-gluon vertex dressings $\lambda^{(1,5,6,7)}$ are flavour-independent while $\lambda^{(2,3,4,8)}$ carry a -mild- flavour-dependence due to $\Delta_{Z_q M_q}$ in $\Delta\lambda^{(2,3,4,8)}$, see  \eq{eq:Delta1-8}.

This concludes our construction of an STI-compatible quark-gluon vertex at finite temperature and density on the basis of the $2$-flavour quark-gluon vertex in the vacuum. We emphasise that the construction is general, and is easily adapted to generic input data not only from the fRG but also from other functional approaches as well as the lattice.

\begin{figure}[t] %[hdbp]
	\includegraphics[width=0.45\textwidth]{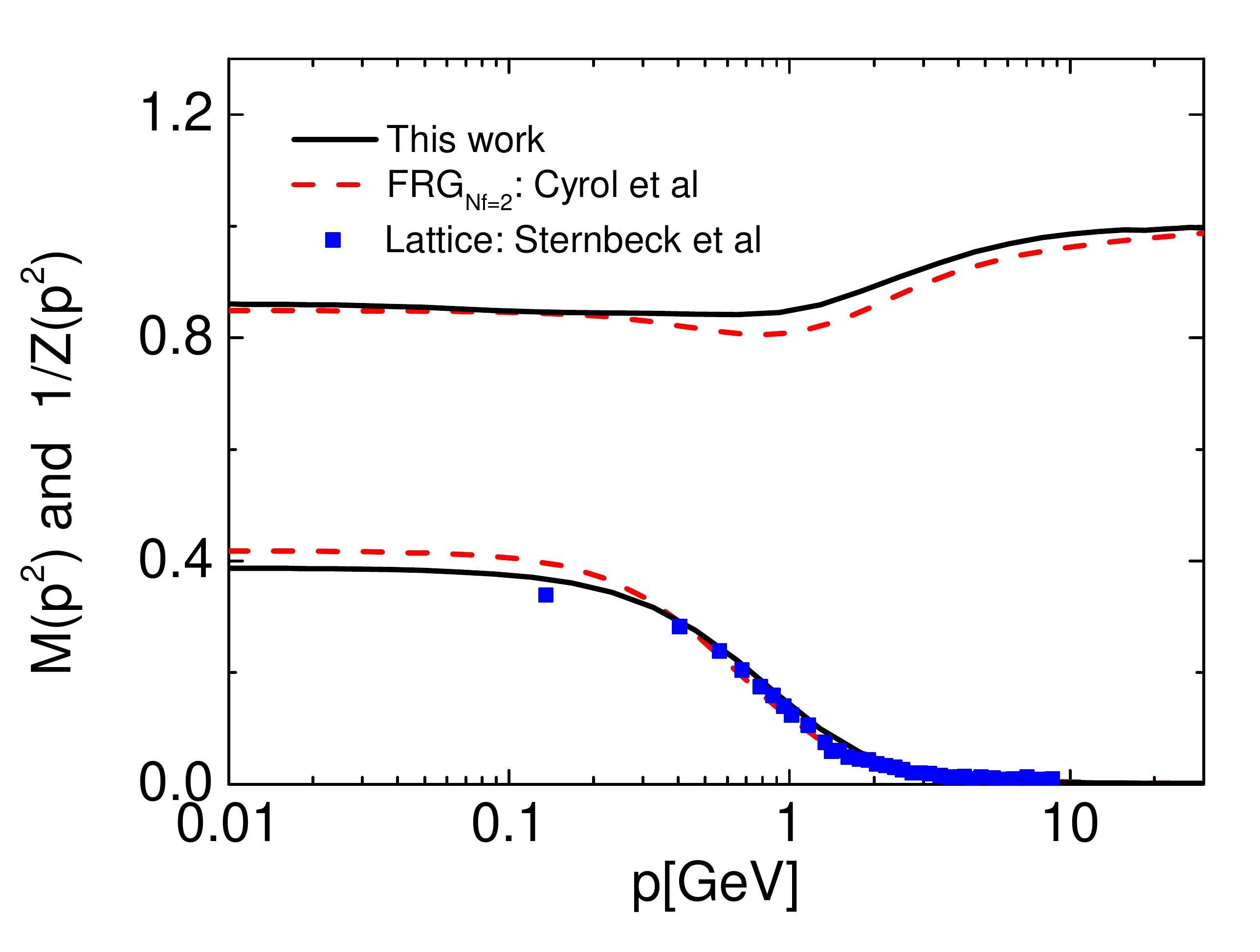}
	\caption{$2$-flavour quark propagator $M(p^2)$  and $Z(p^2)$ in comparison to the fRG results, \cite{Cyrol:2017ewj}, as well as lattice results, \cite{Sternbeck:2012qs}, for $M(p^2)$.}\label{fig:QuarkPropNf2}
\end{figure}
\section{Determination of the fundamental parameters of QCD}\label{sec:PhysParameters}
It is left to determine the fundamental parameters of QCD, the current-quark masses. In the present work we assume isospin symmetry
with identical up and down quark current masses: $m_{u/d}^0=m_l^0$.
Then the current masses are determined by computing suitable observables.
Convenient choices are the vacuum pion decay constant $f_\pi$ or the vacuum pion pole mass $m_\pi$. In the present work we adjust the pion pole mass. Moreover, in the $2+1$-flavour case we also adjust the strange quark mass $m_s^0$ with the relation $m_s^0/m_l^0\approx 27$.
In summary this leads us to
\begin{align}
\label{eq:ml-ms}
m_\pi(m_l^0) = 140\,\textrm{MeV}\,,\quad \quad N_f=2+1:\ \frac{m_s^0}{m_l^0}=27\,,
\end{align}
with the pion decay constant as a first prediction. Also, the quark mass function $M_q(p)$ can be compared with the respective fRG and lattice results. We emphasise that no
phenomenological infrared parameter such as the infrared vertex strength is adjusted, the current setup has the full predictive power of a first principle approach.

\begin{align}\nonumber
f_\pi =&\,\frac{4N_c}{N_\pi}\int_p \frac{M^r_q}{Z_q(p^2+M_q^2)^2}\left[M_q  -\frac{p^2}{2}\frac{\partial M_q}{\partial p^2}\right]\,,\\[2ex]
m^2_\pi=&\,\frac{8N_c}{ f^2_\pi}\int_p \frac{m_l^0 M^r_q}{Z_q(p^2 +M_q^2)}\,,
\label{eq:fpi+mpi}\end{align}
with the normalisation $N_\pi$ of the Bethe-Salpeter wave function of the pion,
\begin{align}\nonumber
N^2_\pi=&\, f_\pi N_\pi\\[1ex]
&\, +2\,N_c\int_p \frac{ (M^r_q)^2 \,\left( p^2\, Z_q Z_q''+2 Z_q\,Z_q'-p^2 Z_q'' \right)
}{Z^2_q(p^2+M^2_q)}\,,
\label{eq:Normpion}\end{align}
where $Z_q'(p^2) = \partial_{p^2}Z_q(p^2)$ and $ Z_q''(p^2) = \partial_{p^2}^2 Z_q(p^2)$ and $M^r_q(p^2)=M(p^2)-m_l^0\frac{\partial M(p^2)}{\partial m_l^0}$. The relations in \eq{eq:fpi+mpi} are reductions of the exact relations with additional form factor that are set to unity in \eq{eq:fpi+mpi}. In particular, the approximation \eq{eq:fpi+mpi} leads to an underestimation of the pion decay constant. In summary, with \eq{eq:fpi+mpi} we can compute the pion mass and decay constant from the
quark propagator. This allows us to determine the respective current-quark masses $m_{u/d}^0=m_l^0$.

\begin{figure}[t] %[hdbp]
	\includegraphics[width=\columnwidth]{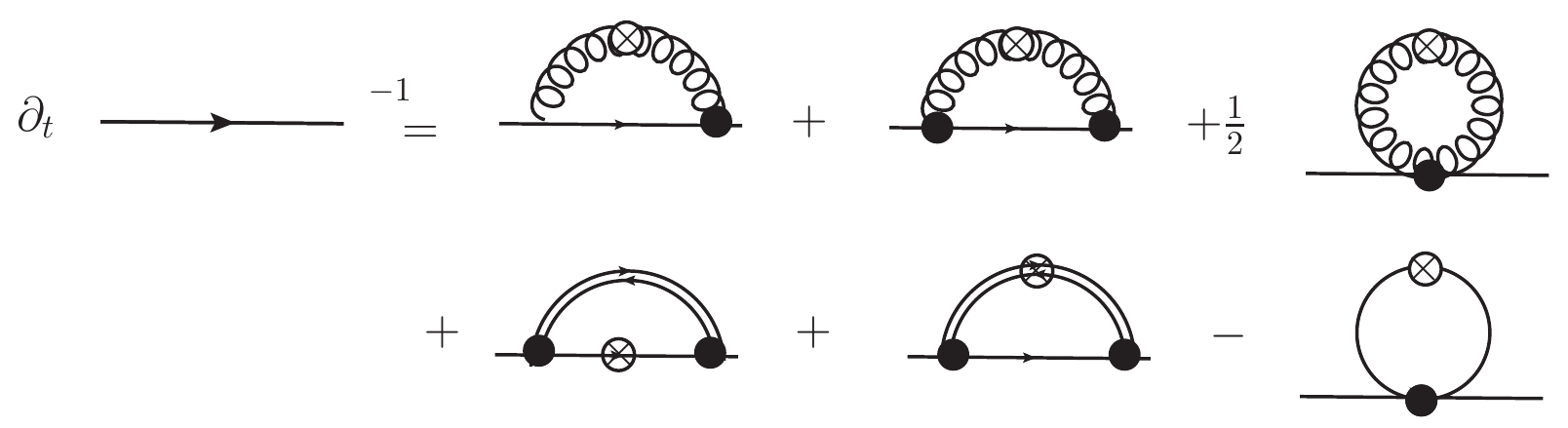}
	\caption{Flow equation for the quark two-point function. All propagators and vertices are full ones, indicated with the blobs. The $\tilde\partial_t$-derivative is taken at fixed $\Gamma_k^{(n)}$, that is $\tilde\partial_t \Gamma^{(n)}_k =0$. It only hits the explicit cutoff dependence in the regulator $R_k(p)$, the latter being a shift in the (classical) mass functions of the fields, $m^2_{\Phi_i}\to m^2_{\Phi_i}+R_{\Phi_i,k}(p)$, for more details see e.g.\ \cite{Pawlowski:2005xe, Cyrol:2017ewj, Fu:2019hdw}. The double lines stand for scalar-pseudoscalar degrees of freedom introduced with dynamical hadronisation, \cite{Gies:2001nw, Gies:2002hq, Pawlowski:2005xe, Floerchinger:2009uf, Braun:2014ata, Mitter:2014wpa, Cyrol:2017ewj, Fu:2019hdw}.}\label{fig:QuarkfRG}
\end{figure}
The pion mass $m_\pi$ and decay constant $f_\pi$ can be computed from the
quark propagator, see e.g.\ \cite{Bender:1997jf,Gao:2017gvf}, for reviews see \cite{Fischer:2006ub, Bashir:2012fs, Eichmann:2016yit}. We use the Pagels-Stoker approximation for the pion decay constant, and the Gell-Mann--Oakes--Renner relation for the pion mass,

This concludes our setup for fRG-assisted DSE computations described in \Sec{sec:FRG-DSE} and the present section \Sec{sec:GapEquation}.  Now we determine the $2$-flavour current-quark masses with \eq{eq:ml-ms}, using \eq{eq:fpi+mpi}. This leads us to the quark propagator depicted in \Fig{fig:QuarkPropNf2} and
\begin{align}
\label{eq:mpi-mq2}
 m_l^0 = 2.5\,\textrm{MeV}: f_\pi=92\,\textrm{MeV}\,,\quad M_q(0)=387\, \textrm{MeV}\,.
\end{align}
With $N_\pi= 88$\,MeV the difference between the pion decay constant and the normalisation of the pion wave function is rather small: $(f_\pi- N_\pi)/f_\pi = 0.043$. For large differences the approximations used in \eq{eq:fpi+mpi} lack reliability, while a small difference serves as a self-consistency check.

An important consistency check is given by the momentum dependence of the quark mass function $M_q(p)$ and, to a lesser extent, by the quark wave function renormalisation $Z_q(p)$. Note that the latter has a relatively strong dependence on the renormalisation scheme. The result is shown in \Fig{fig:QuarkPropNf2}. Our result is consistent with the lattice data and the quark propagator from the fRG computation. In particular we find that the constituent quark mass at vanishing momentum, $M(p^2=0)=387$\,MeV in  \Fig{fig:QuarkPropNf2}, is slightly smaller than that in the fRG computation, and is rather close to the result from lattice QCD.

\begin{figure}[t] %[hdbp]
	\includegraphics[width=0.45\textwidth]{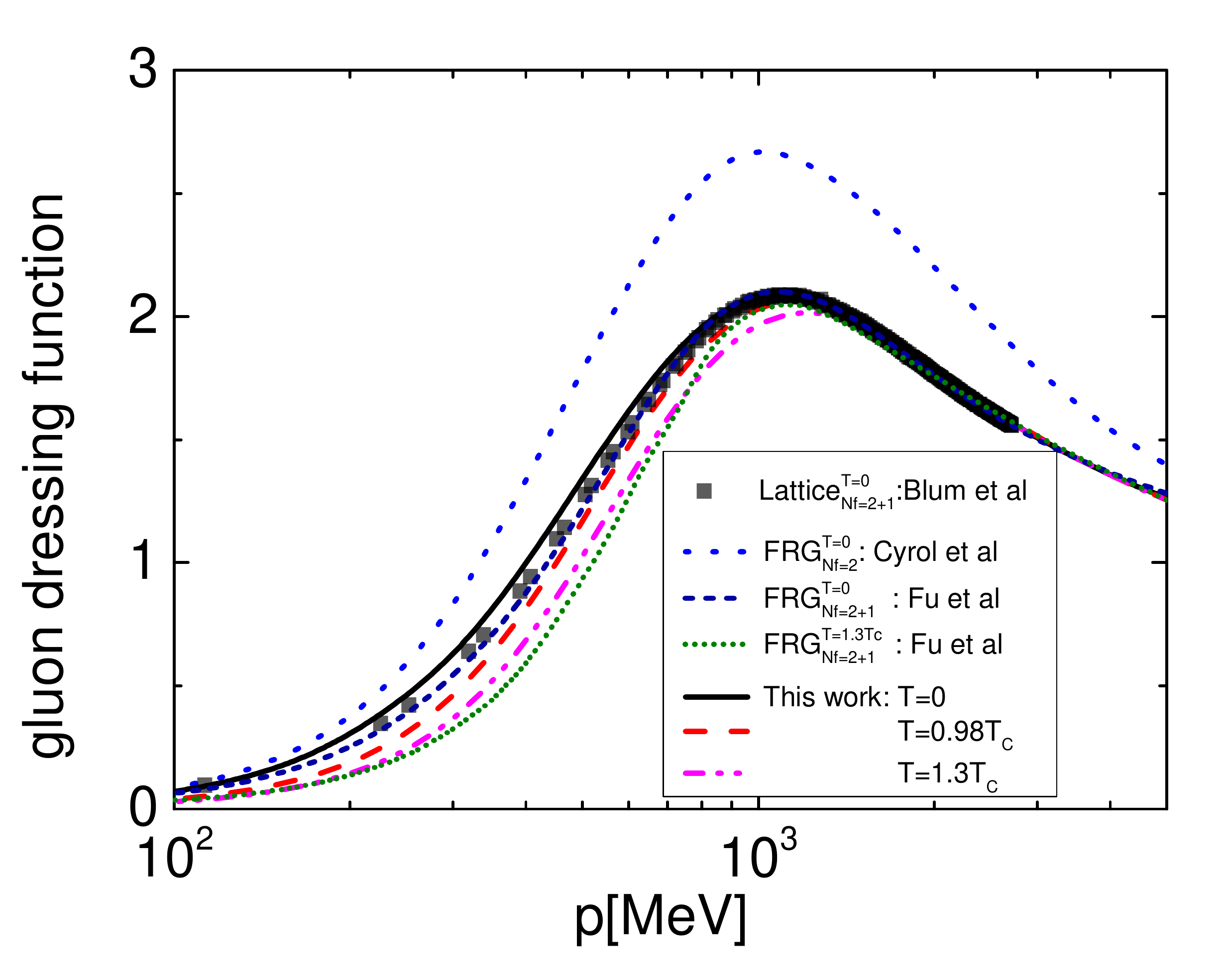}
	\caption{$2+1$-flavour  gluon dressing function $Z^M_A(p^2)$  at vanishing and finite temperature, computed from the $2$-flavour vacuum propagator from \cite{Cyrol:2017ewj}. We also display 2+1-flavour vacuum results from the fRG, \cite{Fu:2019hdw}, and the lattice, \cite{Blum:2014tka}, as well as fRG results at $T=1.3\,T_c$ from \cite{Fu:2019hdw}.}\label{fig:GluonProp2+1}
\end{figure}

In this context we emphasise that the resummation schemes for the DSE and the fRG are working rather differently. The former equation is depicted in \Fig{fig:QuarkDSE}, the latter one is depicted in \Fig{fig:QuarkfRG}, for more details see
\cite{Cyrol:2017ewj}. In comparison, the fRG equation consists  of more diagrams as well as
a different vertex structure: First of all, all vertices are dressed and in particular the
flow diagram with quark-gluon vertices has two dressed quark-gluon vertices. Moreover, the flow equation for the quark two-point function contains tadpole terms with the full two-quark--two-gluon scattering vertex, the two-quark--two-ghost scattering vertex and the quark four-point vertex that are absent in the respective DSE. In particular the latter tadpole is important for the strength of chiral symmetry breaking. It carries in particular resonant meson exchanges of the $\sigma$-mode and pions. In summary, the small fRG-DSE deviations are within the expected systematic error of the approximations used both in the fRG and the DSE, which is a respective non-trivial consistency check for both functional approaches.

\begin{figure}[t] %[hdbp]
	\includegraphics[width=0.45\textwidth]{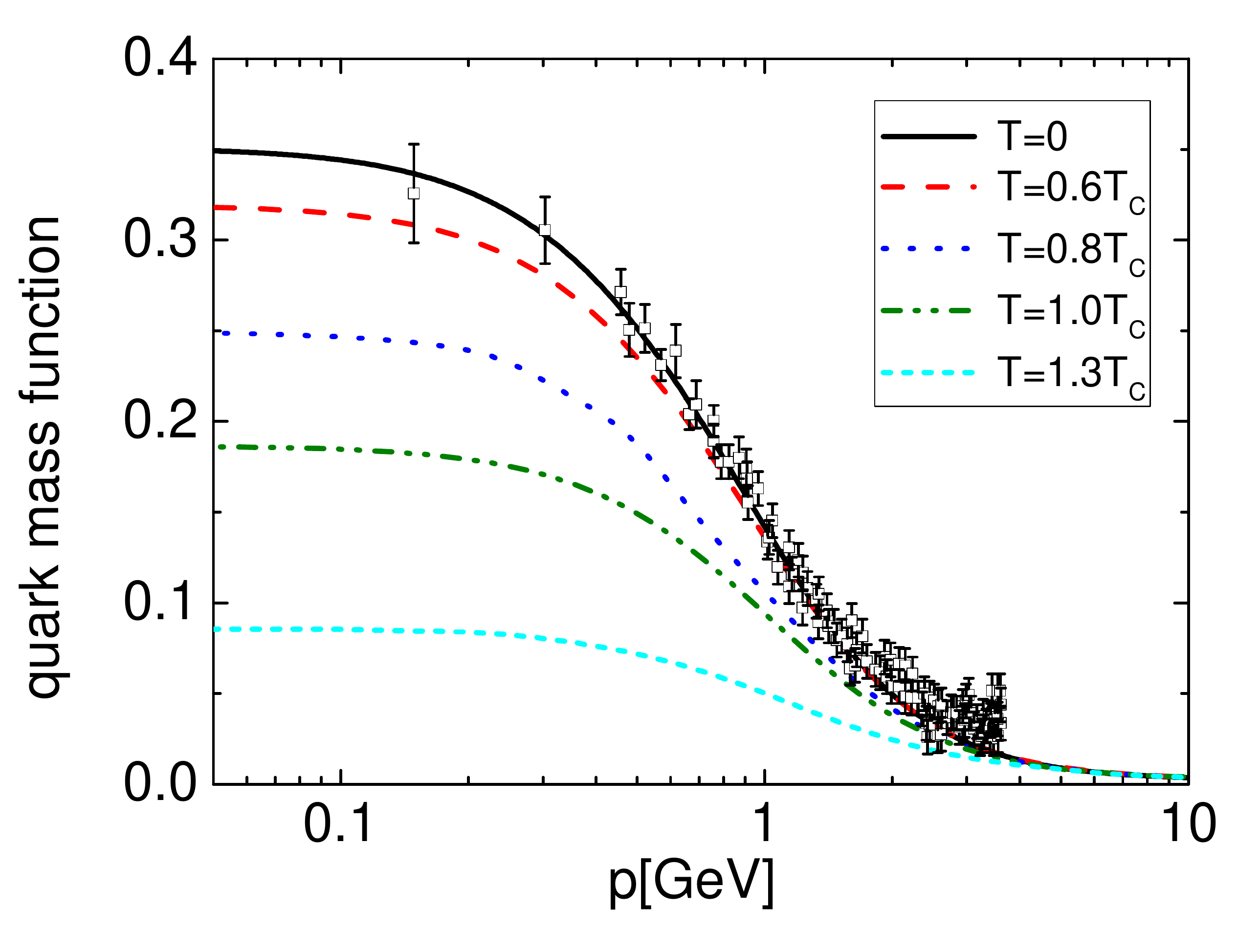}
	\caption{Mass function  $M(p^2)$ of u/d quark propagator
		for 2+1 flavour QCD at different temperatures. We also show the vacuum light quark propagator from  unquenched lattice simulations, \cite{Bowman:2005vx}, which is in quantitative comparison with the present DSE results.}\label{fig:QuarkProp2+1}
\end{figure}

We proceed with the $2+1$-flavour case. The results for the vacuum gluon propagator and quark propagator are depicted in \Fig{fig:GluonProp2+1} and \Fig{fig:QuarkProp2+1} together with finite temperature results that are discussed in the next \Sec{sec:Results}.
As in the $2$-flavour case we determine the current-quark masses with \eq{eq:ml-ms}, using \eq{eq:fpi+mpi},
\begin{align}
\label{eq:mpi-mq2+1}
m_l^0 = 2.7\,\textrm{MeV}: f_\pi=89\,\textrm{MeV}\,, \qquad M_q(0)=351\, \textrm{MeV}\,.
\end{align}
With $N_\pi= 86$\,MeV the difference between the pion decay constant and the normalisation of the pion wave function is rather small: $(f_\pi- N_\pi)/f_\pi = 0.034$. As in the $2$-flavour case this serves as a self-consistency check of the approximations in \eq{eq:fpi+mpi}. The small reduction of the pion decay constant in comparison to the full one with 93\,MeV is a well-known artefact of the approximations in \eq{eq:fpi+mpi}.

This completes the setup. We emphasise that in the current approach there is no phenomenological infrared parameter, the only input are the fundamental mass parameters of QCD: the light current-quark masses fixed with the pion pole mass, and, in the case of 2+1 flavours, the ratio of the light and strange current-quark masses. Accordingly, the quark propagator and in particular the pion decay constant $f_\pi$, the mass function $M_q(p)$ and the constituent quark mass $M_q(0)$ are already predictions within the current setup. Note also that we could have fixed the physical point with the pion decay constant, then the pion mass would have been a prediction. In summary this amounts to a fully first principle setup, we only have to determine the fundamental parameters and all observables are predictions in the current approach.

\section{Results and discussions}\label{sec:Results}

In this Section we present our results for $2$- and $2+1$-flavour QCD at vanishing and finite temperature and density. We depict our results for the physical case of $2+1$-flavour QCD, the $2$-flavour results show a similar behaviour. In \Fig{fig:GluonProp2+1}
and \Fig{fig:QuarkProp2+1} we show the 2+1-flavour gluon  and quark propagators in the vacuum and at finite temperature. Again the results are in quantitative agreement with the respective lattice results in the vacuum (gluon: \cite{Blum:2014tka,Abelev:2009ac,Zafeiropoulos:2019flq,Boucaud:2018xup}, quark: \cite{Oliveira:2018lln,Silva:2013foa}) and the functional results at vanishing and finite temperature (gluon and quark: fRG \cite{Fu:2019hdw}, DSE \cite{Fischer:2014ata, Fischer:2018sdj}). At finite temperature the gapping of the gluon propagator increases, consistent with the functional results in \cite{Fu:2019hdw, Fischer:2014ata, Fischer:2018sdj}, reflecting the increase of the thermal screening masses.

\begin{figure}[t] %[hdbp]
	\includegraphics[width=0.48\textwidth]{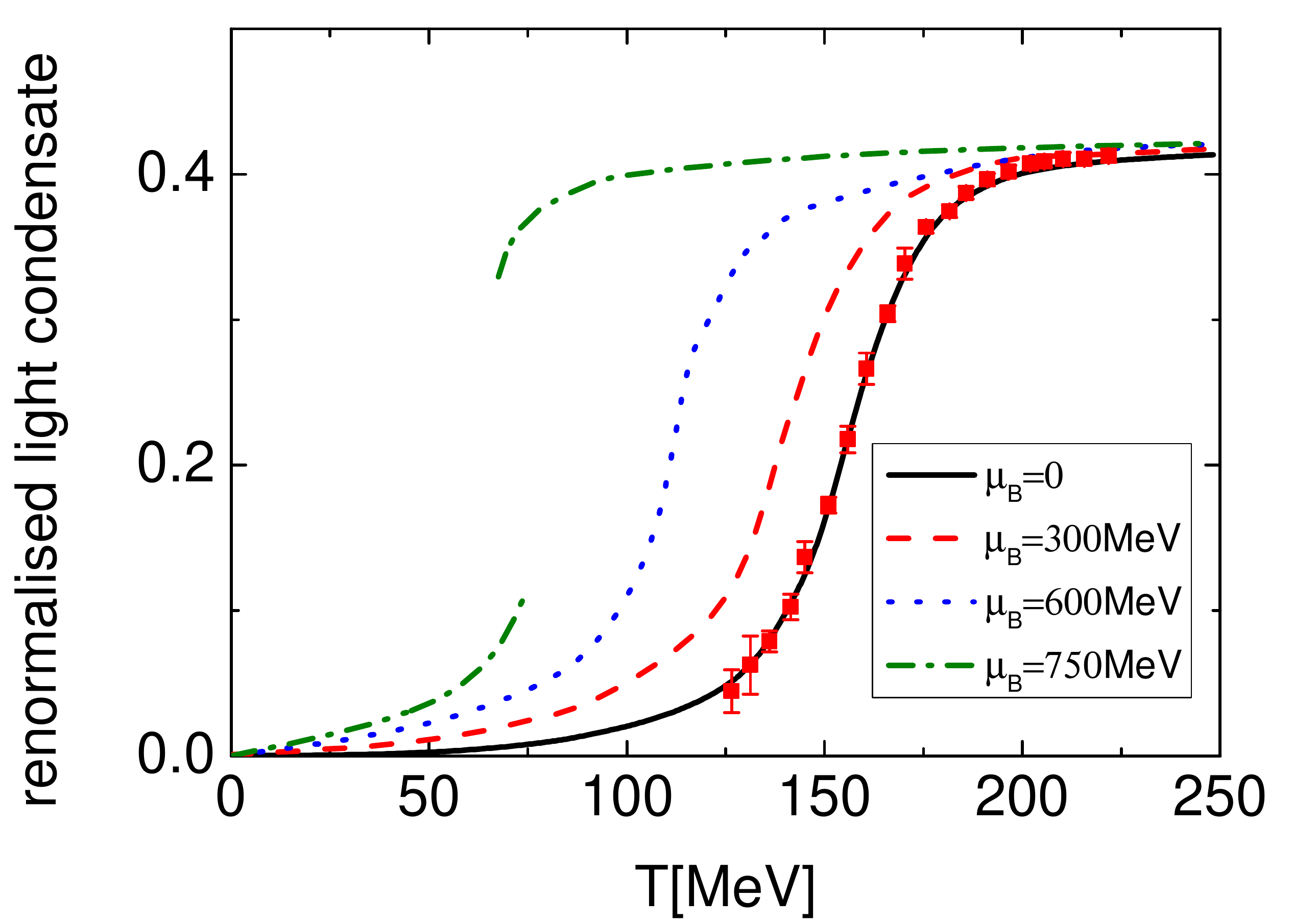}
	\caption{Renormalised light chiral condensate  for 2+1 flavour QCD as a function of temperature for different baryon-chemical potentials. The lattice data at vanishing chemical potential are from \cite{Borsanyi:2010bp}. The crossover steepens, finally leading to 2nd order CEP and a first order regime. A comparison with the fRG results in \cite{Fu:2019hdw} is done in \App{app:conden}. }\label{fig:conden}
\end{figure}

These results readily extend to finite baryon-chemical potential. We are interested in particular in the chiral phase structure at finite temperature and density. In the present work we employ the definition of the chiral transition temperature $T_c(\mu_B)$ given by the thermal susceptibility of the renormalised light chiral condensate $\Delta_{l,R}$. Up to renormalisation terms the quark condensate $\Delta_{q_i}$ is defined as
\begin{align}\label{eq:chiralcondG}
\Delta_{q_i}\simeq  - m_{q_i}^0
T\sum_{n\in\mathbb{Z}} \int \frac{d^3 q}{(2 \pi)^3}
\tr \,G_{q_i\bar q_i} (q)\,,
\end{align}
with $q_i=u,d,s$. The renormalised light chiral condensate comprises the thermal and density part of the chiral condensate. In particular, the renormalised light chiral condensate is given by
\begin{align}\label{eq:chiralcondren}
\Delta_{l,R} = \frac{1}{{2 \cal N}_R}\sum_{q=u,d}\Bigl[\Delta_{q}(T,\mu_B) -
\Delta_{q}(0,0)\Bigr]\,.
\end{align}
 Where ${\cal N}_R$ is a convenient normalization which leads $\Delta_{l,R}$ dimensionless, and here we choose ${\cal N}_R=m^4_\pi$. The subtraction of the vacuum condensate eliminates the necessity of explicitly discussing the renormalisation of \eq{eq:chiralcondG}. In \Fig{fig:conden} we show the renormalised light chiral condensate obtained from the current computation in comparison to lattice data at vanishing chemical potential, \cite{Borsanyi:2010bp}. \Fig{fig:conden} also contains results for different baryon-chemical potentials.
\begin{figure}[t] %[hdbp]
	\includegraphics[width=0.48\textwidth]{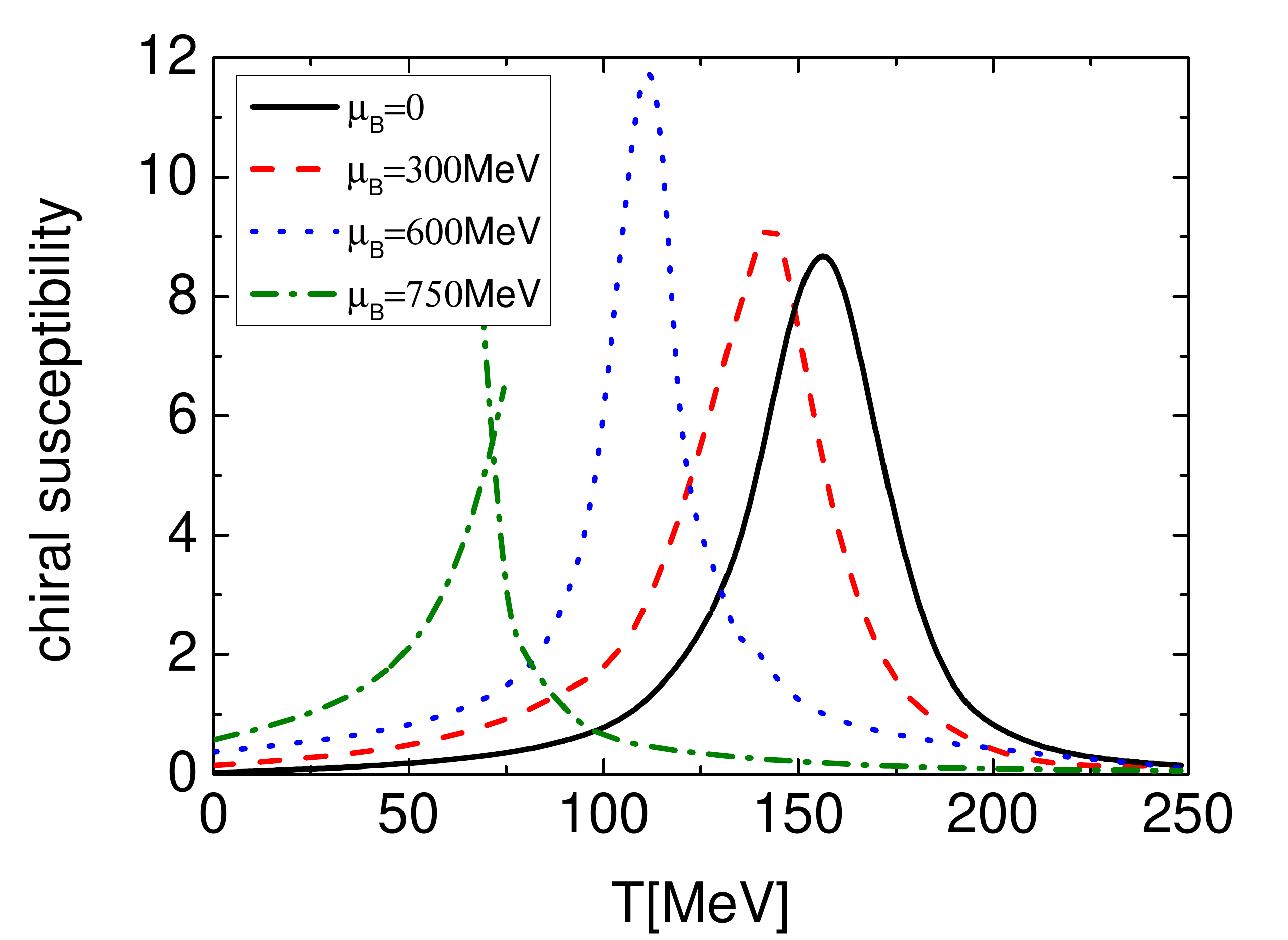}
	\caption{Thermal susceptibility of the renormalised light chiral condensate for 2+1 flavour QCD as a function of temperature for different baryon-chemical potentials. }\label{fig:ChiralSus}
\end{figure}
\begin{figure}[b] %[hdbp]
	\includegraphics[width=0.48\textwidth]{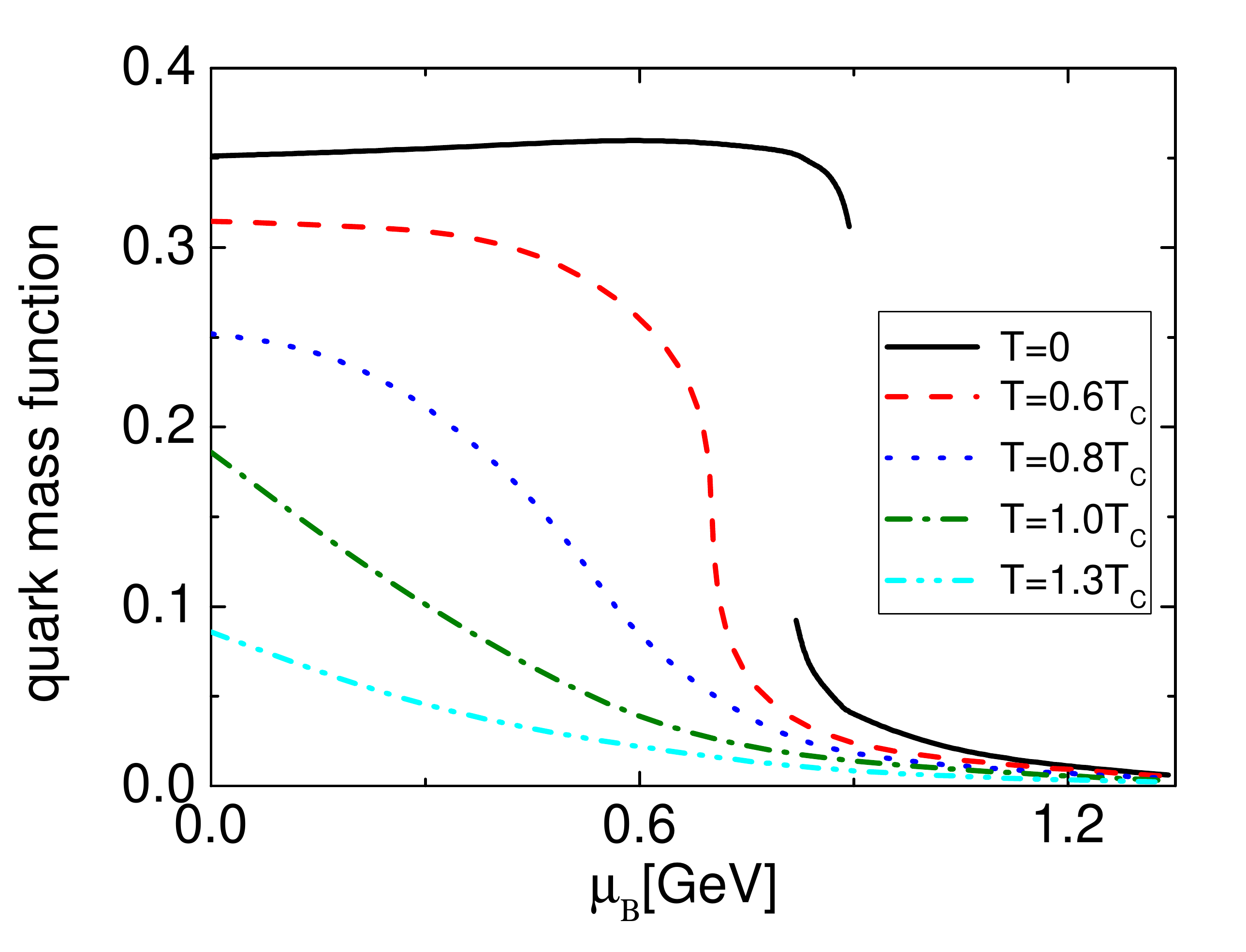}
	\caption{Constituent $u/d$ quark mass at zero momentum $M(0)$
		for 2+1 flavour QCD as a function of $\mu_B$ at different temperature. With temperature decreasing, the phase transition changes from crossover to first order.}\label{fig:QuarkProp2+1Tmu}
\end{figure}
The peak of the thermal susceptibility, $\partial_T \Delta_{l,R}$, provides the chiral transition temperature. The susceptibilities are shown in \Fig{fig:ChiralSus} for different baryon-chemical potentials. The crossover temperatures for $2$- and $2+1$-flavour QCD at vanishing chemical potential are computed as
\begin{align}
\label{eq:Tcmu=0}
T_{c,N_f=2}= 166\,\textrm{MeV}\,,\qquad T_{c,N_f=2+1}= 154\,\textrm{MeV}\,.
\end{align}

\begin{figure}[t] %[hdbp]
	\vspace{-.3cm}
	
	\includegraphics[width=1\columnwidth]{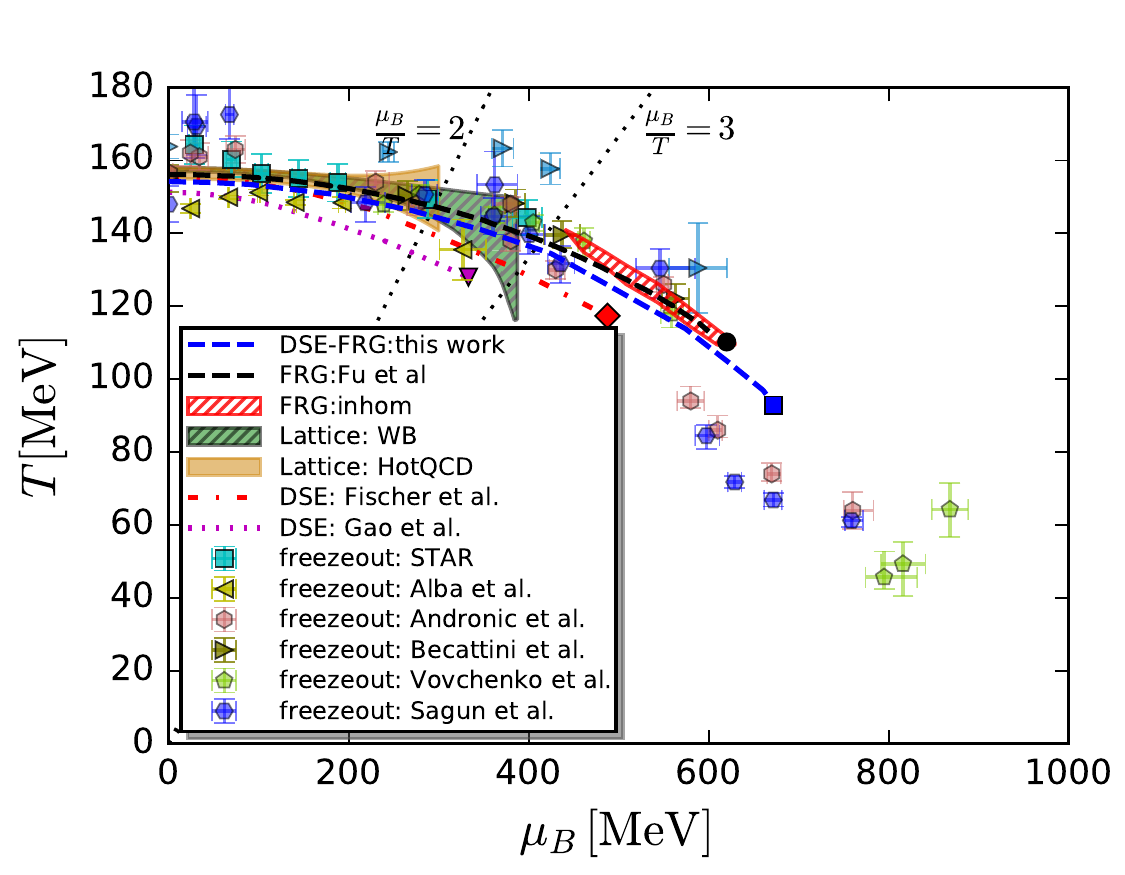}
	\caption{phase diagram for 2+1 flavour QCD in comparison to other theoretical results and phenomenological freeze-out data. The blue dashed line displays the results from the current work. The results agree well with the fRG and lattice results for small chemical potential. In the red hatched regime the fRG results from \cite{Fu:2019hdw} show a minimum of the pion dispersion at nonvanishing spatial momentum together with a sizable chiral condensate. This may indicate an inhomogeneous regime. For a comparison of the  curvature at small baryon-chemical potential see
		\Tab{tab:kappas}.\\
		\textit{Other theoretical results}: fRG for QCD with dynamical hadronisation \cite{Fu:2019hdw} (Fu {\it et al.}), lattice QCD with analytic
		continuation from imaginary chemical potential
		\cite{Borsanyi:2020fev} (WB), lattice QCD with Taylor
		expansion at vanishing chemical potential \cite{Bazavov:2018mes} (HotQCD), DSE
		approach with backcoupled quarks and a dressed vertex
		\cite{Fischer:2014ata} (Fischer {\it et al.}), and DSE calculations
		with a gluon model
		\cite{Gao:2015kea} (Gao {\it et al.}).\\[0.5ex]
		\textit{Freeze-out data}: \cite{Adamczyk:2017iwn} (STAR),
		\cite{Alba:2014eba} (Alba {\it et al.}), \cite{Andronic:2017pug}
		(Andronic {\it et al.}), \cite{Becattini:2016xct} (Becattini {\it et
			al.}), \cite{Vovchenko:2015idt} (Vovchenko {\it et al.}), and
		\cite{Sagun:2017eye} (Sagun {\it et al.}). Note that freeze-out data
		from Becattini {\it et al}.\ with (light blue) and without (dark
		green) afterburner-corrections are shown in two different
		colors.}\label{fig:finalphase}
\end{figure}

The chiral transition temperatures in \eq{eq:Tcmu=0} are in quantitative agreement with the lattice, \cite{Borsanyi:2010bp,Bazavov:2017dus,Cheng:2007jq}, and fRG results, \cite{Fu:2019hdw}.
In \App{app:conden} we also compare the temperature dependence of the renormalised light chiral condensate computed in the present work and that from  \cite{Fu:2019hdw} for different chemical potential.
For small chemical potential they are in quantitative agreement. In turn, they show significant differences for $\mu_B/T\gtrsim 3$, triggered by the different locations of the CEP in both computations, see \Fig{fig:finalphase}.

Now we proceed to the full phase structure at finite density. Before we present the respective results, we briefly discuss the self-consistency of our present approximation at finite baryon-chemical potential. At vanishing temperature and below the baryonic onset $\mu_B<\mu_B^*$, no correlation function has an explicit dependence on $\mu_B$,
\begin{align}
\label{eq:SilverBlaze}
\Gamma^{(n)}_{\Phi_1\cdots \Phi_n}(p_1,...,p_n;\mu_B)= \Gamma^{(n)}_{\Phi_1\cdots \Phi_n}(\tilde p_1,...,\tilde p_n;0)\,,
\end{align}
with $\tilde p$ defined in \eq{eq:tildep}. \Eq{eq:SilverBlaze} entails that the whole $\mu_B$-dependence of a correlation function is carried by the frequency arguments. This is the Silver Blaze property for correlation functions as discussed in \cite{Khan:2015puu,Gunkel:2019xnh}. In the present approximation we only consider the explicit $\mu_B$-dependence in $\gamma_0 \tilde p_0$ and ignore the above subtleties in the dressing functions $M_q(\tilde p^2)\approx M_q(p^2)$ and $Z_q(\tilde p^2)\approx Z_q(p^2)$. In \Fig{fig:QuarkProp2+1Tmu} we show the mass function $M_q(p^2=0)$, that is at
$\tilde p^2 = - \mu_B^2/9$. For $T=0$ the $\mu_B$-dependence is rather small and simply reflects the mild momentum dependence of the quark propagator for $0 \leq p \lesssim 0.5$\,GeV for momenta $p^2 <0$. Accordingly, the violation of the Silver Blaze property is very small.

\begin{figure}[t] %[hdbp]
	\includegraphics[width=0.95\columnwidth]{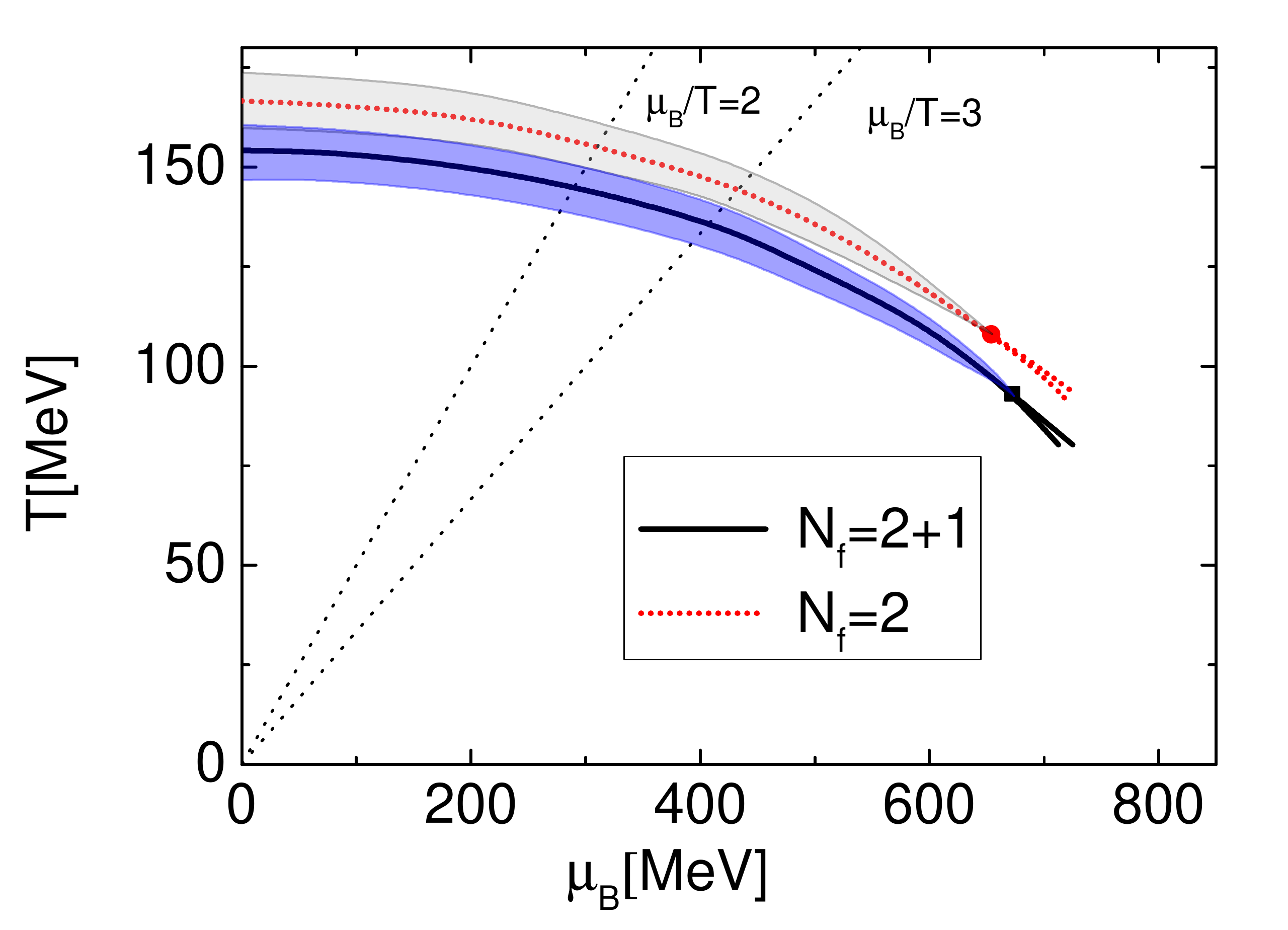}
	\caption{Phase diagram in the plane of the temperature and the
		baryon-chemical potential. The grey and blue bands denote the
		crossover transitions for the $2$- and 2+1 flavour QCD,
		respectively; and the red star and black circle are their relevant
		CEP. The bands are determined with 80$\%$ peak height of
		$ \partial \Delta_{l,R}/\partial T$ at fixed $\mu_B$. The black
		and red dashed lines depict the peak positions for $N_f=2+1$
		and $N_f=2$, respectively. We also depict the two lines with
		$\mu_B/T=2,3$ related to reliability bounds for both lattice and
		functional methods.  }
	\label{fig:phaseNf2-Nf2+1}
\end{figure}
This leads us to one of our main results, the phase structure of $2$- and $2+1$-flavour
QCD summarised in
\Fig{fig:phaseNf2-Nf2+1}. The transition temperature $T_c(\mu_B)$ in \Fig{fig:phaseNf2-Nf2+1} decreases with increasing  baryon-chemical potential. The curvature of the transition temperature line $T_c(\mu_B)$ at vanishing chemical potential
is determined within the expansion about $\mu_B=0$,
\begin{align}\label{eq:curv}
\frac{T_c(\mu_B)}{T_c}=1-\kappa\, \left(\frac{\mu_B}{T_c}\right)^2+\lambda\,\left(\frac{\mu_B}{T_c}\right)^4+\cdots\,,
\end{align}
with $T_c=T_c(\mu_B=0)$. From the results for the thermal susceptibility we are led to
\begin{align}\label{eq:kappaNf2}
\kappa_{N_f=2}=0.0179(8)\,,
\end{align}
in the two flavour case, and
\begin{align}\label{eq:kappaNf2+1}
\kappa_{N_f=2+1}=0.0150(7)\,.
\end{align}
Both results are compatible with the respective fRG results in \cite{Fu:2019hdw}. While the $2$-flavour  lattice results show a wide spread, the $2+1$-flavour lattice results for the curvature are in quantitative agreement with the present results. A comparison of
all results is provided in \Tab{tab:kappas}.

\begin{table}[t]
	\begin{center}
		\begin{tabular}{|c || c | c |}
			\hline  & & \\[-2ex]
			\backslashbox{reference}{curvature $\kappa$}  & $N_f=2$
			& $N_f=2+1$  \\[1ex]
			\hline &  & \\[-1ex]
			fRG-DSE:  this work   & 0.0179(8) & 0.0150(7)  \\[1ex]
			\hline &  & \\[-1ex]
			fRG:  \cite{Fu:2019hdw}   & 0.0176(1) & 0.0142(2)  \\[1ex]
			\hline &  & \\[-1ex]
			lattice: \cite{Borsanyi:2020fev}  &
			\noindent\rule{1cm}{0.4pt}
			& 0.0153(18)  \\[1ex]
			\hline &  &  \\[-1ex]
			lattice:  \cite{Bonati:2018nut}
			& \noindent\rule{1cm}{0.4pt} & 0.0144(26)
			\\[1ex]
			\hline &  &  \\[-1ex]
			lattice:  \cite{Bazavov:2018mes}
			& \noindent\rule{1cm}{0.4pt} & 0.015(4)
			\\[1ex]\hline &  &  \\[-1ex]
				DSE:  \cite{Gao:2015kea}
			& \noindent\rule{1cm}{0.4pt} & 0.038
			\\[1ex] \hline &  &  \\[-1ex]
			DSE:  \cite{Fischer:2012vc,Fischer:2014ata,Fischer:2018sdj}
			& 0.0456 & 0.0238
			\\[1ex]
			\hline && \\[-1ex]
			 fRG:  \cite{Braun:2019aow}   & 0.0051 & \noindent\rule{1cm}{0.4pt}  \\[1ex]
			\hline &  & \\[-1ex]
			lattice:  \cite{Allton:2002zi}
			& 0.0078(39)
			& \noindent\rule{1cm}{0.4pt}   \\[1ex] \hline &  &  \\[-1ex]
			lattice:  \cite{deForcrand:2002hgr}
			& 0.0056(6)
			& \noindent\rule{1cm}{0.4pt}  \\[1ex]
			\hline
		\end{tabular}
		\caption{Curvature coefficients $\kappa$, see \eq{eq:curv}: fRG-DSE: this work;
			fRG: \cite{Fu:2019hdw} (Fu {\it et al.}), \cite{Braun:2019aow} (Braun {\it et al.}); Lattice collaborations: \cite{Borsanyi:2020fev}
			(WB), \cite{Bonati:2018nut} (Bonati {\it et al.}),
			\cite{Bazavov:2018mes} (hotQCD), \cite{Allton:2002zi} (Allton {\it et al.}),
			\cite{deForcrand:2002hgr} (Forcrand and Philipsen), Lattice overviews
			\cite{Philipsen:2007rj,DElia:2018fjp}; DSE: \cite{Gao:2015kea} (Gao {\it et al.}), \cite{Fischer:2012vc,Fischer:2014ata,Fischer:2018sdj} (Fischer {\it et
				al.}), DSE overview
			\cite{Fischer:2018sdj}.}
		\label{tab:kappas}
	\end{center}
\end{table}
We now move to the discussion of the phase structure at large baryon-chemical potential.
This requires a discussion of the reliability bounds of the current approximation: the quark gap equation depends on the gluon propagator and the quark-gluon vertex. As expected, the former correlation function shows little dependence on the baryon-chemical potential in agreement with respective results from the fRG, \cite{Fu:2019hdw} and other DSE studies \cite{Fischer:2018sdj, Gunkel:2019xnh}. Accordingly, the respective systematic error is supposedly small.
In turn, the quark-gluon vertex inherits a strong dependence on the baryon-chemical potential
related to that of the quark propagator via the STI. Moreover, strongly resonant baryon and meson channels are potentially not well-captured by the present STI-construction of the vertex. Based on this analysis and the results in \cite{Fu:2019hdw} on non-trivial momentum dependencies for $\mu_B/T\gtrsim 3$ we conclude that the present approximation has to be upgraded towards solving the DSE for $\Delta\Gamma^{(3)}_{q \bar q A} $ in this regime.

With this caveat in mind we explore the large density regime. For 2-flavour QCD we find a critical end point at large $\mu_B$ with
\begin{align}\label{eq:CEP2}
(T_{_{\tiny{\text{CEP}}}},{\mu_B}_{_{\tiny{\text{CEP}}}})_{_{\tiny{N_f=2}}}
=(108, 654)\,\textrm{MeV}\,.
\end{align}
For 2+1-flavour QCD we find a critical end point with
\begin{align}\label{eq:CEP2+1}
(T_{_{\tiny{\text{CEP}}}},{\mu_B}_{_{\tiny{\text{CEP}}}})_{_{\tiny{N_f=2+1}}}
=(93, 672)\,\textrm{MeV}\,.
\end{align}
Both CEPs are at large ratios $\mu_B/T$,
\begin{align}\nonumber
N_f=2:
&\   \frac{{\mu_B}_{_{\tiny{\text{CEP}}}}}{T_{_{\tiny{\text{CEP}}}}}
= 6.05\,,\\[1ex]
N_f=2+1:
&\  \frac{{\mu_B}_{_{\tiny{\text{CEP}}}}}{T_{_{\tiny{\text{CEP}}}}}
= 7.23\,.
\end{align}
As already mentioned above, in this regime the current approximation has lost its
quantitative reliability. Its systematic improvement is under way: we solve the DSEs for the vertex dressings $\Delta\lambda^{(i)}$ with $i=1-8$ in $\Delta\Gamma^{(3)}_{\bar q q A}$. This analysis also includes scalar-pseudoscalar meson and baryon exchange channel, see \cite{Eichmann:2015kfa}. Note that  the current fRG-DSE setup allows to utilise fRG results on the scalar, pseudoscalar and density four-quark channels as well as higher order scatterings of resonant channels. We hope to report on respective results soon.

We close this section with a comparison of our main result, the phase structure of $N_F=2+1$ flavour QCD with other theoretical results as well as freeze-out data from different  groups, see \Fig{fig:finalphase}. We would like to emphasise again that the current computation relies only on the determination of the fundamental parameters of QCD.  The light current-quark masses are fixed with the pion mass, and the strange current-quark mass is fixed with the ratio of $m_s^{0}/m_l^{0}\approx 27$. All correlation functions, and the renormalised light chiral condensate as well as the critical temperatures $T_c(\mu_B)$ are predictions in first principle functional QCD and compare well to the fRG results and the lattice results at small densities. In summary, the current combined functional setup and respective DSE and fRG studies provide a promising starting point for getting to quantitative predictions in the large density regime including the potential CEP in the next few years.

\section{Summary}

In this work we have evaluated the QCD phase structure at finite temperature and chemical potential with a combination of the functional renormalisation group (fRG) and Dyson-Schwinger (DSE) equations set-up in \Sec{sec:FRG-DSE}: fRG results from \cite{Cyrol:2017ewj} for the $2$-flavour gluon propagator and quark-gluon vertex in the vacuum have been used as an input for the DSEs of $2$- and $2+1$-flavour QCD at finite temperature and density. We have solved the coupled set of DSEs for the quark propagator and for the thermal and density corrections of the gluon propagator for $2$- and $2+1$-flavour QCD. In the latter case we also have computed the strange quark contributions for all temperatures and densities. The respective corrections of the quark-gluon vertex have been deduced from regularity properties of the vertex for momenta $p\gtrsim 1$\,GeV, relations between the dressings of the different tensor structures, and the Slavnov-Taylor identities for the dressing functions, see \Sec{sec:DSEQuarkGluon}. No phenomenological IR-scale has to be tuned and the only parameters are the fundamental ones in QCD, the current-quark masses, see \Sec{sec:PhysParameters}. The latter are determined with the pion pole mass $m_\pi=140$\, MeV, and, in the 2+1 flavour case, with the ratio of light and strange current-quark masses.

The approach has been benchmarked in the vacuum and at finite temperature: the results for gluon propagator, quark propagator and the renormalised light chiral condensate are consistent with lattice results,  e.g.\ \cite{Blum:2014tka,Bowman:2005vx,Buballa:2014tba}, fRG results, \cite{Fu:2019hdw}, and DSE results, e.g.\ \cite{Fischer:2018sdj} and references therein. The respective comparisons are depicted in Figures \ref{fig:QuarkPropNf2}, \ref{fig:GluonProp2+1}, \ref{fig:QuarkProp2+1}, \ref{fig:conden}.

After these successful benchmarks the setup has been used for the computation of the phase structure for $2$- and $2+1$-flavour QCD at finite temperature and density, see \Fig{fig:phaseNf2-Nf2+1}. For $2+1$-flavour QCD the results have been compared with other theoretical results as well as freeze out data, see \Fig{fig:finalphase} and the discussion there. At small chemical potential the present result agrees quantitatively with lattice and fRG results. For $\mu_B/T\lesssim 3$, the present approach can be readily used for the computation of observables such as net baryon number fluctuations relevant for the access to the freeze-out curve as well as for the physics of the QCD epoch in the early universe, see e.g.\ \cite{Wygas:2018otj}.

For larger chemical potential, $\mu_B/T\gtrsim 3$, lattice simulations are obstructed by the sign problem. In this regime the present fRG-assisted DSE results agree quantitatively with the fRG results in \cite{Fu:2019hdw}. This non-trivial agreement enhances the reliability of the respective results, given the very different resummation schemes of DSE and fRG used in particular in the matter sector. Still, the approximation used in FRG computation \cite{Fu:2019hdw} and the present approximation lack quantitative reliability in this regime: the current construction of the thermal and density corrections of the quark-gluon vertex can only be trusted quantitatively as long as it still a correction. This calls for a self-consistent computation of the thermal and density fluctuations that contribute to the quark-gluon vertex in particular in the presence of resonant interactions  triggered by the finite density as well as scaling phenomenon in the vicinity of a potential critical end point. To a smaller degree this reliability analysis also applies to the present linear feedback of temperature and density corrections in the gluon DSE. With the important caveats in mind, the current computation shows a critical endpoint at $(T_\textrm{\tiny{CEP}},{\mu_{B}}_{\textrm{\tiny{CEP}}})=(93,672)$\,MeV, see \eq{eq:CEP2+1}.

We emphasise that the approximations used here do not pose conceptual or computational challenges and can be resolved in an extension of the current work. In order to consolidate the large density results, the approximation is currently systematically improved, as well that used in the respective fRG computation from \cite{Fu:2019hdw}. These combined works should finally allow for a quantitative access to the large density regime including the position of the potential CEP. \\

\noindent{\bf Acknowledgements}\\[1ex]
We thank J.~Braun, G.~Eichmann, W.-j.~Fu, J.~Papavassiliou, F.Rennecke, B.-J.~Schaefer and N.~Wink for discussions.
F.~Gao is supported by the Alexander von Humboldt foundation.
This work is supported by EMMI and the
BMBF grant 05P18VHFCA. It is part of and supported by the DFG
Collaborative Research Centre SFB 1225 (ISOQUANT) and the
DFG under Germany's Excellence Strategy EXC - 2181/1 - 390900948 (the
Heidelberg Excellence Cluster STRUCTURES).

\appendix

\section{Quark-gluon vertex from gauge invariant quark-gluon terms, STIs and regularity}\label{app:STI}

In this appendix we provide the details and numerical checks for the construction of a quark-gluon vertex at finite temperature and density. In particular we utilise the derivation of the complete set of tensor structures from local gauge invariant operators
and the STIs, as well as regularity constraints for $p \gtrsim 1$\,GeV.

\subsection{Relations from gauge invariant quark-gluon interactions and regularity}\label{app:STI1}
The full quark-gluon vertex can be expanded in a tensor basis $\{
{\cal T}_{q\bar q A}^{i}\}$ with $i=1,...,12$. A tensor basis for the transverse part of the vertex, \eq{eq:FullGqbarqA}, has been obtained from the \eq{eq:TensorsQuarkGluon} with $i=1,...,8$ via transverse projections. The four remaining tensors are obtained from the longitudinal projections of these tensors. These projections only contain four linearly independent tensors, and we choose
\begin{align}\label{eq:barqAqlong_tensors}
&\left[{\cal T}^{(9,10,11,12)}_{\bar q q A}\right]_\mu(p,q)=\Pi^\lt_{\mu\nu}(k_+)\left[{\cal T}^{(1,2,6,8)}_{\bar q q A}\right]_\nu(p,q)\,.
\end{align}
The dressing functions of these tensor structures can be determined via the STI for the quark-gluon vertex, which relates
the dressings to the dressings of the quark, gluon and ghost propagators as well as scattering kernels. A very detailed analysis of the respective STI-relation for the longitudinal dressings $\lambda^{(9-12)}_{q\bar q A}$ is given in \cite{Cyrol:2017ewj}, Appendix D.

Regularity for perturbative and semi-perturbative momenta relates in particular $\lambda^{(1)}_{q\bar q A}$ and $\lambda^{(9)}_{q\bar q A}$ with
\begin{align}\label{eq:Reg19}
\lambda^{(9)}_{q\bar q A}(p,q)\simeq \lambda^{(1)}_{q\bar q A}(p,q)\,, \quad \textrm{for}\quad \bar p\gtrsim 1\,\textrm{GeV}\,.
\end{align}%
For deriving \eq{eq:Reg19} we first remark that $\lambda^{(1)}_{q\bar q A}$ is the dressing of the transverse projection of ${\cal T}_{q\bar q A}^{1}$, while $\lambda^{(9)}_{q\bar q A}$ is the dressing of its longitudinal projection. The full relation also involves $\lambda^{(5,7)}_{q\bar q A}$ and can be found in \cite{Cyrol:2017ewj}, Appendix D. Then \eq{eq:Reg19} follows with a further relation,
\begin{align}\label{eq:lambda57}
\frac{\lambda^{(7)}_{q\bar q A}(p,q)}{2} - \lambda^{(5)}_{q\bar q A}(p,q)\approx 0\quad \textrm{for}\quad \bar p\gtrsim 1\,\textrm{GeV}\,.
\end{align}
The cancellation between $\lambda^{(5)}_{q\bar q A}$ and $\lambda^{(7)}_{q\bar q A}$ is suggested by the relation of the respective tensor structures: both can be derived from
the gauge invariant quark-gluon interaction term,
\begin{align}
\bar q\, D_\mu \{ [\gamma_{\mu}\,,\, \gamma_\nu]\, ,\,\dr \}\,D_\nu\, q\,,
\end{align}
related to $\bar q \dr^3 q$, see \cite{Mitter:2014wpa, Cyrol:2017ewj}. There,  \eq{eq:lambda57} has also been checked numerically. For the last of these tensor structures related to $\bar q \dr^3 q$ it has been found numerically,
\begin{align}\label{eq:lambda67}
\frac{\lambda^{(7)}_{q\bar q A}(p,q)}{12} - \lambda^{(6)}_{q\bar q A}(p,q)\approx 0\quad \textrm{for}\quad \bar p\gtrsim 1\,\textrm{GeV}\,,
\end{align}
Note that such relations also hold for
the dressings of the tensor structures $\lambda^{(2,4)}_{q\bar q A}$ derived from $\bar q \dr^2 q$,
\begin{align}\label{eq:lambda24}
\frac{\lambda^{(2)}_{q\bar q A}(p,q)}{2} \approx  \lambda^{(4)}_{q\bar q A}(p,q)\quad \textrm{for}\quad \bar p\gtrsim 1\,\textrm{GeV}\,.
\end{align}
This closes the discussion of relations between dressing functions from gauge invariant quark-gluon interaction terms and regularity.

\subsection{STI-construction}\label{app:STI2}
Now we utilise the STIs for the dressings $\lambda^{(9-12)}_{q\bar q A}$ for the determination of the transverse dressings for finite $T, \mu_B$. With the relations  relations \eq{eq:lambda57}, \eq{eq:lambda67}, \eq{eq:lambda24} this only has to be done for
$i=1,4,7,8$. First we discuss the dressing of the classical tensor structure: with \eq{eq:Reg19} we can use the STI for the longitudinal dressing $\lambda^{(9)}_{q\bar q A}$ also for the transverse dressing $\lambda^{(1)}_{q\bar q A}$. This leads us to
\begin{align}\label{eq:STI_Regular}
\tilde\lambda^{(1)}_{q\bar q A}(p,q) \approx \left( \frac{\lambda^{(1)}_{q\bar q A}(p,q)}{\Sigma_{Z_q}(p,q)}\right)^{\textrm{(in)}}\approx  \left(\frac{\lambda^{(1)}_{c \bar c A}(p,q)}{Z_c(k_+)}\right)^{\textrm{(in)}}\,,
\end{align}
for $\bar p \gtrsim 1\,\textrm{GeV}$. Here, $\lambda^{(1)}_{c \bar c A}$ is the dressing of the transverse part of the ghost-gluon vertex, and the differences $\Delta_{X}$ and sums $\Sigma_X$ of the quark dressings defined in \eq{eq:SigDel} carry the momentum and RG-behaviour of the quark two-point functions. Moreover, in the present work the superscript ${}^{\textrm{(in)}}$ indicate the vacuum $N_f=2$-flavour data. The dressing $\tilde\lambda^{(1)}_{q\bar q A}(p,q)$ is nothing but the full scattering kernel in the STI (up to the convenient rescaling with $Z_c(k_+)$).

\begin{figure}[t] %[hdbp]
	\includegraphics[width=0.9\columnwidth]{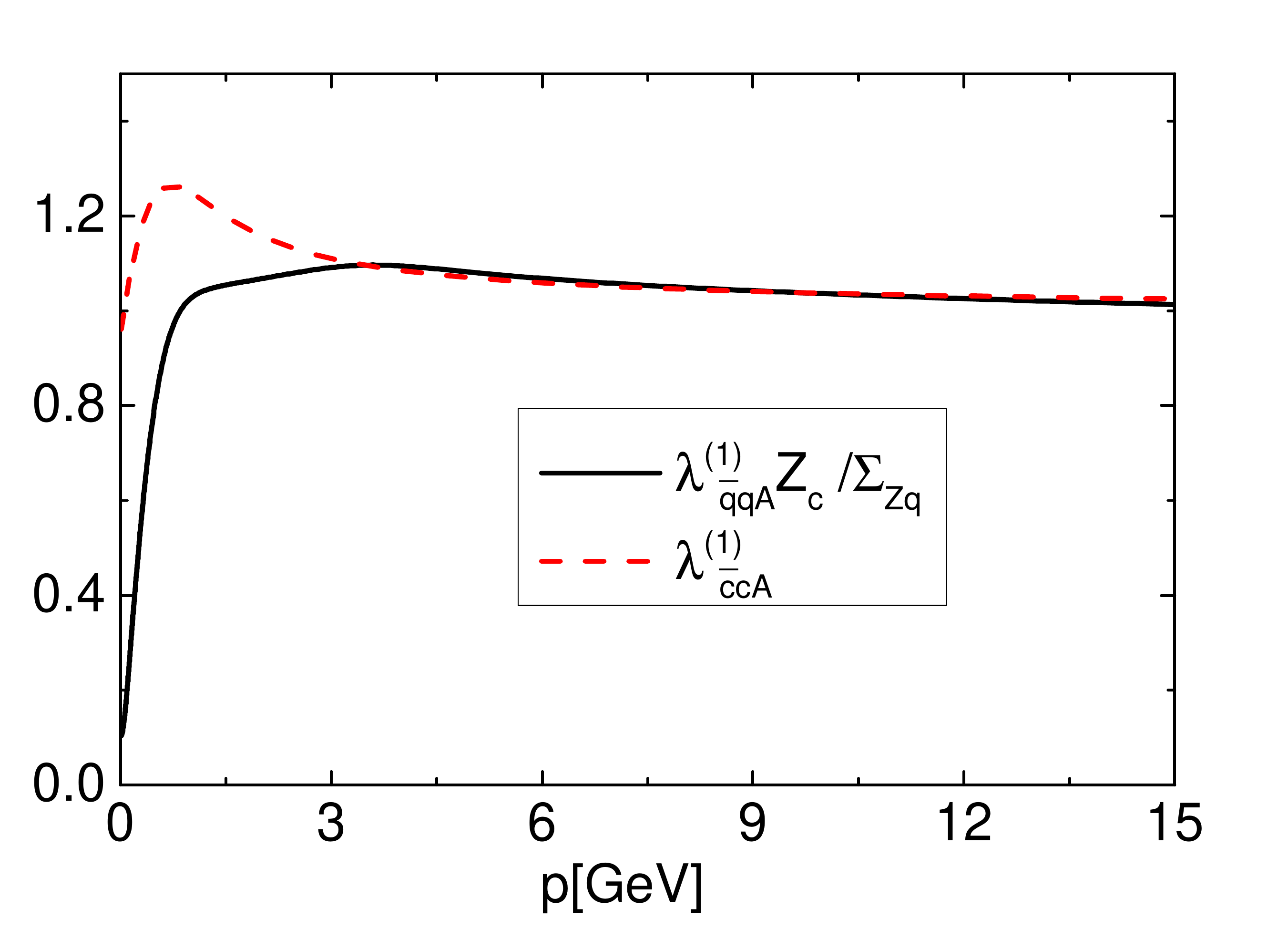}	
	\caption{The ratio $\lambda^{(1)}_{q\bar q A} Z_c /\Sigma_{Z_q}$ in comparison to $\lambda^{(1)}_{c\bar c A}$, the ghost-gluon dressing related to the quark-ghost scattering kernel, see \eq{eq:STI_Regular}, at the symmetric point as a function of momentum. }\label{fig:STI-Regular}
\end{figure}
The ratios in \eq{eq:STI_Regular} are approximately constant for $\bar p\gtrsim 1$\,GeV in line with regularity. In turn,
for $\bar p\lesssim 1$\,GeV regularity fails, see \cite{Cyrol:2017ewj}, which is reflected in a non-trivial momentum dependence of the ratios, see \Fig{fig:STI-Regular}. Importantly, the
STI links the dressing $\lambda^{(1)}_{q\bar q A}$ of the classical tensor structure to ghost-gluon correlation functions and the quark wave function renormalisation. It is well-known that $Z_c$ and $\lambda^{(1)}_{c \bar c A}$ are nearly insensitive to finite temperature and small chemical potentials, as well as the strange quark.

Note also that the non-regularity of vertices is in one-to-one correspondence to the occurrence of a gluon mass gap which signals confinement, for a detailed discussion see \cite{Cyrol:2017ewj, Cyrol:2016tym}. Accordingly, the irregularity in the quark-gluon vertex is triggered by the necessary one in the glue system.
In summary this suggests the following parametrisation of the vertex: the non-trivial deviation from the STI-construction based on the regularity assumption is triggered by the gluon sector and hence is well-captured by the $N_f=2$ flavour vertex. In turn, the strange quark and finite temperature and density fluctuations are regular, potential corrections to the irregularities only come indirectly via the gluon correlation functions in the diagrams for strange, $T$ and $\mu_B$ fluctuations. These subleading effects are discarded here, but are evaluated in a forthcoming publication.

\begin{figure}[t] %[hdbp]
	\includegraphics[width=.95\columnwidth]{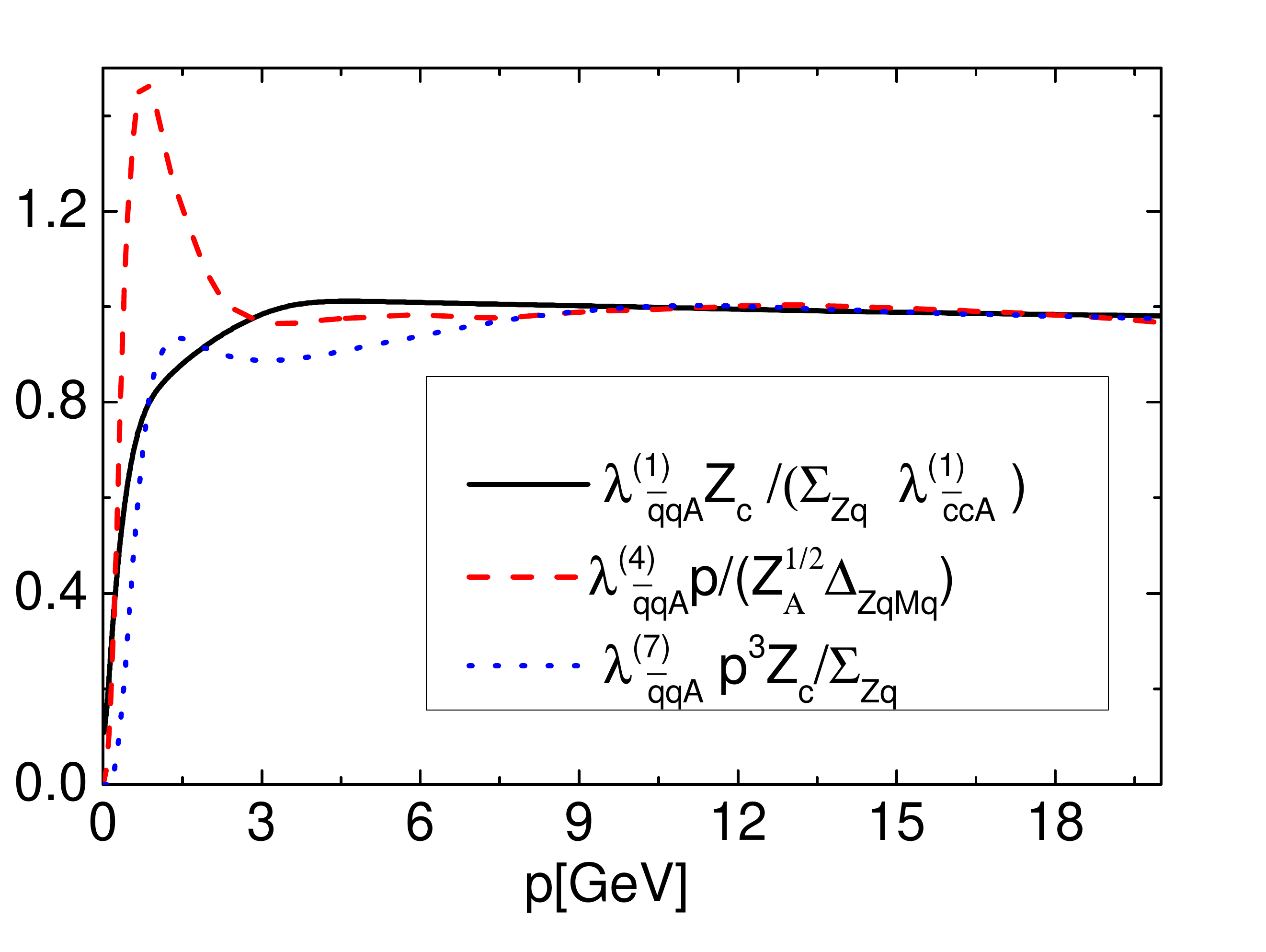}	
	\caption{Momentum-flat combinations of quark-gluon vertex dressings $\lambda^{(1,4,7)}_{q\bar q A}$ with quark, ghost and gluon dressing functions at the symmetric point $\bar p$. Up to approximately $s, T,\mu_B$-independent factors such as $Z_c$ and $\lambda^{(1)}_{c\bar c A}$ these combinations are the $\tilde \lambda^{(1,4,7)}_{q\bar q A}$ defined in \eq{eq:tildel1}, \eq{eq:tildel4}, \eq{eq:tildel7}.
	}\label{fig:Vertexdressings}
\end{figure}

Importantly, under the assumption of regularity of the vertex in the vanishing gluon momentum limit $k_+\to 0$ one can
relate combinations of the transverse dressings $\lambda^{(1-8)}_{q \bar q A}$ to the longitudinal ones, hence fixing the former by the STIs. It has also been shown in \cite{Cyrol:2017ewj, Mitter:2014wpa}, that the regularity of the quark-gluon vertex holds for momenta $k_+^2\gtrsim 1$\,GeV${}^2$, but fails below.
Moreover, also the scattering kernels only deviate from unity in the infrared. In \Fig{fig:Vertexdressings} this is shown for the most relevant dressings
$\lambda^{(1,4,7)}_{q \bar q A}$ at the symmetric point \eq{eq:SymPoint}. Given the small angular dependence, the momentum behaviour at the symmetric point gives a very good estimate of the validity of the STI-regularity relations in the loops. This will provide us with the dressing functions $\Delta \lambda^{(i)}_{q\bar q A}$, $i=1,...,8$ with
\begin{align} \label{eq:Deltalambda}
\lambda^{(i)}_{q\bar q A}=(\lambda^{(i)}_{q\bar q A})^{(\textrm{in})}+\Delta\lambda^{(i)}_{q\bar q A}\,,
\end{align}
where the momentum dependence has been suppressed, and the superscript ${}^{(\textrm{in})}$ indicates the input, here 2-flavour vacuum QCD data.
We start with the generalised version of \eq{eq:Delta1Sym} for $\lambda^{(1)}_{q\bar q A}$, and  are led to
\begin{align}
\label{eq:Deltalambda1}
\Delta\lambda^{(1)}_{q\bar q A}(p,q)= \tilde \lambda^{(1)}_{q\bar q A}(p,q)\left[\Sigma_{Z_q}-\Sigma_{Z_q}^{\textrm{(in)}}\right](p,q)\,.
\end{align}
with
\begin{align} \label{eq:tildel1}
\tilde \lambda^{(1)}_{q\bar q A}(p,q) = \left(\frac{\lambda^{(1)}_{q\bar q A}(p,q)}{\Sigma_{Z_q}(p,q)}\right)^{\textrm{(in)}}\,,
\end{align}
with $\Sigma_X$ and $\Delta_X$ as defined in \eq{eq:SigDel}. At the symmetric point \eq{eq:Deltalambda1} reduces to \eq{eq:Delta1Sym}. As an estimate for the systematic error we shall use both approximations for $\tilde \lambda^{(1)}_{q\bar q A}$ in \eq{eq:STI_Regular}.

It is left to extend the transverse dressings $\lambda^{(4,7,8)}_{q\bar q A}$ to finite $T,\mu_B$. While $\lambda^{(8)}_{q\bar q A}$ is subleading, the other two are relevant, and are depicted in \Fig{fig:Dressing}.

We first discuss the dressing $\lambda^{(4)}_{q\bar q A}$ of the chirally breaking tensor structure ${\cal T}^{(4)}_{q\bar q A}$. Naturally it has to vanish for $M_q\equiv 0$ and a consistent extension to $T,\mu_B\neq 0$ should reflect this property. In the momentum regime with regularity it is related to the longitudinal dressing $\lambda^{(10)}_{q\bar q A}$ but none of the other longitudinal ones. This suggests a construction with
$\Delta_{Z_q M_q}$. Moreover, RG-consistency either enforces a dressing with $Z_A^{1/2}$ or with
$1/Z_c$, the difference being the momentum running of the (ghost-gluon propagator-) coupling.
This leads us to
\begin{align}
\label{eq:tildel4} \tilde\lambda^{(4)}_{q\bar q A}(p,q)= \left(\frac{\lambda^{(4)}_{q\bar q A}(p,q)}{Z_A^{1/2}(p+q)\Delta_{Z_q M_q}(p,q)}\right)^{(\textrm{in})}\,,
\end{align}
depicted in \Fig{fig:Vertexdressings}.

\begin{figure}[t] %[hdbp]
	\includegraphics[width=0.95\columnwidth]{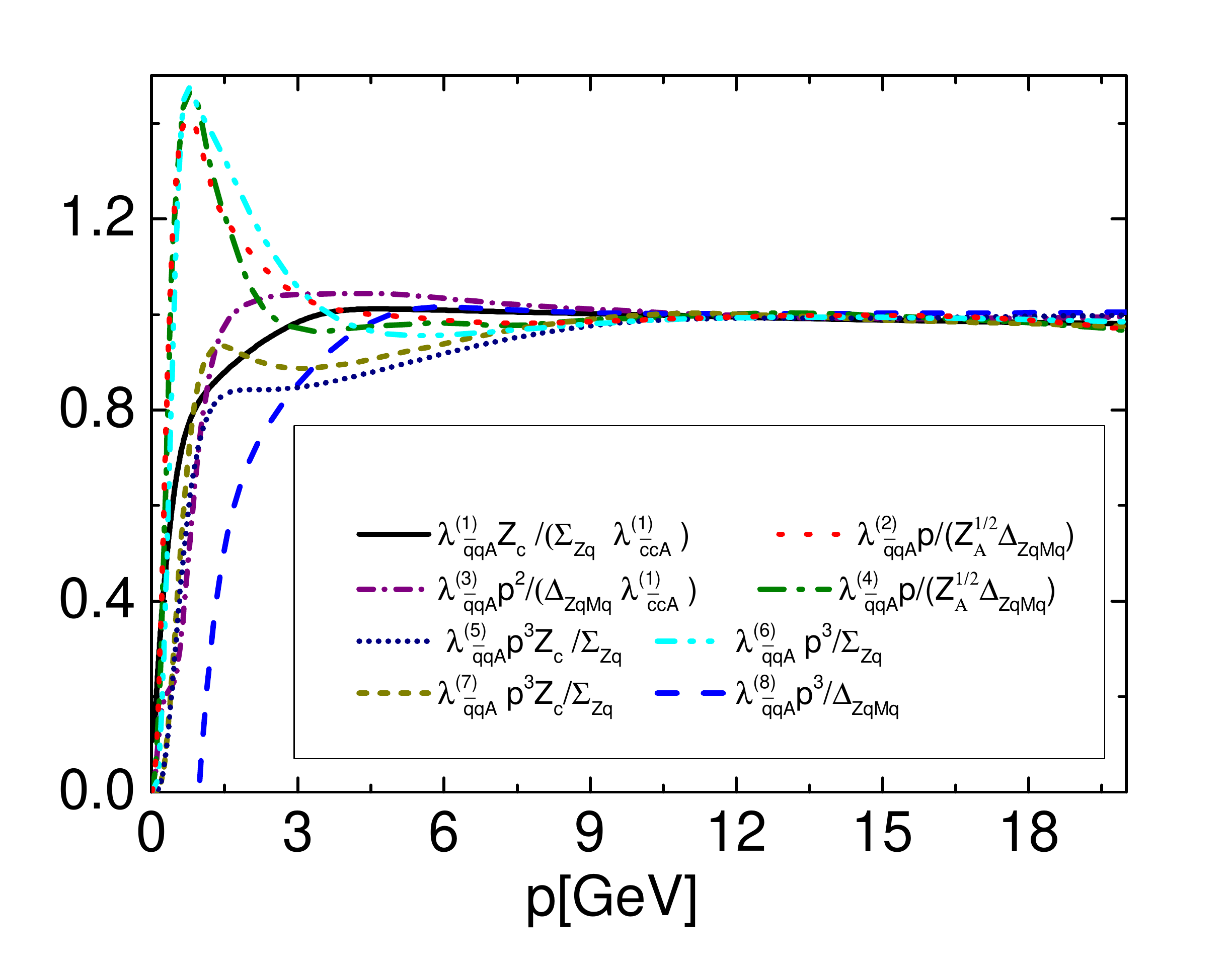}	
	\caption{Momentum-flat combinations of quark-gluon vertex dressings $\lambda^{(1-8)}_{q\bar q A}$  with quark, ghost and gluon dressing functions at the symmetric point $\bar p$. They are related to $\tilde \lambda_{q\bar q A}^{(1-8)}$ defined in \eq{eq:tildel1}, \eq{eq:tildel4}, \eq{eq:tildel7} and the relations \eq{eq:lambda57}, \eq{eq:lambda67}, \eq{eq:lambda24} excluding $i=3$. They are used in the vertex corrections $\Delta\lambda^{(1-8)}_{q\bar q A}$ in \eq{eq:Delta1-8}.
	}\label{fig:AllVertexdressings}
\end{figure}
For $\lambda^{(7)}_{q\bar q A}$ we use its occurrence in the regularity relation for $\lambda^{(9)}_{q\bar q A}$ and the
STI scaling of the latter. This leads us to
\begin{align}
\label{eq:tildel7} \tilde\lambda^{(7)}_{q\bar q A}(p,q)= \left(\frac{\lambda^{(7)}_{q\bar q A}(p,q)}{\Delta_{Z_q}(p,q)} \right)^{(\textrm{in})}\,.
\end{align}
Finally, for $\lambda^{(8)}$ we use its occurrence in the regularity relation for $\lambda^{(12)}_{q\bar q A}$ and the STI scaling of the latter,
\begin{align} \label{eq:tildel8}
\tilde\lambda^{(8)}_{q\bar q A}(p,q)=
\left(\frac{\lambda^{(8)}_{q\bar q A}(p,q)}{\Delta_{Z_q M_q}(p,q)} \right)^{(\textrm{in})}\,.
\end{align}
We note that \eq{eq:tildel8}, as the other relations for the dressing factors $\tilde\lambda^{(i)}_{q\bar q A}$, captures the perturbative limit and hence the qualitative dependence on $T$ and $\mu_B$.

We also emphasise that while the other STI relations used in the present work related to $\lambda_{12}^{(1)}$, the scalar tensor structure in the ghost-quark scattering kernel, that for $\lambda^{(12)}_{q\bar q A}$ also depends on $\lambda_{12}^{(4)}$ proportional to $\sigma_{\mu\nu}$. This may explain the lack of $Z_c$ in the relation above. In any case it simply hints at a more complicated relation for $\lambda^{(8)}_{q \bar q A}$. Here we refrain from a more detailed discussion as $\lambda^{(8)}_{q \bar q A}$ is subleading.

The relations \eq{eq:tildel4}, \eq{eq:tildel7}, \eq{eq:tildel8} lead us to
\begin{align}\nonumber
\Delta\lambda^{(4)}_{q\bar q A}(p,q)= &\tilde \lambda^{(4)}_{q\bar q A}(p,q)\Biggl[Z^{1/2}_A(k_+)\,\Delta_{Z_qM_q} \\[1ex]\nonumber
& \, \hspace{.9cm}
-\left( Z^{1/2}_A(k_+)\,\Delta_{Z_qM_q}\right)^{(\textrm{in})}\Biggr](p,q)\,,\notag\\[1ex]
\Delta\lambda^{(7)}_{q\bar q A}(p,q)= &\tilde \lambda^{(7)}_{q\bar q A}(p,q)\left[\Sigma_{Z_q}-\Sigma_{Z_q}^{(\textrm{in})}\right](p,q)\,,\notag\\[1ex]
\Delta\lambda^{(8)}_{q\bar q A}(p,q)= &\tilde \lambda^{(8)}_{q\bar q A}(p,q)\left[\Delta_{Z_qM_q}-\Delta_{Z_qM_q}^{(\textrm{in})}\right](p,q)\,.
\label{eq:transvere}\end{align}
The respective ratios related to the $\tilde \lambda^{(i)}_{q\bar q A}$ with $i=1,...,8$
are depicted in \Fig{fig:AllVertexdressings}, and the complete list of $T,\mu_B$-dependent dressing is summarised in \eq{eq:Delta1-8}.
\begin{figure}[t]
	\includegraphics[width=0.95\columnwidth]{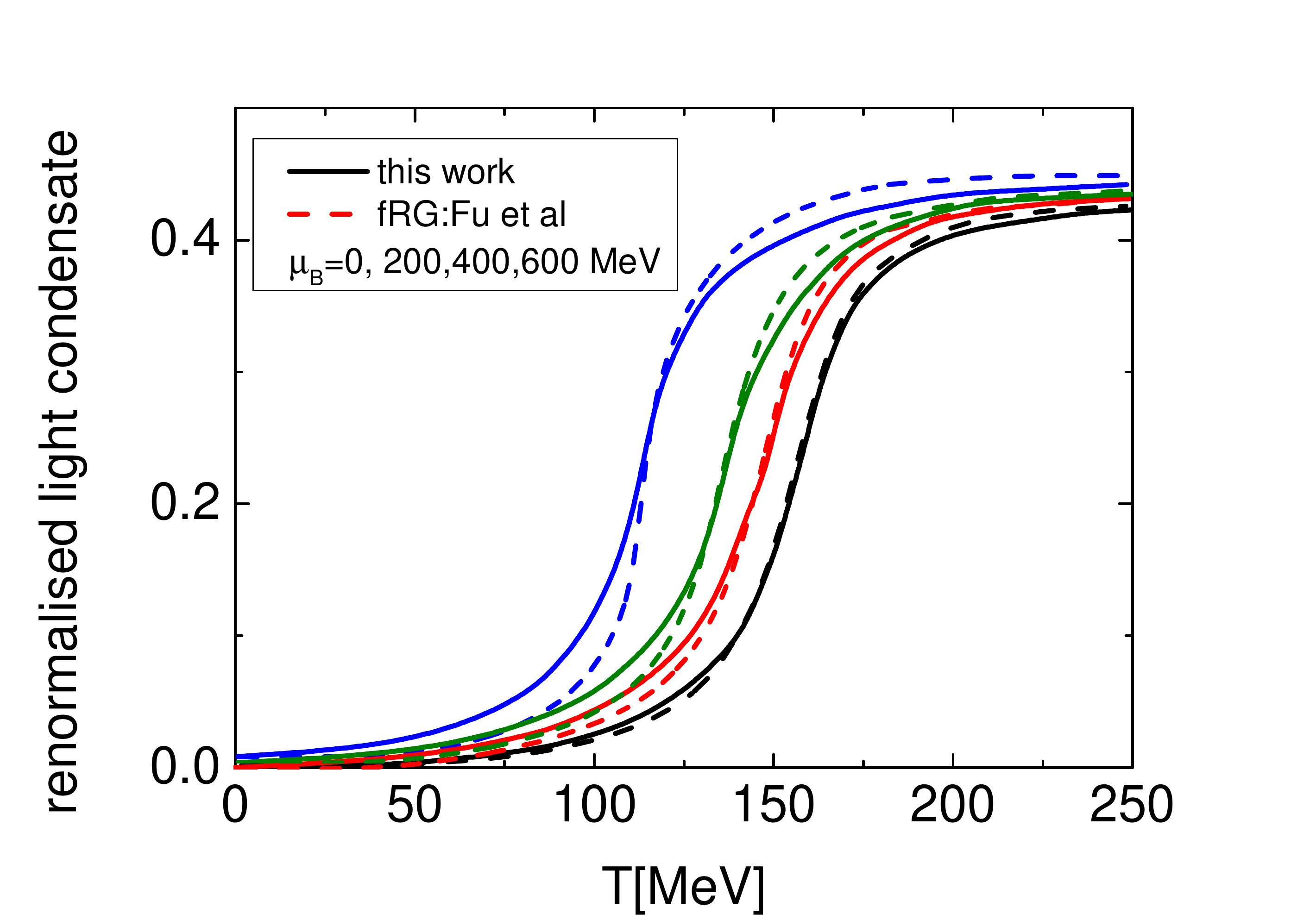}	
	\caption{The renormalised light chiral condensate at $\mu_B=0,\,200,\,400,\,600$\,MeV (\emph{black, red, olive and blue curves, respectively}) as a function of temperature $T$, compared with fRG results from  \cite{Fu:2019hdw}. }\label{fig:CondenMu}
	\vspace{.2cm}
	
\end{figure}

\section{Chiral condensate at finite $T$ and $\mu_B$}\label{app:conden}

Here we compare the temperature dependence of the renormalised light chiral condensate in 2+1 flavour QCD for different baryon-chemical potential with the fRG results from \cite{Fu:2019hdw}. The results are depicted in \Fig{fig:CondenMu}. With increasing baryon-chemical potential the deviations grow larger. The relatively small deviations at $\mu_B/T \lesssim 3$ or $\mu_B\lesssim 400$\,MeV provide an estimate of the systematic error of the respective works. They can be traced back to the different approximation schemes in DSE and fRG. For larger baryon-chemical potential, $\mu_B/T \gtrsim 3$, we approach the CEP in the fRG computation at $\mu_B= 635$\,MeV, and the CEP in the present work is approached for larger chemical potential, $\mu_B=672$\,MeV. Hence, while the different locations of the CEPs also provide systematic errors for the prediction of the CEP, the larger deviations of the condensates does not indicate a qualitative failure of the approximations. Still, for  $\mu_B/T \gtrsim 3$ the respective approximations have to be systematically improved for narrowing down the location (and even presence) of the CEP.

\bibliography{../ref-lib}

\begin{thebibliography}{95}%
\makeatletter
\providecommand \@ifxundefined [1]{%
 \@ifx{#1\undefined}
}%
\providecommand \@ifnum [1]{%
 \ifnum #1\expandafter \@firstoftwo
 \else \expandafter \@secondoftwo
 \fi
}%
\providecommand \@ifx [1]{%
 \ifx #1\expandafter \@firstoftwo
 \else \expandafter \@secondoftwo
 \fi
}%
\providecommand \natexlab [1]{#1}%
\providecommand \enquote  [1]{``#1''}%
\providecommand \bibnamefont  [1]{#1}%
\providecommand \bibfnamefont [1]{#1}%
\providecommand \citenamefont [1]{#1}%
\providecommand \href@noop [0]{\@secondoftwo}%
\providecommand \href [0]{\begingroup \@sanitize@url \@href}%
\providecommand \@href[1]{\@@startlink{#1}\@@href}%
\providecommand \@@href[1]{\endgroup#1\@@endlink}%
\providecommand \@sanitize@url [0]{\catcode `\\12\catcode `\$12\catcode
  `\&12\catcode `\#12\catcode `\^12\catcode `\_12\catcode `\%12\relax}%
\providecommand \@@startlink[1]{}%
\providecommand \@@endlink[0]{}%
\providecommand \url  [0]{\begingroup\@sanitize@url \@url }%
\providecommand \@url [1]{\endgroup\@href {#1}{\urlprefix }}%
\providecommand \urlprefix  [0]{URL }%
\providecommand \Eprint [0]{\href }%
\providecommand \doibase [0]{http://dx.doi.org/}%
\providecommand \selectlanguage [0]{\@gobble}%
\providecommand \bibinfo  [0]{\@secondoftwo}%
\providecommand \bibfield  [0]{\@secondoftwo}%
\providecommand \translation [1]{[#1]}%
\providecommand \BibitemOpen [0]{}%
\providecommand \bibitemStop [0]{}%
\providecommand \bibitemNoStop [0]{.\EOS\space}%
\providecommand \EOS [0]{\spacefactor3000\relax}%
\providecommand \BibitemShut  [1]{\csname bibitem#1\endcsname}%
\let\auto@bib@innerbib\@empty
%</preamble>
\bibitem [{\citenamefont {Cyrol}\ \emph
  {et~al.}(2018{\natexlab{a}})\citenamefont {Cyrol}, \citenamefont {Mitter},
  \citenamefont {Pawlowski},\ and\ \citenamefont {Strodthoff}}]{Cyrol:2017ewj}%
  \BibitemOpen
  \bibfield  {author} {\bibinfo {author} {\bibfnamefont {A.~K.}\ \bibnamefont
  {Cyrol}}, \bibinfo {author} {\bibfnamefont {M.}~\bibnamefont {Mitter}},
  \bibinfo {author} {\bibfnamefont {J.~M.}\ \bibnamefont {Pawlowski}}, \ and\
  \bibinfo {author} {\bibfnamefont {N.}~\bibnamefont {Strodthoff}},\ }\href
  {\doibase 10.1103/PhysRevD.97.054006} {\bibfield  {journal} {\bibinfo
  {journal} {Phys. Rev.}\ }\textbf {\bibinfo {volume} {D97}},\ \bibinfo {pages}
  {054006} (\bibinfo {year} {2018}{\natexlab{a}})},\ \Eprint
  {http://arxiv.org/abs/1706.06326} {arXiv:1706.06326 [hep-ph]} \BibitemShut
  {NoStop}%
%%CITATION = ARXIV:1706.06326;%%
\bibitem [{\citenamefont {Luo}\ and\ \citenamefont {Xu}(2017)}]{Luo:2017faz}%
  \BibitemOpen
  \bibfield  {author} {\bibinfo {author} {\bibfnamefont {X.}~\bibnamefont
  {Luo}}\ and\ \bibinfo {author} {\bibfnamefont {N.}~\bibnamefont {Xu}},\
  }\href {\doibase 10.1007/s41365-017-0257-0} {\bibfield  {journal} {\bibinfo
  {journal} {Nucl. Sci. Tech.}\ }\textbf {\bibinfo {volume} {28}},\ \bibinfo
  {pages} {112} (\bibinfo {year} {2017})},\ \Eprint
  {http://arxiv.org/abs/1701.02105} {arXiv:1701.02105 [nucl-ex]} \BibitemShut
  {NoStop}%
%%CITATION = ARXIV:1701.02105;%%
\bibitem [{\citenamefont {Adamczyk}\ \emph {et~al.}(2017)\citenamefont
  {Adamczyk} \emph {et~al.}}]{Adamczyk:2017iwn}%
  \BibitemOpen
  \bibfield  {author} {\bibinfo {author} {\bibfnamefont {L.}~\bibnamefont
  {Adamczyk}} \emph {et~al.} (\bibinfo {collaboration} {STAR}),\ }\href
  {\doibase 10.1103/PhysRevC.96.044904} {\bibfield  {journal} {\bibinfo
  {journal} {Phys. Rev.}\ }\textbf {\bibinfo {volume} {C96}},\ \bibinfo {pages}
  {044904} (\bibinfo {year} {2017})},\ \Eprint
  {http://arxiv.org/abs/1701.07065} {arXiv:1701.07065 [nucl-ex]} \BibitemShut
  {NoStop}%
%%CITATION = ARXIV:1701.07065;%%
\bibitem [{\citenamefont {Andronic}\ \emph {et~al.}(2018)\citenamefont
  {Andronic}, \citenamefont {Braun-Munzinger}, \citenamefont {Redlich},\ and\
  \citenamefont {Stachel}}]{Andronic:2017pug}%
  \BibitemOpen
  \bibfield  {author} {\bibinfo {author} {\bibfnamefont {A.}~\bibnamefont
  {Andronic}}, \bibinfo {author} {\bibfnamefont {P.}~\bibnamefont
  {Braun-Munzinger}}, \bibinfo {author} {\bibfnamefont {K.}~\bibnamefont
  {Redlich}}, \ and\ \bibinfo {author} {\bibfnamefont {J.}~\bibnamefont
  {Stachel}},\ }\href {\doibase 10.1038/s41586-018-0491-6} {\bibfield
  {journal} {\bibinfo  {journal} {Nature}\ }\textbf {\bibinfo {volume} {561}},\
  \bibinfo {pages} {321} (\bibinfo {year} {2018})},\ \Eprint
  {http://arxiv.org/abs/1710.09425} {arXiv:1710.09425 [nucl-th]} \BibitemShut
  {NoStop}%
%%CITATION = ARXIV:1710.09425;%%
\bibitem [{\citenamefont {Stephanov}(2006)}]{Stephanov:2007fk}%
  \BibitemOpen
  \bibfield  {author} {\bibinfo {author} {\bibfnamefont {M.~A.}\ \bibnamefont
  {Stephanov}},\ }\bibfield  {booktitle} {\emph {\bibinfo {booktitle}
  {{Proceedings, 24th International Symposium on Lattice Field Theory (Lattice
  2006): Tucson, USA, July 23-28, 2006}}},\ }\href@noop {} {\bibfield
  {journal} {\bibinfo  {journal} {PoS}\ }\textbf {\bibinfo {volume}
  {LAT2006}},\ \bibinfo {pages} {024} (\bibinfo {year} {2006})},\ \Eprint
  {http://arxiv.org/abs/hep-lat/0701002} {arXiv:hep-lat/0701002 [hep-lat]}
  \BibitemShut {NoStop}%
%%CITATION = HEP-LAT/0701002;%%
\bibitem [{\citenamefont {Andersen}\ \emph {et~al.}(2016)\citenamefont
  {Andersen}, \citenamefont {Naylor},\ and\ \citenamefont
  {Tranberg}}]{Andersen:2014xxa}%
  \BibitemOpen
  \bibfield  {author} {\bibinfo {author} {\bibfnamefont {J.~O.}\ \bibnamefont
  {Andersen}}, \bibinfo {author} {\bibfnamefont {W.~R.}\ \bibnamefont
  {Naylor}}, \ and\ \bibinfo {author} {\bibfnamefont {A.}~\bibnamefont
  {Tranberg}},\ }\href {\doibase 10.1103/RevModPhys.88.025001} {\bibfield
  {journal} {\bibinfo  {journal} {Rev. Mod. Phys.}\ }\textbf {\bibinfo {volume}
  {88}},\ \bibinfo {pages} {025001} (\bibinfo {year} {2016})},\ \Eprint
  {http://arxiv.org/abs/1411.7176} {arXiv:1411.7176 [hep-ph]} \BibitemShut
  {NoStop}%
%%CITATION = ARXIV:1411.7176;%%
\bibitem [{\citenamefont {Shuryak}(2017)}]{Shuryak:2014zxa}%
  \BibitemOpen
  \bibfield  {author} {\bibinfo {author} {\bibfnamefont {E.}~\bibnamefont
  {Shuryak}},\ }\href {\doibase 10.1103/RevModPhys.89.035001} {\bibfield
  {journal} {\bibinfo  {journal} {Rev. Mod. Phys.}\ }\textbf {\bibinfo {volume}
  {89}},\ \bibinfo {pages} {035001} (\bibinfo {year} {2017})},\ \Eprint
  {http://arxiv.org/abs/1412.8393} {arXiv:1412.8393 [hep-ph]} \BibitemShut
  {NoStop}%
%%CITATION = ARXIV:1412.8393;%%
\bibitem [{\citenamefont {Pawlowski}(2014)}]{Pawlowski:2014aha}%
  \BibitemOpen
  \bibfield  {author} {\bibinfo {author} {\bibfnamefont {J.~M.}\ \bibnamefont
  {Pawlowski}},\ }\bibfield  {booktitle} {\emph {\bibinfo {booktitle}
  {{Proceedings, 24th International Conference on Ultra-Relativistic
  Nucleus-Nucleus Collisions (Quark Matter 2014): Darmstadt, Germany, May
  19-24, 2014}}},\ }\href {\doibase 10.1016/j.nuclphysa.2014.09.074} {\bibfield
   {journal} {\bibinfo  {journal} {Nucl. Phys.}\ }\textbf {\bibinfo {volume}
  {A931}},\ \bibinfo {pages} {113} (\bibinfo {year} {2014})}\BibitemShut
  {NoStop}%
%%CITATION = NUPHA,A931,113;%%
\bibitem [{\citenamefont {Roberts}\ and\ \citenamefont
  {Schmidt}(2000)}]{Roberts:2000aa}%
  \BibitemOpen
  \bibfield  {author} {\bibinfo {author} {\bibfnamefont {C.~D.}\ \bibnamefont
  {Roberts}}\ and\ \bibinfo {author} {\bibfnamefont {S.~M.}\ \bibnamefont
  {Schmidt}},\ }\href {\doibase 10.1016/S0146-6410(00)90011-5} {\bibfield
  {journal} {\bibinfo  {journal} {Prog. Part. Nucl. Phys.}\ }\textbf {\bibinfo
  {volume} {45}},\ \bibinfo {pages} {S1} (\bibinfo {year} {2000})},\ \Eprint
  {http://arxiv.org/abs/nucl-th/0005064} {arXiv:nucl-th/0005064 [nucl-th]}
  \BibitemShut {NoStop}%
%%CITATION = NUCL-TH/0005064;%%
\bibitem [{\citenamefont {Fischer}(2019)}]{Fischer:2018sdj}%
  \BibitemOpen
  \bibfield  {author} {\bibinfo {author} {\bibfnamefont {C.~S.}\ \bibnamefont
  {Fischer}},\ }\href {\doibase 10.1016/j.ppnp.2019.01.002} {\bibfield
  {journal} {\bibinfo  {journal} {Prog. Part. Nucl. Phys.}\ }\textbf {\bibinfo
  {volume} {105}},\ \bibinfo {pages} {1} (\bibinfo {year} {2019})},\ \Eprint
  {http://arxiv.org/abs/1810.12938} {arXiv:1810.12938 [hep-ph]} \BibitemShut
  {NoStop}%
%%CITATION = ARXIV:1810.12938;%%
\bibitem [{\citenamefont {Yin}(2018)}]{Yin:2018ejt}%
  \BibitemOpen
  \bibfield  {author} {\bibinfo {author} {\bibfnamefont {Y.}~\bibnamefont
  {Yin}},\ }\href@noop {} {\  (\bibinfo {year} {2018})},\ \Eprint
  {http://arxiv.org/abs/1811.06519} {arXiv:1811.06519 [nucl-th]} \BibitemShut
  {NoStop}%
%%CITATION = ARXIV:1811.06519;%%
\bibitem [{\citenamefont {Braun}(2009)}]{Braun:2008pi}%
  \BibitemOpen
  \bibfield  {author} {\bibinfo {author} {\bibfnamefont {J.}~\bibnamefont
  {Braun}},\ }\href {\doibase 10.1140/epjc/s10052-009-1136-6} {\bibfield
  {journal} {\bibinfo  {journal} {Eur. Phys. J.}\ }\textbf {\bibinfo {volume}
  {C64}},\ \bibinfo {pages} {459} (\bibinfo {year} {2009})},\ \Eprint
  {http://arxiv.org/abs/0810.1727} {arXiv:0810.1727 [hep-ph]} \BibitemShut
  {NoStop}%
%%CITATION = ARXIV:0810.1727;%%
\bibitem [{\citenamefont {Braun}\ \emph {et~al.}(2011)\citenamefont {Braun},
  \citenamefont {Haas}, \citenamefont {Marhauser},\ and\ \citenamefont
  {Pawlowski}}]{Braun:2009gm}%
  \BibitemOpen
  \bibfield  {author} {\bibinfo {author} {\bibfnamefont {J.}~\bibnamefont
  {Braun}}, \bibinfo {author} {\bibfnamefont {L.~M.}\ \bibnamefont {Haas}},
  \bibinfo {author} {\bibfnamefont {F.}~\bibnamefont {Marhauser}}, \ and\
  \bibinfo {author} {\bibfnamefont {J.~M.}\ \bibnamefont {Pawlowski}},\ }\href
  {\doibase 10.1103/PhysRevLett.106.022002} {\bibfield  {journal} {\bibinfo
  {journal} {Phys. Rev. Lett.}\ }\textbf {\bibinfo {volume} {106}},\ \bibinfo
  {pages} {022002} (\bibinfo {year} {2011})},\ \Eprint
  {http://arxiv.org/abs/0908.0008} {arXiv:0908.0008 [hep-ph]} \BibitemShut
  {NoStop}%
%%CITATION = ARXIV:0908.0008;%%
\bibitem [{\citenamefont {Fister}\ and\ \citenamefont
  {Pawlowski}(2011)}]{Fister:2011uw}%
  \BibitemOpen
  \bibfield  {author} {\bibinfo {author} {\bibfnamefont {L.}~\bibnamefont
  {Fister}}\ and\ \bibinfo {author} {\bibfnamefont {J.~M.}\ \bibnamefont
  {Pawlowski}},\ }\href@noop {} {\  (\bibinfo {year} {2011})},\ \Eprint
  {http://arxiv.org/abs/1112.5440} {arXiv:1112.5440 [hep-ph]} \BibitemShut
  {NoStop}%
%%CITATION = ARXIV:1112.5440;%%
\bibitem [{\citenamefont {Mitter}\ \emph {et~al.}(2015)\citenamefont {Mitter},
  \citenamefont {Pawlowski},\ and\ \citenamefont
  {Strodthoff}}]{Mitter:2014wpa}%
  \BibitemOpen
  \bibfield  {author} {\bibinfo {author} {\bibfnamefont {M.}~\bibnamefont
  {Mitter}}, \bibinfo {author} {\bibfnamefont {J.~M.}\ \bibnamefont
  {Pawlowski}}, \ and\ \bibinfo {author} {\bibfnamefont {N.}~\bibnamefont
  {Strodthoff}},\ }\href {\doibase 10.1103/PhysRevD.91.054035} {\bibfield
  {journal} {\bibinfo  {journal} {Phys. Rev.}\ }\textbf {\bibinfo {volume}
  {D91}},\ \bibinfo {pages} {054035} (\bibinfo {year} {2015})},\ \Eprint
  {http://arxiv.org/abs/1411.7978} {arXiv:1411.7978 [hep-ph]} \BibitemShut
  {NoStop}%
%%CITATION = ARXIV:1411.7978;%%
\bibitem [{\citenamefont {Braun}\ \emph {et~al.}(2016)\citenamefont {Braun},
  \citenamefont {Fister}, \citenamefont {Pawlowski},\ and\ \citenamefont
  {Rennecke}}]{Braun:2014ata}%
  \BibitemOpen
  \bibfield  {author} {\bibinfo {author} {\bibfnamefont {J.}~\bibnamefont
  {Braun}}, \bibinfo {author} {\bibfnamefont {L.}~\bibnamefont {Fister}},
  \bibinfo {author} {\bibfnamefont {J.~M.}\ \bibnamefont {Pawlowski}}, \ and\
  \bibinfo {author} {\bibfnamefont {F.}~\bibnamefont {Rennecke}},\ }\href
  {\doibase 10.1103/PhysRevD.94.034016} {\bibfield  {journal} {\bibinfo
  {journal} {Phys. Rev.}\ }\textbf {\bibinfo {volume} {D94}},\ \bibinfo {pages}
  {034016} (\bibinfo {year} {2016})},\ \Eprint {http://arxiv.org/abs/1412.1045}
  {arXiv:1412.1045 [hep-ph]} \BibitemShut {NoStop}%
%%CITATION = ARXIV:1412.1045;%%
\bibitem [{\citenamefont {Rennecke}(2015)}]{Rennecke:2015eba}%
  \BibitemOpen
  \bibfield  {author} {\bibinfo {author} {\bibfnamefont {F.}~\bibnamefont
  {Rennecke}},\ }\href {\doibase 10.1103/PhysRevD.92.076012} {\bibfield
  {journal} {\bibinfo  {journal} {Phys. Rev.}\ }\textbf {\bibinfo {volume}
  {D92}},\ \bibinfo {pages} {076012} (\bibinfo {year} {2015})},\ \Eprint
  {http://arxiv.org/abs/1504.03585} {arXiv:1504.03585 [hep-ph]} \BibitemShut
  {NoStop}%
%%CITATION = ARXIV:1504.03585;%%
\bibitem [{\citenamefont {Fu}\ \emph {et~al.}(2016)\citenamefont {Fu},
  \citenamefont {Pawlowski}, \citenamefont {Rennecke},\ and\ \citenamefont
  {Schaefer}}]{Fu:2016tey}%
  \BibitemOpen
  \bibfield  {author} {\bibinfo {author} {\bibfnamefont {W.-j.}\ \bibnamefont
  {Fu}}, \bibinfo {author} {\bibfnamefont {J.~M.}\ \bibnamefont {Pawlowski}},
  \bibinfo {author} {\bibfnamefont {F.}~\bibnamefont {Rennecke}}, \ and\
  \bibinfo {author} {\bibfnamefont {B.-J.}\ \bibnamefont {Schaefer}},\ }\href
  {\doibase 10.1103/PhysRevD.94.116020} {\bibfield  {journal} {\bibinfo
  {journal} {Phys. Rev.}\ }\textbf {\bibinfo {volume} {D94}},\ \bibinfo {pages}
  {116020} (\bibinfo {year} {2016})},\ \Eprint
  {http://arxiv.org/abs/1608.04302} {arXiv:1608.04302 [hep-ph]} \BibitemShut
  {NoStop}%
%%CITATION = ARXIV:1608.04302;%%
\bibitem [{\citenamefont {Rennecke}\ and\ \citenamefont
  {Schaefer}(2017)}]{Rennecke:2016tkm}%
  \BibitemOpen
  \bibfield  {author} {\bibinfo {author} {\bibfnamefont {F.}~\bibnamefont
  {Rennecke}}\ and\ \bibinfo {author} {\bibfnamefont {B.-J.}\ \bibnamefont
  {Schaefer}},\ }\href {\doibase 10.1103/PhysRevD.96.016009} {\bibfield
  {journal} {\bibinfo  {journal} {Phys. Rev.}\ }\textbf {\bibinfo {volume}
  {D96}},\ \bibinfo {pages} {016009} (\bibinfo {year} {2017})},\ \Eprint
  {http://arxiv.org/abs/1610.08748} {arXiv:1610.08748 [hep-ph]} \BibitemShut
  {NoStop}%
%%CITATION = ARXIV:1610.08748;%%
\bibitem [{\citenamefont {Cyrol}\ \emph {et~al.}(2016)\citenamefont {Cyrol},
  \citenamefont {Fister}, \citenamefont {Mitter}, \citenamefont {Pawlowski},\
  and\ \citenamefont {Strodthoff}}]{Cyrol:2016tym}%
  \BibitemOpen
  \bibfield  {author} {\bibinfo {author} {\bibfnamefont {A.~K.}\ \bibnamefont
  {Cyrol}}, \bibinfo {author} {\bibfnamefont {L.}~\bibnamefont {Fister}},
  \bibinfo {author} {\bibfnamefont {M.}~\bibnamefont {Mitter}}, \bibinfo
  {author} {\bibfnamefont {J.~M.}\ \bibnamefont {Pawlowski}}, \ and\ \bibinfo
  {author} {\bibfnamefont {N.}~\bibnamefont {Strodthoff}},\ }\href {\doibase
  10.1103/PhysRevD.94.054005} {\bibfield  {journal} {\bibinfo  {journal} {Phys.
  Rev.}\ }\textbf {\bibinfo {volume} {D94}},\ \bibinfo {pages} {054005}
  (\bibinfo {year} {2016})},\ \Eprint {http://arxiv.org/abs/1605.01856}
  {arXiv:1605.01856 [hep-ph]} \BibitemShut {NoStop}%
%%CITATION = ARXIV:1605.01856;%%
\bibitem [{\citenamefont {Cyrol}\ \emph
  {et~al.}(2018{\natexlab{b}})\citenamefont {Cyrol}, \citenamefont {Mitter},
  \citenamefont {Pawlowski},\ and\ \citenamefont {Strodthoff}}]{Cyrol:2017qkl}%
  \BibitemOpen
  \bibfield  {author} {\bibinfo {author} {\bibfnamefont {A.~K.}\ \bibnamefont
  {Cyrol}}, \bibinfo {author} {\bibfnamefont {M.}~\bibnamefont {Mitter}},
  \bibinfo {author} {\bibfnamefont {J.~M.}\ \bibnamefont {Pawlowski}}, \ and\
  \bibinfo {author} {\bibfnamefont {N.}~\bibnamefont {Strodthoff}},\ }\href
  {\doibase 10.1103/PhysRevD.97.054015} {\bibfield  {journal} {\bibinfo
  {journal} {Phys. Rev.}\ }\textbf {\bibinfo {volume} {D97}},\ \bibinfo {pages}
  {054015} (\bibinfo {year} {2018}{\natexlab{b}})},\ \Eprint
  {http://arxiv.org/abs/1708.03482} {arXiv:1708.03482 [hep-ph]} \BibitemShut
  {NoStop}%
%%CITATION = ARXIV:1708.03482;%%
\bibitem [{\citenamefont {Fu}\ \emph {et~al.}(2018)\citenamefont {Fu},
  \citenamefont {Pawlowski},\ and\ \citenamefont {Rennecke}}]{Fu:2018qsk}%
  \BibitemOpen
  \bibfield  {author} {\bibinfo {author} {\bibfnamefont {W.-j.}\ \bibnamefont
  {Fu}}, \bibinfo {author} {\bibfnamefont {J.~M.}\ \bibnamefont {Pawlowski}}, \
  and\ \bibinfo {author} {\bibfnamefont {F.}~\bibnamefont {Rennecke}},\
  }\href@noop {} {\  (\bibinfo {year} {2018})},\ \Eprint
  {http://arxiv.org/abs/1808.00410} {arXiv:1808.00410 [hep-ph]} \BibitemShut
  {NoStop}%
%%CITATION = ARXIV:1808.00410;%%
\bibitem [{\citenamefont {Fu}\ \emph {et~al.}(2019)\citenamefont {Fu},
  \citenamefont {Pawlowski},\ and\ \citenamefont {Rennecke}}]{Fu:2019hdw}%
  \BibitemOpen
  \bibfield  {author} {\bibinfo {author} {\bibfnamefont {W.-j.}\ \bibnamefont
  {Fu}}, \bibinfo {author} {\bibfnamefont {J.~M.}\ \bibnamefont {Pawlowski}}, \
  and\ \bibinfo {author} {\bibfnamefont {F.}~\bibnamefont {Rennecke}},\
  }\href@noop {} {\  (\bibinfo {year} {2019})},\ \Eprint
  {http://arxiv.org/abs/1909.02991} {arXiv:1909.02991 [hep-ph]} \BibitemShut
  {NoStop}%
%%CITATION = ARXIV:1909.02991;%%
\bibitem [{\citenamefont {Leonhardt}\ \emph {et~al.}(2019)\citenamefont
  {Leonhardt}, \citenamefont {Pospiech}, \citenamefont {Schallmo},
  \citenamefont {Braun}, \citenamefont {Drischler}, \citenamefont {Hebeler},\
  and\ \citenamefont {Schwenk}}]{Leonhardt:2019fua}%
  \BibitemOpen
  \bibfield  {author} {\bibinfo {author} {\bibfnamefont {M.}~\bibnamefont
  {Leonhardt}}, \bibinfo {author} {\bibfnamefont {M.}~\bibnamefont {Pospiech}},
  \bibinfo {author} {\bibfnamefont {B.}~\bibnamefont {Schallmo}}, \bibinfo
  {author} {\bibfnamefont {J.}~\bibnamefont {Braun}}, \bibinfo {author}
  {\bibfnamefont {C.}~\bibnamefont {Drischler}}, \bibinfo {author}
  {\bibfnamefont {K.}~\bibnamefont {Hebeler}}, \ and\ \bibinfo {author}
  {\bibfnamefont {A.}~\bibnamefont {Schwenk}},\ }\href@noop {} {\  (\bibinfo
  {year} {2019})},\ \Eprint {http://arxiv.org/abs/1907.05814} {arXiv:1907.05814
  [nucl-th]} \BibitemShut {NoStop}%
%%CITATION = ARXIV:1907.05814;%%
\bibitem [{\citenamefont {Braun}\ \emph
  {et~al.}(2020{\natexlab{a}})\citenamefont {Braun}, \citenamefont
  {Leonhardt},\ and\ \citenamefont {Pospiech}}]{Braun:2019aow}%
  \BibitemOpen
  \bibfield  {author} {\bibinfo {author} {\bibfnamefont {J.}~\bibnamefont
  {Braun}}, \bibinfo {author} {\bibfnamefont {M.}~\bibnamefont {Leonhardt}}, \
  and\ \bibinfo {author} {\bibfnamefont {M.}~\bibnamefont {Pospiech}},\ }\href
  {\doibase 10.1103/PhysRevD.101.036004} {\bibfield  {journal} {\bibinfo
  {journal} {Phys. Rev. D}\ }\textbf {\bibinfo {volume} {101}},\ \bibinfo
  {pages} {036004} (\bibinfo {year} {2020}{\natexlab{a}})},\ \Eprint
  {http://arxiv.org/abs/1909.06298} {arXiv:1909.06298 [hep-ph]} \BibitemShut
  {NoStop}%
\bibitem [{\citenamefont {Braun}\ \emph
  {et~al.}(2020{\natexlab{b}})\citenamefont {Braun}, \citenamefont {Fu},
  \citenamefont {Pawlowski}, \citenamefont {Rennecke}, \citenamefont
  {Rosenblüh},\ and\ \citenamefont {Yin}}]{Braun:2020ada}%
  \BibitemOpen
  \bibfield  {author} {\bibinfo {author} {\bibfnamefont {J.}~\bibnamefont
  {Braun}}, \bibinfo {author} {\bibfnamefont {W.-j.}\ \bibnamefont {Fu}},
  \bibinfo {author} {\bibfnamefont {J.~M.}\ \bibnamefont {Pawlowski}}, \bibinfo
  {author} {\bibfnamefont {F.}~\bibnamefont {Rennecke}}, \bibinfo {author}
  {\bibfnamefont {D.}~\bibnamefont {Rosenblüh}}, \ and\ \bibinfo {author}
  {\bibfnamefont {S.}~\bibnamefont {Yin}},\ }\href@noop {} {\  (\bibinfo {year}
  {2020}{\natexlab{b}})},\ \Eprint {http://arxiv.org/abs/2003.13112}
  {arXiv:2003.13112 [hep-ph]} \BibitemShut {NoStop}%
\bibitem [{\citenamefont {Qin}\ \emph {et~al.}(2011)\citenamefont {Qin},
  \citenamefont {Chang}, \citenamefont {Chen}, \citenamefont {Liu},\ and\
  \citenamefont {Roberts}}]{Qin:2010nq}%
  \BibitemOpen
  \bibfield  {author} {\bibinfo {author} {\bibfnamefont {S.-x.}\ \bibnamefont
  {Qin}}, \bibinfo {author} {\bibfnamefont {L.}~\bibnamefont {Chang}}, \bibinfo
  {author} {\bibfnamefont {H.}~\bibnamefont {Chen}}, \bibinfo {author}
  {\bibfnamefont {Y.-x.}\ \bibnamefont {Liu}}, \ and\ \bibinfo {author}
  {\bibfnamefont {C.~D.}\ \bibnamefont {Roberts}},\ }\href {\doibase
  10.1103/PhysRevLett.106.172301} {\bibfield  {journal} {\bibinfo  {journal}
  {Phys. Rev. Lett.}\ }\textbf {\bibinfo {volume} {106}},\ \bibinfo {pages}
  {172301} (\bibinfo {year} {2011})},\ \Eprint {http://arxiv.org/abs/1011.2876}
  {arXiv:1011.2876 [nucl-th]} \BibitemShut {NoStop}%
%%CITATION = ARXIV:1011.2876;%%
\bibitem [{\citenamefont {Fischer}\ \emph {et~al.}(2011)\citenamefont
  {Fischer}, \citenamefont {Luecker},\ and\ \citenamefont
  {Mueller}}]{Fischer:2011mz}%
  \BibitemOpen
  \bibfield  {author} {\bibinfo {author} {\bibfnamefont {C.~S.}\ \bibnamefont
  {Fischer}}, \bibinfo {author} {\bibfnamefont {J.}~\bibnamefont {Luecker}}, \
  and\ \bibinfo {author} {\bibfnamefont {J.~A.}\ \bibnamefont {Mueller}},\
  }\href {\doibase 10.1016/j.physletb.2011.07.039} {\bibfield  {journal}
  {\bibinfo  {journal} {Phys. Lett.}\ }\textbf {\bibinfo {volume} {B702}},\
  \bibinfo {pages} {438} (\bibinfo {year} {2011})},\ \Eprint
  {http://arxiv.org/abs/1104.1564} {arXiv:1104.1564 [hep-ph]} \BibitemShut
  {NoStop}%
%%CITATION = ARXIV:1104.1564;%%
\bibitem [{\citenamefont {Fischer}\ \emph
  {et~al.}(2014{\natexlab{a}})\citenamefont {Fischer}, \citenamefont {Fister},
  \citenamefont {Luecker},\ and\ \citenamefont {Pawlowski}}]{Fischer:2013eca}%
  \BibitemOpen
  \bibfield  {author} {\bibinfo {author} {\bibfnamefont {C.~S.}\ \bibnamefont
  {Fischer}}, \bibinfo {author} {\bibfnamefont {L.}~\bibnamefont {Fister}},
  \bibinfo {author} {\bibfnamefont {J.}~\bibnamefont {Luecker}}, \ and\
  \bibinfo {author} {\bibfnamefont {J.~M.}\ \bibnamefont {Pawlowski}},\ }\href
  {\doibase 10.1016/j.physletb.2014.03.057} {\bibfield  {journal} {\bibinfo
  {journal} {Phys. Lett.}\ }\textbf {\bibinfo {volume} {B732}},\ \bibinfo
  {pages} {273} (\bibinfo {year} {2014}{\natexlab{a}})},\ \Eprint
  {http://arxiv.org/abs/1306.6022} {arXiv:1306.6022 [hep-ph]} \BibitemShut
  {NoStop}%
%%CITATION = ARXIV:1306.6022;%%
\bibitem [{\citenamefont {Fischer}\ \emph
  {et~al.}(2014{\natexlab{b}})\citenamefont {Fischer}, \citenamefont
  {Luecker},\ and\ \citenamefont {Welzbacher}}]{Fischer:2014ata}%
  \BibitemOpen
  \bibfield  {author} {\bibinfo {author} {\bibfnamefont {C.~S.}\ \bibnamefont
  {Fischer}}, \bibinfo {author} {\bibfnamefont {J.}~\bibnamefont {Luecker}}, \
  and\ \bibinfo {author} {\bibfnamefont {C.~A.}\ \bibnamefont {Welzbacher}},\
  }\href {\doibase 10.1103/PhysRevD.90.034022} {\bibfield  {journal} {\bibinfo
  {journal} {Phys. Rev.}\ }\textbf {\bibinfo {volume} {D90}},\ \bibinfo {pages}
  {034022} (\bibinfo {year} {2014}{\natexlab{b}})},\ \Eprint
  {http://arxiv.org/abs/1405.4762} {arXiv:1405.4762 [hep-ph]} \BibitemShut
  {NoStop}%
%%CITATION = ARXIV:1405.4762;%%
\bibitem [{\citenamefont {Eichmann}\ \emph
  {et~al.}(2016{\natexlab{a}})\citenamefont {Eichmann}, \citenamefont
  {Fischer},\ and\ \citenamefont {Welzbacher}}]{Eichmann:2015kfa}%
  \BibitemOpen
  \bibfield  {author} {\bibinfo {author} {\bibfnamefont {G.}~\bibnamefont
  {Eichmann}}, \bibinfo {author} {\bibfnamefont {C.~S.}\ \bibnamefont
  {Fischer}}, \ and\ \bibinfo {author} {\bibfnamefont {C.~A.}\ \bibnamefont
  {Welzbacher}},\ }\href {\doibase 10.1103/PhysRevD.93.034013} {\bibfield
  {journal} {\bibinfo  {journal} {Phys. Rev.}\ }\textbf {\bibinfo {volume}
  {D93}},\ \bibinfo {pages} {034013} (\bibinfo {year} {2016}{\natexlab{a}})},\
  \Eprint {http://arxiv.org/abs/1509.02082} {arXiv:1509.02082 [hep-ph]}
  \BibitemShut {NoStop}%
%%CITATION = ARXIV:1509.02082;%%
\bibitem [{\citenamefont {Gao}\ \emph {et~al.}(2016)\citenamefont {Gao},
  \citenamefont {Chen}, \citenamefont {Liu}, \citenamefont {Qin}, \citenamefont
  {Roberts},\ and\ \citenamefont {Schmidt}}]{Gao:2015kea}%
  \BibitemOpen
  \bibfield  {author} {\bibinfo {author} {\bibfnamefont {F.}~\bibnamefont
  {Gao}}, \bibinfo {author} {\bibfnamefont {J.}~\bibnamefont {Chen}}, \bibinfo
  {author} {\bibfnamefont {Y.-X.}\ \bibnamefont {Liu}}, \bibinfo {author}
  {\bibfnamefont {S.-X.}\ \bibnamefont {Qin}}, \bibinfo {author} {\bibfnamefont
  {C.~D.}\ \bibnamefont {Roberts}}, \ and\ \bibinfo {author} {\bibfnamefont
  {S.~M.}\ \bibnamefont {Schmidt}},\ }\href {\doibase
  10.1103/PhysRevD.93.094019} {\bibfield  {journal} {\bibinfo  {journal} {Phys.
  Rev.}\ }\textbf {\bibinfo {volume} {D93}},\ \bibinfo {pages} {094019}
  (\bibinfo {year} {2016})},\ \Eprint {http://arxiv.org/abs/1507.00875}
  {arXiv:1507.00875 [nucl-th]} \BibitemShut {NoStop}%
%%CITATION = ARXIV:1507.00875;%%
\bibitem [{\citenamefont {Gao}\ and\ \citenamefont {Liu}(2016)}]{Gao:2016qkh}%
  \BibitemOpen
  \bibfield  {author} {\bibinfo {author} {\bibfnamefont {F.}~\bibnamefont
  {Gao}}\ and\ \bibinfo {author} {\bibfnamefont {Y.-x.}\ \bibnamefont {Liu}},\
  }\href {\doibase 10.1103/PhysRevD.94.076009} {\bibfield  {journal} {\bibinfo
  {journal} {Phys. Rev.}\ }\textbf {\bibinfo {volume} {D94}},\ \bibinfo {pages}
  {076009} (\bibinfo {year} {2016})},\ \Eprint
  {http://arxiv.org/abs/1607.01675} {arXiv:1607.01675 [hep-ph]} \BibitemShut
  {NoStop}%
%%CITATION = ARXIV:1607.01675;%%
\bibitem [{\citenamefont {Tang}\ \emph {et~al.}(2019)\citenamefont {Tang},
  \citenamefont {Gao},\ and\ \citenamefont {Liu}}]{Tang:2019zbk}%
  \BibitemOpen
  \bibfield  {author} {\bibinfo {author} {\bibfnamefont {C.}~\bibnamefont
  {Tang}}, \bibinfo {author} {\bibfnamefont {F.}~\bibnamefont {Gao}}, \ and\
  \bibinfo {author} {\bibfnamefont {Y.-X.}\ \bibnamefont {Liu}},\ }\href
  {\doibase 10.1103/PhysRevD.100.056001} {\bibfield  {journal} {\bibinfo
  {journal} {Phys. Rev. D}\ }\textbf {\bibinfo {volume} {100}},\ \bibinfo
  {pages} {056001} (\bibinfo {year} {2019})},\ \Eprint
  {http://arxiv.org/abs/1902.01679} {arXiv:1902.01679 [hep-ph]} \BibitemShut
  {NoStop}%
\bibitem [{\citenamefont {Gunkel}\ \emph {et~al.}(2019)\citenamefont {Gunkel},
  \citenamefont {Fischer},\ and\ \citenamefont {Isserstedt}}]{Gunkel:2019xnh}%
  \BibitemOpen
  \bibfield  {author} {\bibinfo {author} {\bibfnamefont {P.~J.}\ \bibnamefont
  {Gunkel}}, \bibinfo {author} {\bibfnamefont {C.~S.}\ \bibnamefont {Fischer}},
  \ and\ \bibinfo {author} {\bibfnamefont {P.}~\bibnamefont {Isserstedt}},\
  }\href {\doibase 10.1140/epja/i2019-12868-1} {\bibfield  {journal} {\bibinfo
  {journal} {Eur. Phys. J.}\ }\textbf {\bibinfo {volume} {A55}},\ \bibinfo
  {pages} {169} (\bibinfo {year} {2019})},\ \Eprint
  {http://arxiv.org/abs/1907.08110} {arXiv:1907.08110 [hep-ph]} \BibitemShut
  {NoStop}%
%%CITATION = ARXIV:1907.08110;%%
\bibitem [{\citenamefont {Isserstedt}\ \emph {et~al.}(2019)\citenamefont
  {Isserstedt}, \citenamefont {Buballa}, \citenamefont {Fischer},\ and\
  \citenamefont {Gunkel}}]{Isserstedt:2019pgx}%
  \BibitemOpen
  \bibfield  {author} {\bibinfo {author} {\bibfnamefont {P.}~\bibnamefont
  {Isserstedt}}, \bibinfo {author} {\bibfnamefont {M.}~\bibnamefont {Buballa}},
  \bibinfo {author} {\bibfnamefont {C.~S.}\ \bibnamefont {Fischer}}, \ and\
  \bibinfo {author} {\bibfnamefont {P.~J.}\ \bibnamefont {Gunkel}},\ }\href
  {\doibase 10.1103/PhysRevD.100.074011} {\bibfield  {journal} {\bibinfo
  {journal} {Phys. Rev.}\ }\textbf {\bibinfo {volume} {D100}},\ \bibinfo
  {pages} {074011} (\bibinfo {year} {2019})},\ \Eprint
  {http://arxiv.org/abs/1906.11644} {arXiv:1906.11644 [hep-ph]} \BibitemShut
  {NoStop}%
%%CITATION = ARXIV:1906.11644;%%
\bibitem [{\citenamefont {Reinosa}\ \emph {et~al.}(2015)\citenamefont
  {Reinosa}, \citenamefont {Serreau},\ and\ \citenamefont
  {Tissier}}]{Reinosa:2015oua}%
  \BibitemOpen
  \bibfield  {author} {\bibinfo {author} {\bibfnamefont {U.}~\bibnamefont
  {Reinosa}}, \bibinfo {author} {\bibfnamefont {J.}~\bibnamefont {Serreau}}, \
  and\ \bibinfo {author} {\bibfnamefont {M.}~\bibnamefont {Tissier}},\ }\href
  {\doibase 10.1103/PhysRevD.92.025021} {\bibfield  {journal} {\bibinfo
  {journal} {Phys. Rev.}\ }\textbf {\bibinfo {volume} {D92}},\ \bibinfo {pages}
  {025021} (\bibinfo {year} {2015})},\ \Eprint
  {http://arxiv.org/abs/1504.02916} {arXiv:1504.02916 [hep-th]} \BibitemShut
  {NoStop}%
%%CITATION = ARXIV:1504.02916;%%
\bibitem [{\citenamefont {Reinosa}\ \emph {et~al.}(2017)\citenamefont
  {Reinosa}, \citenamefont {Serreau}, \citenamefont {Tissier},\ and\
  \citenamefont {Tresmontant}}]{Reinosa:2016iml}%
  \BibitemOpen
  \bibfield  {author} {\bibinfo {author} {\bibfnamefont {U.}~\bibnamefont
  {Reinosa}}, \bibinfo {author} {\bibfnamefont {J.}~\bibnamefont {Serreau}},
  \bibinfo {author} {\bibfnamefont {M.}~\bibnamefont {Tissier}}, \ and\
  \bibinfo {author} {\bibfnamefont {A.}~\bibnamefont {Tresmontant}},\ }\href
  {\doibase 10.1103/PhysRevD.95.045014} {\bibfield  {journal} {\bibinfo
  {journal} {Phys. Rev.}\ }\textbf {\bibinfo {volume} {D95}},\ \bibinfo {pages}
  {045014} (\bibinfo {year} {2017})},\ \Eprint
  {http://arxiv.org/abs/1606.08012} {arXiv:1606.08012 [hep-th]} \BibitemShut
  {NoStop}%
%%CITATION = ARXIV:1606.08012;%%
\bibitem [{\citenamefont {Maelger}\ \emph
  {et~al.}(2018{\natexlab{a}})\citenamefont {Maelger}, \citenamefont
  {Reinosa},\ and\ \citenamefont {Serreau}}]{Maelger:2017amh}%
  \BibitemOpen
  \bibfield  {author} {\bibinfo {author} {\bibfnamefont {J.}~\bibnamefont
  {Maelger}}, \bibinfo {author} {\bibfnamefont {U.}~\bibnamefont {Reinosa}}, \
  and\ \bibinfo {author} {\bibfnamefont {J.}~\bibnamefont {Serreau}},\ }\href
  {\doibase 10.1103/PhysRevD.97.074027} {\bibfield  {journal} {\bibinfo
  {journal} {Phys. Rev.}\ }\textbf {\bibinfo {volume} {D97}},\ \bibinfo {pages}
  {074027} (\bibinfo {year} {2018}{\natexlab{a}})},\ \Eprint
  {http://arxiv.org/abs/1710.01930} {arXiv:1710.01930 [hep-ph]} \BibitemShut
  {NoStop}%
%%CITATION = ARXIV:1710.01930;%%
\bibitem [{\citenamefont {Maelger}\ \emph
  {et~al.}(2018{\natexlab{b}})\citenamefont {Maelger}, \citenamefont
  {Reinosa},\ and\ \citenamefont {Serreau}}]{Maelger:2018vow}%
  \BibitemOpen
  \bibfield  {author} {\bibinfo {author} {\bibfnamefont {J.}~\bibnamefont
  {Maelger}}, \bibinfo {author} {\bibfnamefont {U.}~\bibnamefont {Reinosa}}, \
  and\ \bibinfo {author} {\bibfnamefont {J.}~\bibnamefont {Serreau}},\ }\href
  {\doibase 10.1103/PhysRevD.98.094020} {\bibfield  {journal} {\bibinfo
  {journal} {Phys. Rev.}\ }\textbf {\bibinfo {volume} {D98}},\ \bibinfo {pages}
  {094020} (\bibinfo {year} {2018}{\natexlab{b}})},\ \Eprint
  {http://arxiv.org/abs/1805.10015} {arXiv:1805.10015 [hep-th]} \BibitemShut
  {NoStop}%
%%CITATION = ARXIV:1805.10015;%%
\bibitem [{\citenamefont {Maelger}\ \emph {et~al.}(2020)\citenamefont
  {Maelger}, \citenamefont {Reinosa},\ and\ \citenamefont
  {Serreau}}]{Maelger:2019cbk}%
  \BibitemOpen
  \bibfield  {author} {\bibinfo {author} {\bibfnamefont {J.}~\bibnamefont
  {Maelger}}, \bibinfo {author} {\bibfnamefont {U.}~\bibnamefont {Reinosa}}, \
  and\ \bibinfo {author} {\bibfnamefont {J.}~\bibnamefont {Serreau}},\ }\href
  {\doibase 10.1103/PhysRevD.101.014028} {\bibfield  {journal} {\bibinfo
  {journal} {Phys. Rev.}\ }\textbf {\bibinfo {volume} {D101}},\ \bibinfo
  {pages} {014028} (\bibinfo {year} {2020})},\ \Eprint
  {http://arxiv.org/abs/1903.04184} {arXiv:1903.04184 [hep-th]} \BibitemShut
  {NoStop}%
%%CITATION = ARXIV:1903.04184;%%
\bibitem [{\citenamefont {Aguilar}\ \emph {et~al.}(2017)\citenamefont
  {Aguilar}, \citenamefont {Cardona}, \citenamefont {Ferreira},\ and\
  \citenamefont {Papavassiliou}}]{Aguilar:2016lbe}%
  \BibitemOpen
  \bibfield  {author} {\bibinfo {author} {\bibfnamefont {A.~C.}\ \bibnamefont
  {Aguilar}}, \bibinfo {author} {\bibfnamefont {J.~C.}\ \bibnamefont
  {Cardona}}, \bibinfo {author} {\bibfnamefont {M.~N.}\ \bibnamefont
  {Ferreira}}, \ and\ \bibinfo {author} {\bibfnamefont {J.}~\bibnamefont
  {Papavassiliou}},\ }\href {\doibase 10.1103/PhysRevD.96.014029} {\bibfield
  {journal} {\bibinfo  {journal} {Phys. Rev.}\ }\textbf {\bibinfo {volume}
  {D96}},\ \bibinfo {pages} {014029} (\bibinfo {year} {2017})},\ \Eprint
  {http://arxiv.org/abs/1610.06158} {arXiv:1610.06158 [hep-ph]} \BibitemShut
  {NoStop}%
%%CITATION = ARXIV:1610.06158;%%
\bibitem [{\citenamefont {Aguilar}\ \emph
  {et~al.}(2018{\natexlab{a}})\citenamefont {Aguilar}, \citenamefont {Binosi},
  \citenamefont {Figueiredo},\ and\ \citenamefont
  {Papavassiliou}}]{Aguilar:2017dco}%
  \BibitemOpen
  \bibfield  {author} {\bibinfo {author} {\bibfnamefont {A.~C.}\ \bibnamefont
  {Aguilar}}, \bibinfo {author} {\bibfnamefont {D.}~\bibnamefont {Binosi}},
  \bibinfo {author} {\bibfnamefont {C.~T.}\ \bibnamefont {Figueiredo}}, \ and\
  \bibinfo {author} {\bibfnamefont {J.}~\bibnamefont {Papavassiliou}},\ }\href
  {\doibase 10.1140/epjc/s10052-018-5679-2} {\bibfield  {journal} {\bibinfo
  {journal} {Eur. Phys. J.}\ }\textbf {\bibinfo {volume} {C78}},\ \bibinfo
  {pages} {181} (\bibinfo {year} {2018}{\natexlab{a}})},\ \Eprint
  {http://arxiv.org/abs/1712.06926} {arXiv:1712.06926 [hep-ph]} \BibitemShut
  {NoStop}%
%%CITATION = ARXIV:1712.06926;%%
\bibitem [{\citenamefont {Aguilar}\ \emph
  {et~al.}(2018{\natexlab{b}})\citenamefont {Aguilar}, \citenamefont {Cardona},
  \citenamefont {Ferreira},\ and\ \citenamefont
  {Papavassiliou}}]{Aguilar:2018epe}%
  \BibitemOpen
  \bibfield  {author} {\bibinfo {author} {\bibfnamefont {A.~C.}\ \bibnamefont
  {Aguilar}}, \bibinfo {author} {\bibfnamefont {J.~C.}\ \bibnamefont
  {Cardona}}, \bibinfo {author} {\bibfnamefont {M.~N.}\ \bibnamefont
  {Ferreira}}, \ and\ \bibinfo {author} {\bibfnamefont {J.}~\bibnamefont
  {Papavassiliou}},\ }\href {\doibase 10.1103/PhysRevD.98.014002} {\bibfield
  {journal} {\bibinfo  {journal} {Phys. Rev.}\ }\textbf {\bibinfo {volume}
  {D98}},\ \bibinfo {pages} {014002} (\bibinfo {year} {2018}{\natexlab{b}})},\
  \Eprint {http://arxiv.org/abs/1804.04229} {arXiv:1804.04229 [hep-ph]}
  \BibitemShut {NoStop}%
%%CITATION = ARXIV:1804.04229;%%
\bibitem [{\citenamefont {Bazavov}\ \emph {et~al.}(2012)\citenamefont {Bazavov}
  \emph {et~al.}}]{Bazavov:2012vg}%
  \BibitemOpen
  \bibfield  {author} {\bibinfo {author} {\bibfnamefont {A.}~\bibnamefont
  {Bazavov}} \emph {et~al.},\ }\href {\doibase 10.1103/PhysRevLett.109.192302}
  {\bibfield  {journal} {\bibinfo  {journal} {Phys. Rev. Lett.}\ }\textbf
  {\bibinfo {volume} {109}},\ \bibinfo {pages} {192302} (\bibinfo {year}
  {2012})},\ \Eprint {http://arxiv.org/abs/1208.1220} {arXiv:1208.1220
  [hep-lat]} \BibitemShut {NoStop}%
%%CITATION = ARXIV:1208.1220;%%
\bibitem [{\citenamefont {Borsanyi}\ \emph {et~al.}(2013)\citenamefont
  {Borsanyi}, \citenamefont {Fodor}, \citenamefont {Katz}, \citenamefont
  {Krieg}, \citenamefont {Ratti},\ and\ \citenamefont
  {Szabo}}]{Borsanyi:2013hza}%
  \BibitemOpen
  \bibfield  {author} {\bibinfo {author} {\bibfnamefont {S.}~\bibnamefont
  {Borsanyi}}, \bibinfo {author} {\bibfnamefont {Z.}~\bibnamefont {Fodor}},
  \bibinfo {author} {\bibfnamefont {S.~D.}\ \bibnamefont {Katz}}, \bibinfo
  {author} {\bibfnamefont {S.}~\bibnamefont {Krieg}}, \bibinfo {author}
  {\bibfnamefont {C.}~\bibnamefont {Ratti}}, \ and\ \bibinfo {author}
  {\bibfnamefont {K.~K.}\ \bibnamefont {Szabo}},\ }\href {\doibase
  10.1103/PhysRevLett.111.062005} {\bibfield  {journal} {\bibinfo  {journal}
  {Phys. Rev. Lett.}\ }\textbf {\bibinfo {volume} {111}},\ \bibinfo {pages}
  {062005} (\bibinfo {year} {2013})},\ \Eprint {http://arxiv.org/abs/1305.5161}
  {arXiv:1305.5161 [hep-lat]} \BibitemShut {NoStop}%
%%CITATION = ARXIV:1305.5161;%%
\bibitem [{\citenamefont {Borsanyi}\ \emph {et~al.}(2014)\citenamefont
  {Borsanyi}, \citenamefont {Fodor}, \citenamefont {Katz}, \citenamefont
  {Krieg}, \citenamefont {Ratti},\ and\ \citenamefont
  {Szabo}}]{Borsanyi:2014ewa}%
  \BibitemOpen
  \bibfield  {author} {\bibinfo {author} {\bibfnamefont {S.}~\bibnamefont
  {Borsanyi}}, \bibinfo {author} {\bibfnamefont {Z.}~\bibnamefont {Fodor}},
  \bibinfo {author} {\bibfnamefont {S.~D.}\ \bibnamefont {Katz}}, \bibinfo
  {author} {\bibfnamefont {S.}~\bibnamefont {Krieg}}, \bibinfo {author}
  {\bibfnamefont {C.}~\bibnamefont {Ratti}}, \ and\ \bibinfo {author}
  {\bibfnamefont {K.~K.}\ \bibnamefont {Szabo}},\ }\href {\doibase
  10.1103/PhysRevLett.113.052301} {\bibfield  {journal} {\bibinfo  {journal}
  {Phys. Rev. Lett.}\ }\textbf {\bibinfo {volume} {113}},\ \bibinfo {pages}
  {052301} (\bibinfo {year} {2014})},\ \Eprint {http://arxiv.org/abs/1403.4576}
  {arXiv:1403.4576 [hep-lat]} \BibitemShut {NoStop}%
%%CITATION = ARXIV:1403.4576;%%
\bibitem [{\citenamefont {Bonati}\ \emph {et~al.}(2015)\citenamefont {Bonati},
  \citenamefont {D'Elia}, \citenamefont {Mariti}, \citenamefont {Mesiti},
  \citenamefont {Negro},\ and\ \citenamefont {Sanfilippo}}]{Bonati:2015bha}%
  \BibitemOpen
  \bibfield  {author} {\bibinfo {author} {\bibfnamefont {C.}~\bibnamefont
  {Bonati}}, \bibinfo {author} {\bibfnamefont {M.}~\bibnamefont {D'Elia}},
  \bibinfo {author} {\bibfnamefont {M.}~\bibnamefont {Mariti}}, \bibinfo
  {author} {\bibfnamefont {M.}~\bibnamefont {Mesiti}}, \bibinfo {author}
  {\bibfnamefont {F.}~\bibnamefont {Negro}}, \ and\ \bibinfo {author}
  {\bibfnamefont {F.}~\bibnamefont {Sanfilippo}},\ }\href {\doibase
  10.1103/PhysRevD.92.054503} {\bibfield  {journal} {\bibinfo  {journal} {Phys.
  Rev.}\ }\textbf {\bibinfo {volume} {D92}},\ \bibinfo {pages} {054503}
  (\bibinfo {year} {2015})},\ \Eprint {http://arxiv.org/abs/1507.03571}
  {arXiv:1507.03571 [hep-lat]} \BibitemShut {NoStop}%
%%CITATION = ARXIV:1507.03571;%%
\bibitem [{\citenamefont {Bellwied}\ \emph {et~al.}(2015)\citenamefont
  {Bellwied}, \citenamefont {Borsanyi}, \citenamefont {Fodor}, \citenamefont
  {Guenther}, \citenamefont {Katz}, \citenamefont {Ratti},\ and\ \citenamefont
  {Szabo}}]{Bellwied:2015rza}%
  \BibitemOpen
  \bibfield  {author} {\bibinfo {author} {\bibfnamefont {R.}~\bibnamefont
  {Bellwied}}, \bibinfo {author} {\bibfnamefont {S.}~\bibnamefont {Borsanyi}},
  \bibinfo {author} {\bibfnamefont {Z.}~\bibnamefont {Fodor}}, \bibinfo
  {author} {\bibfnamefont {J.}~\bibnamefont {Guenther}}, \bibinfo {author}
  {\bibfnamefont {S.~D.}\ \bibnamefont {Katz}}, \bibinfo {author}
  {\bibfnamefont {C.}~\bibnamefont {Ratti}}, \ and\ \bibinfo {author}
  {\bibfnamefont {K.~K.}\ \bibnamefont {Szabo}},\ }\href {\doibase
  10.1016/j.physletb.2015.11.011} {\bibfield  {journal} {\bibinfo  {journal}
  {Phys. Lett.}\ }\textbf {\bibinfo {volume} {B751}},\ \bibinfo {pages} {559}
  (\bibinfo {year} {2015})},\ \Eprint {http://arxiv.org/abs/1507.07510}
  {arXiv:1507.07510 [hep-lat]} \BibitemShut {NoStop}%
%%CITATION = ARXIV:1507.07510;%%
\bibitem [{\citenamefont {Bazavov}\ \emph
  {et~al.}(2017{\natexlab{a}})\citenamefont {Bazavov} \emph
  {et~al.}}]{Bazavov:2017dus}%
  \BibitemOpen
  \bibfield  {author} {\bibinfo {author} {\bibfnamefont {A.}~\bibnamefont
  {Bazavov}} \emph {et~al.},\ }\href {\doibase 10.1103/PhysRevD.95.054504}
  {\bibfield  {journal} {\bibinfo  {journal} {Phys. Rev.}\ }\textbf {\bibinfo
  {volume} {D95}},\ \bibinfo {pages} {054504} (\bibinfo {year}
  {2017}{\natexlab{a}})},\ \Eprint {http://arxiv.org/abs/1701.04325}
  {arXiv:1701.04325 [hep-lat]} \BibitemShut {NoStop}%
%%CITATION = ARXIV:1701.04325;%%
\bibitem [{\citenamefont {Bazavov}\ \emph
  {et~al.}(2017{\natexlab{b}})\citenamefont {Bazavov} \emph
  {et~al.}}]{Bazavov:2017tot}%
  \BibitemOpen
  \bibfield  {author} {\bibinfo {author} {\bibfnamefont {A.}~\bibnamefont
  {Bazavov}} \emph {et~al.} (\bibinfo {collaboration} {HotQCD}),\ }\href
  {\doibase 10.1103/PhysRevD.96.074510} {\bibfield  {journal} {\bibinfo
  {journal} {Phys. Rev.}\ }\textbf {\bibinfo {volume} {D96}},\ \bibinfo {pages}
  {074510} (\bibinfo {year} {2017}{\natexlab{b}})},\ \Eprint
  {http://arxiv.org/abs/1708.04897} {arXiv:1708.04897 [hep-lat]} \BibitemShut
  {NoStop}%
%%CITATION = ARXIV:1708.04897;%%
\bibitem [{\citenamefont {Bonati}\ \emph {et~al.}(2018)\citenamefont {Bonati},
  \citenamefont {D'Elia}, \citenamefont {Negro}, \citenamefont {Sanfilippo},\
  and\ \citenamefont {Zambello}}]{Bonati:2018nut}%
  \BibitemOpen
  \bibfield  {author} {\bibinfo {author} {\bibfnamefont {C.}~\bibnamefont
  {Bonati}}, \bibinfo {author} {\bibfnamefont {M.}~\bibnamefont {D'Elia}},
  \bibinfo {author} {\bibfnamefont {F.}~\bibnamefont {Negro}}, \bibinfo
  {author} {\bibfnamefont {F.}~\bibnamefont {Sanfilippo}}, \ and\ \bibinfo
  {author} {\bibfnamefont {K.}~\bibnamefont {Zambello}},\ }\href {\doibase
  10.1103/PhysRevD.98.054510} {\bibfield  {journal} {\bibinfo  {journal} {Phys.
  Rev.}\ }\textbf {\bibinfo {volume} {D98}},\ \bibinfo {pages} {054510}
  (\bibinfo {year} {2018})},\ \Eprint {http://arxiv.org/abs/1805.02960}
  {arXiv:1805.02960 [hep-lat]} \BibitemShut {NoStop}%
%%CITATION = ARXIV:1805.02960;%%
\bibitem [{\citenamefont {Borsanyi}\ \emph {et~al.}(2018)\citenamefont
  {Borsanyi}, \citenamefont {Fodor}, \citenamefont {Guenther}, \citenamefont
  {Katz}, \citenamefont {Szabo}, \citenamefont {Pasztor}, \citenamefont
  {Portillo},\ and\ \citenamefont {Ratti}}]{Borsanyi:2018grb}%
  \BibitemOpen
  \bibfield  {author} {\bibinfo {author} {\bibfnamefont {S.}~\bibnamefont
  {Borsanyi}}, \bibinfo {author} {\bibfnamefont {Z.}~\bibnamefont {Fodor}},
  \bibinfo {author} {\bibfnamefont {J.~N.}\ \bibnamefont {Guenther}}, \bibinfo
  {author} {\bibfnamefont {S.~K.}\ \bibnamefont {Katz}}, \bibinfo {author}
  {\bibfnamefont {K.~K.}\ \bibnamefont {Szabo}}, \bibinfo {author}
  {\bibfnamefont {A.}~\bibnamefont {Pasztor}}, \bibinfo {author} {\bibfnamefont
  {I.}~\bibnamefont {Portillo}}, \ and\ \bibinfo {author} {\bibfnamefont
  {C.}~\bibnamefont {Ratti}},\ }\href {\doibase 10.1007/JHEP10(2018)205}
  {\bibfield  {journal} {\bibinfo  {journal} {JHEP}\ }\textbf {\bibinfo
  {volume} {10}},\ \bibinfo {pages} {205} (\bibinfo {year} {2018})},\ \Eprint
  {http://arxiv.org/abs/1805.04445} {arXiv:1805.04445 [hep-lat]} \BibitemShut
  {NoStop}%
%%CITATION = ARXIV:1805.04445;%%
\bibitem [{\citenamefont {Bazavov}\ \emph {et~al.}(2019)\citenamefont {Bazavov}
  \emph {et~al.}}]{Bazavov:2018mes}%
  \BibitemOpen
  \bibfield  {author} {\bibinfo {author} {\bibfnamefont {A.}~\bibnamefont
  {Bazavov}} \emph {et~al.} (\bibinfo {collaboration} {HotQCD}),\ }\href
  {\doibase 10.1016/j.physletb.2019.05.013} {\bibfield  {journal} {\bibinfo
  {journal} {Phys. Lett.}\ }\textbf {\bibinfo {volume} {B795}},\ \bibinfo
  {pages} {15} (\bibinfo {year} {2019})},\ \Eprint
  {http://arxiv.org/abs/1812.08235} {arXiv:1812.08235 [hep-lat]} \BibitemShut
  {NoStop}%
%%CITATION = ARXIV:1812.08235;%%
\bibitem [{\citenamefont {Guenther}\ \emph {et~al.}(2018)\citenamefont
  {Guenther}, \citenamefont {Borsanyi}, \citenamefont {Fodor}, \citenamefont
  {Katz}, \citenamefont {Szab}, \citenamefont {Pasztor}, \citenamefont
  {Portillo},\ and\ \citenamefont {Ratti}}]{Guenther:2018flo}%
  \BibitemOpen
  \bibfield  {author} {\bibinfo {author} {\bibfnamefont {J.~N.}\ \bibnamefont
  {Guenther}}, \bibinfo {author} {\bibfnamefont {S.}~\bibnamefont {Borsanyi}},
  \bibinfo {author} {\bibfnamefont {Z.}~\bibnamefont {Fodor}}, \bibinfo
  {author} {\bibfnamefont {S.~K.}\ \bibnamefont {Katz}}, \bibinfo {author}
  {\bibfnamefont {K.~K.}\ \bibnamefont {Szab}}, \bibinfo {author}
  {\bibfnamefont {A.}~\bibnamefont {Pasztor}}, \bibinfo {author} {\bibfnamefont
  {I.}~\bibnamefont {Portillo}}, \ and\ \bibinfo {author} {\bibfnamefont
  {C.}~\bibnamefont {Ratti}},\ }\bibfield  {booktitle} {\emph {\bibinfo
  {booktitle} {{Proceedings, 34th Winter Workshop on Nuclear Dynamics (WWND
  2018): Guadeloupe, French West Indies, March 25-31, 2018}}},\ }\href
  {\doibase 10.1088/1742-6596/1070/1/012002} {\bibfield  {journal} {\bibinfo
  {journal} {J. Phys. Conf. Ser.}\ }\textbf {\bibinfo {volume} {1070}},\
  \bibinfo {pages} {012002} (\bibinfo {year} {2018})}\BibitemShut {NoStop}%
%%CITATION = 00462,1070,012002;%%
\bibitem [{\citenamefont {Ding}\ \emph {et~al.}(2019)\citenamefont {Ding} \emph
  {et~al.}}]{Ding:2019prx}%
  \BibitemOpen
  \bibfield  {author} {\bibinfo {author} {\bibfnamefont {H.~T.}\ \bibnamefont
  {Ding}} \emph {et~al.},\ }\href {\doibase 10.1103/PhysRevLett.123.062002}
  {\bibfield  {journal} {\bibinfo  {journal} {Phys. Rev. Lett.}\ }\textbf
  {\bibinfo {volume} {123}},\ \bibinfo {pages} {062002} (\bibinfo {year}
  {2019})},\ \Eprint {http://arxiv.org/abs/1903.04801} {arXiv:1903.04801
  [hep-lat]} \BibitemShut {NoStop}%
%%CITATION = ARXIV:1903.04801;%%
\bibitem [{\citenamefont {Borsanyi}\ \emph {et~al.}(2020)\citenamefont
  {Borsanyi}, \citenamefont {Fodor}, \citenamefont {Guenther}, \citenamefont
  {Kara}, \citenamefont {Katz}, \citenamefont {Parotto}, \citenamefont
  {Pasztor}, \citenamefont {Ratti},\ and\ \citenamefont
  {Szabo}}]{Borsanyi:2020fev}%
  \BibitemOpen
  \bibfield  {author} {\bibinfo {author} {\bibfnamefont {S.}~\bibnamefont
  {Borsanyi}}, \bibinfo {author} {\bibfnamefont {Z.}~\bibnamefont {Fodor}},
  \bibinfo {author} {\bibfnamefont {J.~N.}\ \bibnamefont {Guenther}}, \bibinfo
  {author} {\bibfnamefont {R.}~\bibnamefont {Kara}}, \bibinfo {author}
  {\bibfnamefont {S.~D.}\ \bibnamefont {Katz}}, \bibinfo {author}
  {\bibfnamefont {P.}~\bibnamefont {Parotto}}, \bibinfo {author} {\bibfnamefont
  {A.}~\bibnamefont {Pasztor}}, \bibinfo {author} {\bibfnamefont
  {C.}~\bibnamefont {Ratti}}, \ and\ \bibinfo {author} {\bibfnamefont {K.~K.}\
  \bibnamefont {Szabo}},\ }\href@noop {} {\  (\bibinfo {year} {2020})},\
  \Eprint {http://arxiv.org/abs/2002.02821} {arXiv:2002.02821 [hep-lat]}
  \BibitemShut {NoStop}%
\bibitem [{\citenamefont {Braun}\ \emph {et~al.}(2010)\citenamefont {Braun},
  \citenamefont {Gies},\ and\ \citenamefont {Pawlowski}}]{Braun:2007bx}%
  \BibitemOpen
  \bibfield  {author} {\bibinfo {author} {\bibfnamefont {J.}~\bibnamefont
  {Braun}}, \bibinfo {author} {\bibfnamefont {H.}~\bibnamefont {Gies}}, \ and\
  \bibinfo {author} {\bibfnamefont {J.~M.}\ \bibnamefont {Pawlowski}},\ }\href
  {\doibase 10.1016/j.physletb.2010.01.009} {\bibfield  {journal} {\bibinfo
  {journal} {Phys.Lett.}\ }\textbf {\bibinfo {volume} {B684}},\ \bibinfo
  {pages} {262} (\bibinfo {year} {2010})},\ \Eprint
  {http://arxiv.org/abs/0708.2413} {arXiv:0708.2413 [hep-th]} \BibitemShut
  {NoStop}%
\bibitem [{\citenamefont {Fister}\ and\ \citenamefont
  {Pawlowski}(2013)}]{Fister:2013bh}%
  \BibitemOpen
  \bibfield  {author} {\bibinfo {author} {\bibfnamefont {L.}~\bibnamefont
  {Fister}}\ and\ \bibinfo {author} {\bibfnamefont {J.~M.}\ \bibnamefont
  {Pawlowski}},\ }\href {\doibase 10.1103/PhysRevD.88.045010} {\bibfield
  {journal} {\bibinfo  {journal} {Phys.Rev.}\ }\textbf {\bibinfo {volume}
  {D88}},\ \bibinfo {pages} {045010} (\bibinfo {year} {2013})},\ \Eprint
  {http://arxiv.org/abs/1301.4163} {arXiv:1301.4163 [hep-ph]} \BibitemShut
  {NoStop}%
%%CITATION = ARXIV:1301.4163;%%
\bibitem [{\citenamefont {Cucchieri}\ \emph {et~al.}(2007)\citenamefont
  {Cucchieri}, \citenamefont {Maas},\ and\ \citenamefont
  {Mendes}}]{Cucchieri:2007ta}%
  \BibitemOpen
  \bibfield  {author} {\bibinfo {author} {\bibfnamefont {A.}~\bibnamefont
  {Cucchieri}}, \bibinfo {author} {\bibfnamefont {A.}~\bibnamefont {Maas}}, \
  and\ \bibinfo {author} {\bibfnamefont {T.}~\bibnamefont {Mendes}},\ }\href
  {\doibase 10.1103/PhysRevD.75.076003} {\bibfield  {journal} {\bibinfo
  {journal} {Phys. Rev.}\ }\textbf {\bibinfo {volume} {D75}},\ \bibinfo {pages}
  {076003} (\bibinfo {year} {2007})},\ \Eprint
  {http://arxiv.org/abs/hep-lat/0702022} {arXiv:hep-lat/0702022 [hep-lat]}
  \BibitemShut {NoStop}%
%%CITATION = HEP-LAT/0702022;%%
\bibitem [{\citenamefont {Ilgenfritz}\ \emph {et~al.}(2007)\citenamefont
  {Ilgenfritz}, \citenamefont {Muller-Preussker}, \citenamefont {Sternbeck},
  \citenamefont {Schiller},\ and\ \citenamefont
  {Bogolubsky}}]{Ilgenfritz:2006he}%
  \BibitemOpen
  \bibfield  {author} {\bibinfo {author} {\bibfnamefont {E.-M.}\ \bibnamefont
  {Ilgenfritz}}, \bibinfo {author} {\bibfnamefont {M.}~\bibnamefont
  {Muller-Preussker}}, \bibinfo {author} {\bibfnamefont {A.}~\bibnamefont
  {Sternbeck}}, \bibinfo {author} {\bibfnamefont {A.}~\bibnamefont {Schiller}},
  \ and\ \bibinfo {author} {\bibfnamefont {I.}~\bibnamefont {Bogolubsky}},\
  }\href {\doibase 10.1590/S0103-97332007000200006} {\bibfield  {journal}
  {\bibinfo  {journal} {Braz. J. Phys.}\ }\textbf {\bibinfo {volume} {37}},\
  \bibinfo {pages} {193} (\bibinfo {year} {2007})},\ \Eprint
  {http://arxiv.org/abs/hep-lat/0609043} {arXiv:hep-lat/0609043} \BibitemShut
  {NoStop}%
\bibitem [{\citenamefont {Williams}(2015)}]{Williams:2014iea}%
  \BibitemOpen
  \bibfield  {author} {\bibinfo {author} {\bibfnamefont {R.}~\bibnamefont
  {Williams}},\ }\href {\doibase 10.1140/epja/i2015-15057-4} {\bibfield
  {journal} {\bibinfo  {journal} {Eur. Phys. J.}\ }\textbf {\bibinfo {volume}
  {A51}},\ \bibinfo {pages} {57} (\bibinfo {year} {2015})},\ \Eprint
  {http://arxiv.org/abs/1404.2545} {arXiv:1404.2545 [hep-ph]} \BibitemShut
  {NoStop}%
%%CITATION = ARXIV:1404.2545;%%
\bibitem [{\citenamefont {Williams}\ \emph {et~al.}(2016)\citenamefont
  {Williams}, \citenamefont {Fischer},\ and\ \citenamefont
  {Heupel}}]{Williams:2015cvx}%
  \BibitemOpen
  \bibfield  {author} {\bibinfo {author} {\bibfnamefont {R.}~\bibnamefont
  {Williams}}, \bibinfo {author} {\bibfnamefont {C.~S.}\ \bibnamefont
  {Fischer}}, \ and\ \bibinfo {author} {\bibfnamefont {W.}~\bibnamefont
  {Heupel}},\ }\href {\doibase 10.1103/PhysRevD.93.034026} {\bibfield
  {journal} {\bibinfo  {journal} {Phys. Rev.}\ }\textbf {\bibinfo {volume}
  {D93}},\ \bibinfo {pages} {034026} (\bibinfo {year} {2016})},\ \Eprint
  {http://arxiv.org/abs/1512.00455} {arXiv:1512.00455 [hep-ph]} \BibitemShut
  {NoStop}%
%%CITATION = ARXIV:1512.00455;%%
\bibitem [{\citenamefont {Maris}\ and\ \citenamefont
  {Tandy}(1999)}]{Maris:1999nt}%
  \BibitemOpen
  \bibfield  {author} {\bibinfo {author} {\bibfnamefont {P.}~\bibnamefont
  {Maris}}\ and\ \bibinfo {author} {\bibfnamefont {P.~C.}\ \bibnamefont
  {Tandy}},\ }\href {\doibase 10.1103/PhysRevC.60.055214} {\bibfield  {journal}
  {\bibinfo  {journal} {Phys. Rev.}\ }\textbf {\bibinfo {volume} {C60}},\
  \bibinfo {pages} {055214} (\bibinfo {year} {1999})},\ \Eprint
  {http://arxiv.org/abs/nucl-th/9905056} {arXiv:nucl-th/9905056 [nucl-th]}
  \BibitemShut {NoStop}%
%%CITATION = NUCL-TH/9905056;%%
\bibitem [{\citenamefont {Eichmann}\ \emph
  {et~al.}(2016{\natexlab{b}})\citenamefont {Eichmann}, \citenamefont
  {Sanchis-Alepuz}, \citenamefont {Williams}, \citenamefont {Alkofer},\ and\
  \citenamefont {Fischer}}]{Eichmann:2016yit}%
  \BibitemOpen
  \bibfield  {author} {\bibinfo {author} {\bibfnamefont {G.}~\bibnamefont
  {Eichmann}}, \bibinfo {author} {\bibfnamefont {H.}~\bibnamefont
  {Sanchis-Alepuz}}, \bibinfo {author} {\bibfnamefont {R.}~\bibnamefont
  {Williams}}, \bibinfo {author} {\bibfnamefont {R.}~\bibnamefont {Alkofer}}, \
  and\ \bibinfo {author} {\bibfnamefont {C.~S.}\ \bibnamefont {Fischer}},\
  }\href {\doibase 10.1016/j.ppnp.2016.07.001} {\bibfield  {journal} {\bibinfo
  {journal} {Prog. Part. Nucl. Phys.}\ }\textbf {\bibinfo {volume} {91}},\
  \bibinfo {pages} {1} (\bibinfo {year} {2016}{\natexlab{b}})},\ \Eprint
  {http://arxiv.org/abs/1606.09602} {arXiv:1606.09602 [hep-ph]} \BibitemShut
  {NoStop}%
%%CITATION = ARXIV:1606.09602;%%
\bibitem [{\citenamefont {Sternbeck}\ \emph {et~al.}(2012)\citenamefont
  {Sternbeck}, \citenamefont {Maltman}, \citenamefont {M{\"u}ller-Preussker},\
  and\ \citenamefont {von Smekal}}]{Sternbeck:2012qs}%
  \BibitemOpen
  \bibfield  {author} {\bibinfo {author} {\bibfnamefont {A.}~\bibnamefont
  {Sternbeck}}, \bibinfo {author} {\bibfnamefont {K.}~\bibnamefont {Maltman}},
  \bibinfo {author} {\bibfnamefont {M.}~\bibnamefont {M{\"u}ller-Preussker}}, \
  and\ \bibinfo {author} {\bibfnamefont {L.}~\bibnamefont {von Smekal}},\
  }\bibfield  {booktitle} {\emph {\bibinfo {booktitle} {{Proceedings, 30th
  International Symposium on Lattice Field Theory (Lattice 2012): Cairns,
  Australia, June 24-29, 2012}}},\ }\href {\doibase 10.22323/1.164.0243}
  {\bibfield  {journal} {\bibinfo  {journal} {PoS}\ }\textbf {\bibinfo {volume}
  {LATTICE2012}},\ \bibinfo {pages} {243} (\bibinfo {year} {2012})},\ \Eprint
  {http://arxiv.org/abs/1212.2039} {arXiv:1212.2039 [hep-lat]} \BibitemShut
  {NoStop}%
%%CITATION = ARXIV:1212.2039;%%
\bibitem [{\citenamefont {Pawlowski}(2007)}]{Pawlowski:2005xe}%
  \BibitemOpen
  \bibfield  {author} {\bibinfo {author} {\bibfnamefont {J.~M.}\ \bibnamefont
  {Pawlowski}},\ }\href {\doibase 10.1016/j.aop.2007.01.007} {\bibfield
  {journal} {\bibinfo  {journal} {Annals Phys.}\ }\textbf {\bibinfo {volume}
  {322}},\ \bibinfo {pages} {2831} (\bibinfo {year} {2007})},\ \Eprint
  {http://arxiv.org/abs/hep-th/0512261} {arXiv:hep-th/0512261 [hep-th]}
  \BibitemShut {NoStop}%
%%CITATION = HEP-TH/0512261;%%
\bibitem [{\citenamefont {Gies}\ and\ \citenamefont
  {Wetterich}(2002)}]{Gies:2001nw}%
  \BibitemOpen
  \bibfield  {author} {\bibinfo {author} {\bibfnamefont {H.}~\bibnamefont
  {Gies}}\ and\ \bibinfo {author} {\bibfnamefont {C.}~\bibnamefont
  {Wetterich}},\ }\href {\doibase 10.1103/PhysRevD.65.065001} {\bibfield
  {journal} {\bibinfo  {journal} {Phys. Rev.}\ }\textbf {\bibinfo {volume}
  {D65}},\ \bibinfo {pages} {065001} (\bibinfo {year} {2002})},\ \Eprint
  {http://arxiv.org/abs/hep-th/0107221} {arXiv:hep-th/0107221 [hep-th]}
  \BibitemShut {NoStop}%
%%CITATION = HEP-TH/0107221;%%
\bibitem [{\citenamefont {Gies}\ and\ \citenamefont
  {Wetterich}(2004)}]{Gies:2002hq}%
  \BibitemOpen
  \bibfield  {author} {\bibinfo {author} {\bibfnamefont {H.}~\bibnamefont
  {Gies}}\ and\ \bibinfo {author} {\bibfnamefont {C.}~\bibnamefont
  {Wetterich}},\ }\href {\doibase 10.1103/PhysRevD.69.025001} {\bibfield
  {journal} {\bibinfo  {journal} {Phys. Rev.}\ }\textbf {\bibinfo {volume}
  {D69}},\ \bibinfo {pages} {025001} (\bibinfo {year} {2004})},\ \Eprint
  {http://arxiv.org/abs/hep-th/0209183} {arXiv:hep-th/0209183 [hep-th]}
  \BibitemShut {NoStop}%
%%CITATION = HEP-TH/0209183;%%
\bibitem [{\citenamefont {Floerchinger}\ and\ \citenamefont
  {Wetterich}(2009)}]{Floerchinger:2009uf}%
  \BibitemOpen
  \bibfield  {author} {\bibinfo {author} {\bibfnamefont {S.}~\bibnamefont
  {Floerchinger}}\ and\ \bibinfo {author} {\bibfnamefont {C.}~\bibnamefont
  {Wetterich}},\ }\href {\doibase 10.1016/j.physletb.2009.09.014} {\bibfield
  {journal} {\bibinfo  {journal} {Phys. Lett.}\ }\textbf {\bibinfo {volume}
  {B680}},\ \bibinfo {pages} {371} (\bibinfo {year} {2009})},\ \Eprint
  {http://arxiv.org/abs/0905.0915} {arXiv:0905.0915 [hep-th]} \BibitemShut
  {NoStop}%
%%CITATION = ARXIV:0905.0915;%%
\bibitem [{\citenamefont {Bender}\ \emph {et~al.}(1998)\citenamefont {Bender},
  \citenamefont {Poulis}, \citenamefont {Roberts}, \citenamefont {Schmidt},\
  and\ \citenamefont {Thomas}}]{Bender:1997jf}%
  \BibitemOpen
  \bibfield  {author} {\bibinfo {author} {\bibfnamefont {A.}~\bibnamefont
  {Bender}}, \bibinfo {author} {\bibfnamefont {G.~I.}\ \bibnamefont {Poulis}},
  \bibinfo {author} {\bibfnamefont {C.~D.}\ \bibnamefont {Roberts}}, \bibinfo
  {author} {\bibfnamefont {S.~M.}\ \bibnamefont {Schmidt}}, \ and\ \bibinfo
  {author} {\bibfnamefont {A.~W.}\ \bibnamefont {Thomas}},\ }\href {\doibase
  10.1016/S0370-2693(98)00546-2} {\bibfield  {journal} {\bibinfo  {journal}
  {Phys. Lett.}\ }\textbf {\bibinfo {volume} {B431}},\ \bibinfo {pages} {263}
  (\bibinfo {year} {1998})},\ \Eprint {http://arxiv.org/abs/nucl-th/9710069}
  {arXiv:nucl-th/9710069 [nucl-th]} \BibitemShut {NoStop}%
%%CITATION = NUCL-TH/9710069;%%
\bibitem [{\citenamefont {Gao}\ and\ \citenamefont {Liu}(2018)}]{Gao:2017gvf}%
  \BibitemOpen
  \bibfield  {author} {\bibinfo {author} {\bibfnamefont {F.}~\bibnamefont
  {Gao}}\ and\ \bibinfo {author} {\bibfnamefont {Y.-x.}\ \bibnamefont {Liu}},\
  }\href {\doibase 10.1103/PhysRevD.97.056011} {\bibfield  {journal} {\bibinfo
  {journal} {Phys. Rev.}\ }\textbf {\bibinfo {volume} {D97}},\ \bibinfo {pages}
  {056011} (\bibinfo {year} {2018})},\ \Eprint
  {http://arxiv.org/abs/1702.01420} {arXiv:1702.01420 [hep-ph]} \BibitemShut
  {NoStop}%
%%CITATION = ARXIV:1702.01420;%%
\bibitem [{\citenamefont {Fischer}(2006)}]{Fischer:2006ub}%
  \BibitemOpen
  \bibfield  {author} {\bibinfo {author} {\bibfnamefont {C.~S.}\ \bibnamefont
  {Fischer}},\ }\href {\doibase 10.1088/0954-3899/32/8/R02} {\bibfield
  {journal} {\bibinfo  {journal} {J. Phys.}\ }\textbf {\bibinfo {volume}
  {G32}},\ \bibinfo {pages} {R253} (\bibinfo {year} {2006})},\ \Eprint
  {http://arxiv.org/abs/hep-ph/0605173} {arXiv:hep-ph/0605173 [hep-ph]}
  \BibitemShut {NoStop}%
%%CITATION = HEP-PH/0605173;%%
\bibitem [{\citenamefont {Bashir}\ \emph {et~al.}(2012)\citenamefont {Bashir},
  \citenamefont {Chang}, \citenamefont {Cloet}, \citenamefont {El-Bennich},
  \citenamefont {Liu}, \citenamefont {Roberts},\ and\ \citenamefont
  {Tandy}}]{Bashir:2012fs}%
  \BibitemOpen
  \bibfield  {author} {\bibinfo {author} {\bibfnamefont {A.}~\bibnamefont
  {Bashir}}, \bibinfo {author} {\bibfnamefont {L.}~\bibnamefont {Chang}},
  \bibinfo {author} {\bibfnamefont {I.~C.}\ \bibnamefont {Cloet}}, \bibinfo
  {author} {\bibfnamefont {B.}~\bibnamefont {El-Bennich}}, \bibinfo {author}
  {\bibfnamefont {Y.-X.}\ \bibnamefont {Liu}}, \bibinfo {author} {\bibfnamefont
  {C.~D.}\ \bibnamefont {Roberts}}, \ and\ \bibinfo {author} {\bibfnamefont
  {P.~C.}\ \bibnamefont {Tandy}},\ }\href {\doibase 10.1088/0253-6102/58/1/16}
  {\bibfield  {journal} {\bibinfo  {journal} {Commun. Theor. Phys.}\ }\textbf
  {\bibinfo {volume} {58}},\ \bibinfo {pages} {79} (\bibinfo {year} {2012})},\
  \Eprint {http://arxiv.org/abs/1201.3366} {arXiv:1201.3366 [nucl-th]}
  \BibitemShut {NoStop}%
%%CITATION = ARXIV:1201.3366;%%
\bibitem [{\citenamefont {Blum}\ \emph {et~al.}(2016)\citenamefont {Blum} \emph
  {et~al.}}]{Blum:2014tka}%
  \BibitemOpen
  \bibfield  {author} {\bibinfo {author} {\bibfnamefont {T.}~\bibnamefont
  {Blum}} \emph {et~al.} (\bibinfo {collaboration} {RBC, UKQCD}),\ }\href
  {\doibase 10.1103/PhysRevD.93.074505} {\bibfield  {journal} {\bibinfo
  {journal} {Phys. Rev.}\ }\textbf {\bibinfo {volume} {D93}},\ \bibinfo {pages}
  {074505} (\bibinfo {year} {2016})},\ \Eprint {http://arxiv.org/abs/1411.7017}
  {arXiv:1411.7017 [hep-lat]} \BibitemShut {NoStop}%
%%CITATION = ARXIV:1411.7017;%%
\bibitem [{\citenamefont {Bowman}\ \emph {et~al.}(2005)\citenamefont {Bowman},
  \citenamefont {Heller}, \citenamefont {Leinweber}, \citenamefont
  {Parappilly}, \citenamefont {Williams},\ and\ \citenamefont
  {Zhang}}]{Bowman:2005vx}%
  \BibitemOpen
  \bibfield  {author} {\bibinfo {author} {\bibfnamefont {P.~O.}\ \bibnamefont
  {Bowman}}, \bibinfo {author} {\bibfnamefont {U.~M.}\ \bibnamefont {Heller}},
  \bibinfo {author} {\bibfnamefont {D.~B.}\ \bibnamefont {Leinweber}}, \bibinfo
  {author} {\bibfnamefont {M.~B.}\ \bibnamefont {Parappilly}}, \bibinfo
  {author} {\bibfnamefont {A.~G.}\ \bibnamefont {Williams}}, \ and\ \bibinfo
  {author} {\bibfnamefont {J.-b.}\ \bibnamefont {Zhang}},\ }\href {\doibase
  10.1103/PhysRevD.71.054507} {\bibfield  {journal} {\bibinfo  {journal} {Phys.
  Rev.}\ }\textbf {\bibinfo {volume} {D71}},\ \bibinfo {pages} {054507}
  (\bibinfo {year} {2005})},\ \Eprint {http://arxiv.org/abs/hep-lat/0501019}
  {arXiv:hep-lat/0501019 [hep-lat]} \BibitemShut {NoStop}%
%%CITATION = HEP-LAT/0501019;%%
\bibitem [{\citenamefont {Abelev}\ \emph {et~al.}(2009)\citenamefont {Abelev}
  \emph {et~al.}}]{Abelev:2009ac}%
  \BibitemOpen
  \bibfield  {author} {\bibinfo {author} {\bibfnamefont {B.~I.}\ \bibnamefont
  {Abelev}} \emph {et~al.} (\bibinfo {collaboration} {STAR}),\ }\href {\doibase
  10.1103/PhysRevLett.103.251601} {\bibfield  {journal} {\bibinfo  {journal}
  {Phys. Rev. Lett.}\ }\textbf {\bibinfo {volume} {103}},\ \bibinfo {pages}
  {251601} (\bibinfo {year} {2009})},\ \Eprint {http://arxiv.org/abs/0909.1739}
  {arXiv:0909.1739 [nucl-ex]} \BibitemShut {NoStop}%
%%CITATION = ARXIV:0909.1739;%%
\bibitem [{\citenamefont {Zafeiropoulos}\ \emph {et~al.}(2019)\citenamefont
  {Zafeiropoulos}, \citenamefont {Boucaud}, \citenamefont {De~Soto},
  \citenamefont {Rodr{\'\i}guez-Quintero},\ and\ \citenamefont
  {Segovia}}]{Zafeiropoulos:2019flq}%
  \BibitemOpen
  \bibfield  {author} {\bibinfo {author} {\bibfnamefont {S.}~\bibnamefont
  {Zafeiropoulos}}, \bibinfo {author} {\bibfnamefont {P.}~\bibnamefont
  {Boucaud}}, \bibinfo {author} {\bibfnamefont {F.}~\bibnamefont {De~Soto}},
  \bibinfo {author} {\bibfnamefont {J.}~\bibnamefont
  {Rodr{\'\i}guez-Quintero}}, \ and\ \bibinfo {author} {\bibfnamefont
  {J.}~\bibnamefont {Segovia}},\ }\href {\doibase
  10.1103/PhysRevLett.122.162002} {\bibfield  {journal} {\bibinfo  {journal}
  {Phys. Rev. Lett.}\ }\textbf {\bibinfo {volume} {122}},\ \bibinfo {pages}
  {162002} (\bibinfo {year} {2019})},\ \Eprint
  {http://arxiv.org/abs/1902.08148} {arXiv:1902.08148 [hep-ph]} \BibitemShut
  {NoStop}%
%%CITATION = ARXIV:1902.08148;%%
\bibitem [{\citenamefont {Boucaud}\ \emph {et~al.}(2018)\citenamefont
  {Boucaud}, \citenamefont {De~Soto}, \citenamefont {Raya}, \citenamefont
  {Rodr{\'\i}guez-Quintero},\ and\ \citenamefont
  {Zafeiropoulos}}]{Boucaud:2018xup}%
  \BibitemOpen
  \bibfield  {author} {\bibinfo {author} {\bibfnamefont {P.}~\bibnamefont
  {Boucaud}}, \bibinfo {author} {\bibfnamefont {F.}~\bibnamefont {De~Soto}},
  \bibinfo {author} {\bibfnamefont {K.}~\bibnamefont {Raya}}, \bibinfo {author}
  {\bibfnamefont {J.}~\bibnamefont {Rodr{\'\i}guez-Quintero}}, \ and\ \bibinfo
  {author} {\bibfnamefont {S.}~\bibnamefont {Zafeiropoulos}},\ }\href {\doibase
  10.1103/PhysRevD.98.114515} {\bibfield  {journal} {\bibinfo  {journal} {Phys.
  Rev.}\ }\textbf {\bibinfo {volume} {D98}},\ \bibinfo {pages} {114515}
  (\bibinfo {year} {2018})},\ \Eprint {http://arxiv.org/abs/1809.05776}
  {arXiv:1809.05776 [hep-ph]} \BibitemShut {NoStop}%
%%CITATION = ARXIV:1809.05776;%%
\bibitem [{\citenamefont {Oliveira}\ \emph {et~al.}(2019)\citenamefont
  {Oliveira}, \citenamefont {Silva}, \citenamefont {Skullerud},\ and\
  \citenamefont {Sternbeck}}]{Oliveira:2018lln}%
  \BibitemOpen
  \bibfield  {author} {\bibinfo {author} {\bibfnamefont {O.}~\bibnamefont
  {Oliveira}}, \bibinfo {author} {\bibfnamefont {P.~J.}\ \bibnamefont {Silva}},
  \bibinfo {author} {\bibfnamefont {J.-I.}\ \bibnamefont {Skullerud}}, \ and\
  \bibinfo {author} {\bibfnamefont {A.}~\bibnamefont {Sternbeck}},\ }\href
  {\doibase 10.1103/PhysRevD.99.094506} {\bibfield  {journal} {\bibinfo
  {journal} {Phys. Rev.}\ }\textbf {\bibinfo {volume} {D99}},\ \bibinfo {pages}
  {094506} (\bibinfo {year} {2019})},\ \Eprint
  {http://arxiv.org/abs/1809.02541} {arXiv:1809.02541 [hep-lat]} \BibitemShut
  {NoStop}%
%%CITATION = ARXIV:1809.02541;%%
\bibitem [{\citenamefont {Silva}\ \emph {et~al.}(2014)\citenamefont {Silva},
  \citenamefont {Dudal},\ and\ \citenamefont {Oliveira}}]{Silva:2013foa}%
  \BibitemOpen
  \bibfield  {author} {\bibinfo {author} {\bibfnamefont {P.~J.}\ \bibnamefont
  {Silva}}, \bibinfo {author} {\bibfnamefont {D.}~\bibnamefont {Dudal}}, \ and\
  \bibinfo {author} {\bibfnamefont {O.}~\bibnamefont {Oliveira}},\ }\bibfield
  {booktitle} {\emph {\bibinfo {booktitle} {{Proceedings, 31st International
  Symposium on Lattice Field Theory (Lattice 2013): Mainz, Germany, July
  29-August 3, 2013}}},\ }\href {\doibase 10.22323/1.187.0366} {\bibfield
  {journal} {\bibinfo  {journal} {PoS}\ }\textbf {\bibinfo {volume}
  {LATTICE2013}},\ \bibinfo {pages} {366} (\bibinfo {year} {2014})},\ \Eprint
  {http://arxiv.org/abs/1311.3643} {arXiv:1311.3643 [hep-lat]} \BibitemShut
  {NoStop}%
%%CITATION = ARXIV:1311.3643;%%
\bibitem [{\citenamefont {Borsanyi}\ \emph {et~al.}(2010)\citenamefont
  {Borsanyi}, \citenamefont {Fodor}, \citenamefont {Hoelbling}, \citenamefont
  {Katz}, \citenamefont {Krieg}, \citenamefont {Ratti},\ and\ \citenamefont
  {Szabo}}]{Borsanyi:2010bp}%
  \BibitemOpen
  \bibfield  {author} {\bibinfo {author} {\bibfnamefont {S.}~\bibnamefont
  {Borsanyi}}, \bibinfo {author} {\bibfnamefont {Z.}~\bibnamefont {Fodor}},
  \bibinfo {author} {\bibfnamefont {C.}~\bibnamefont {Hoelbling}}, \bibinfo
  {author} {\bibfnamefont {S.~D.}\ \bibnamefont {Katz}}, \bibinfo {author}
  {\bibfnamefont {S.}~\bibnamefont {Krieg}}, \bibinfo {author} {\bibfnamefont
  {C.}~\bibnamefont {Ratti}}, \ and\ \bibinfo {author} {\bibfnamefont {K.~K.}\
  \bibnamefont {Szabo}} (\bibinfo {collaboration} {Wuppertal-Budapest}),\
  }\href {\doibase 10.1007/JHEP09(2010)073} {\bibfield  {journal} {\bibinfo
  {journal} {JHEP}\ }\textbf {\bibinfo {volume} {09}},\ \bibinfo {pages} {073}
  (\bibinfo {year} {2010})},\ \Eprint {http://arxiv.org/abs/1005.3508}
  {arXiv:1005.3508 [hep-lat]} \BibitemShut {NoStop}%
%%CITATION = ARXIV:1005.3508;%%
\bibitem [{\citenamefont {Cheng}\ \emph {et~al.}(2008)\citenamefont {Cheng}
  \emph {et~al.}}]{Cheng:2007jq}%
  \BibitemOpen
  \bibfield  {author} {\bibinfo {author} {\bibfnamefont {M.}~\bibnamefont
  {Cheng}} \emph {et~al.},\ }\href {\doibase 10.1103/PhysRevD.77.014511}
  {\bibfield  {journal} {\bibinfo  {journal} {Phys. Rev.}\ }\textbf {\bibinfo
  {volume} {D77}},\ \bibinfo {pages} {014511} (\bibinfo {year} {2008})},\
  \Eprint {http://arxiv.org/abs/0710.0354} {arXiv:0710.0354 [hep-lat]}
  \BibitemShut {NoStop}%
%%CITATION = ARXIV:0710.0354;%%
\bibitem [{\citenamefont {Khan}\ \emph {et~al.}(2015)\citenamefont {Khan},
  \citenamefont {Pawlowski}, \citenamefont {Rennecke},\ and\ \citenamefont
  {Scherer}}]{Khan:2015puu}%
  \BibitemOpen
  \bibfield  {author} {\bibinfo {author} {\bibfnamefont {N.}~\bibnamefont
  {Khan}}, \bibinfo {author} {\bibfnamefont {J.~M.}\ \bibnamefont {Pawlowski}},
  \bibinfo {author} {\bibfnamefont {F.}~\bibnamefont {Rennecke}}, \ and\
  \bibinfo {author} {\bibfnamefont {M.~M.}\ \bibnamefont {Scherer}},\
  }\href@noop {} {\  (\bibinfo {year} {2015})},\ \Eprint
  {http://arxiv.org/abs/1512.03673} {arXiv:1512.03673 [hep-ph]} \BibitemShut
  {NoStop}%
%%CITATION = ARXIV:1512.03673;%%
\bibitem [{\citenamefont {Fischer}\ and\ \citenamefont
  {Luecker}(2013)}]{Fischer:2012vc}%
  \BibitemOpen
  \bibfield  {author} {\bibinfo {author} {\bibfnamefont {C.~S.}\ \bibnamefont
  {Fischer}}\ and\ \bibinfo {author} {\bibfnamefont {J.}~\bibnamefont
  {Luecker}},\ }\href {\doibase 10.1016/j.physletb.2012.11.054} {\bibfield
  {journal} {\bibinfo  {journal} {Phys. Lett.}\ }\textbf {\bibinfo {volume}
  {B718}},\ \bibinfo {pages} {1036} (\bibinfo {year} {2013})},\ \Eprint
  {http://arxiv.org/abs/1206.5191} {arXiv:1206.5191 [hep-ph]} \BibitemShut
  {NoStop}%
%%CITATION = ARXIV:1206.5191;%%
\bibitem [{\citenamefont {Allton}\ \emph {et~al.}(2002)\citenamefont {Allton},
  \citenamefont {Ejiri}, \citenamefont {Hands}, \citenamefont {Kaczmarek},
  \citenamefont {Karsch}, \citenamefont {Laermann}, \citenamefont {Schmidt},\
  and\ \citenamefont {Scorzato}}]{Allton:2002zi}%
  \BibitemOpen
  \bibfield  {author} {\bibinfo {author} {\bibfnamefont {C.~R.}\ \bibnamefont
  {Allton}}, \bibinfo {author} {\bibfnamefont {S.}~\bibnamefont {Ejiri}},
  \bibinfo {author} {\bibfnamefont {S.~J.}\ \bibnamefont {Hands}}, \bibinfo
  {author} {\bibfnamefont {O.}~\bibnamefont {Kaczmarek}}, \bibinfo {author}
  {\bibfnamefont {F.}~\bibnamefont {Karsch}}, \bibinfo {author} {\bibfnamefont
  {E.}~\bibnamefont {Laermann}}, \bibinfo {author} {\bibfnamefont
  {C.}~\bibnamefont {Schmidt}}, \ and\ \bibinfo {author} {\bibfnamefont
  {L.}~\bibnamefont {Scorzato}},\ }\href {\doibase 10.1103/PhysRevD.66.074507}
  {\bibfield  {journal} {\bibinfo  {journal} {Phys. Rev.}\ }\textbf {\bibinfo
  {volume} {D66}},\ \bibinfo {pages} {074507} (\bibinfo {year} {2002})},\
  \Eprint {http://arxiv.org/abs/hep-lat/0204010} {arXiv:hep-lat/0204010
  [hep-lat]} \BibitemShut {NoStop}%
%%CITATION = HEP-LAT/0204010;%%
\bibitem [{\citenamefont {de~Forcrand}\ and\ \citenamefont
  {Philipsen}(2002)}]{deForcrand:2002hgr}%
  \BibitemOpen
  \bibfield  {author} {\bibinfo {author} {\bibfnamefont {P.}~\bibnamefont
  {de~Forcrand}}\ and\ \bibinfo {author} {\bibfnamefont {O.}~\bibnamefont
  {Philipsen}},\ }\href {\doibase 10.1016/S0550-3213(02)00626-0} {\bibfield
  {journal} {\bibinfo  {journal} {Nucl. Phys.}\ }\textbf {\bibinfo {volume}
  {B642}},\ \bibinfo {pages} {290} (\bibinfo {year} {2002})},\ \Eprint
  {http://arxiv.org/abs/hep-lat/0205016} {arXiv:hep-lat/0205016 [hep-lat]}
  \BibitemShut {NoStop}%
%%CITATION = HEP-LAT/0205016;%%
\bibitem [{\citenamefont {Philipsen}(2007)}]{Philipsen:2007rj}%
  \BibitemOpen
  \bibfield  {author} {\bibinfo {author} {\bibfnamefont {O.}~\bibnamefont
  {Philipsen}},\ }\bibfield  {booktitle} {\emph {\bibinfo {booktitle}
  {{Conceptual and Numerical Challenges in Femto- and Peta-Scale Physics.
  Proceedings, 45. Internationale Universitätswochen für theoretische Physik:
  Schladming, Austria, February 24-March 3, 2007}}},\ }\href {\doibase
  10.1140/epjst/e2007-00376-3} {\bibfield  {journal} {\bibinfo  {journal} {Eur.
  Phys. J. ST}\ }\textbf {\bibinfo {volume} {152}},\ \bibinfo {pages} {29}
  (\bibinfo {year} {2007})},\ \Eprint {http://arxiv.org/abs/0708.1293}
  {arXiv:0708.1293 [hep-lat]} \BibitemShut {NoStop}%
%%CITATION = ARXIV:0708.1293;%%
\bibitem [{\citenamefont {D'Elia}(2019)}]{DElia:2018fjp}%
  \BibitemOpen
  \bibfield  {author} {\bibinfo {author} {\bibfnamefont {M.}~\bibnamefont
  {D'Elia}},\ }\bibfield  {booktitle} {\emph {\bibinfo {booktitle}
  {{Proceedings, 27th International Conference on Ultrarelativistic
  Nucleus-Nucleus Collisions (Quark Matter 2018): Venice, Italy, May 14-19,
  2018}}},\ }\href {\doibase 10.1016/j.nuclphysa.2018.10.042} {\bibfield
  {journal} {\bibinfo  {journal} {Nucl. Phys.}\ }\textbf {\bibinfo {volume}
  {A982}},\ \bibinfo {pages} {99} (\bibinfo {year} {2019})},\ \Eprint
  {http://arxiv.org/abs/1809.10660} {arXiv:1809.10660 [hep-lat]} \BibitemShut
  {NoStop}%
%%CITATION = ARXIV:1809.10660;%%
\bibitem [{\citenamefont {Alba}\ \emph {et~al.}(2014)\citenamefont {Alba},
  \citenamefont {Alberico}, \citenamefont {Bellwied}, \citenamefont {Bluhm},
  \citenamefont {Mantovani~Sarti}, \citenamefont {Nahrgang},\ and\
  \citenamefont {Ratti}}]{Alba:2014eba}%
  \BibitemOpen
  \bibfield  {author} {\bibinfo {author} {\bibfnamefont {P.}~\bibnamefont
  {Alba}}, \bibinfo {author} {\bibfnamefont {W.}~\bibnamefont {Alberico}},
  \bibinfo {author} {\bibfnamefont {R.}~\bibnamefont {Bellwied}}, \bibinfo
  {author} {\bibfnamefont {M.}~\bibnamefont {Bluhm}}, \bibinfo {author}
  {\bibfnamefont {V.}~\bibnamefont {Mantovani~Sarti}}, \bibinfo {author}
  {\bibfnamefont {M.}~\bibnamefont {Nahrgang}}, \ and\ \bibinfo {author}
  {\bibfnamefont {C.}~\bibnamefont {Ratti}},\ }\href {\doibase
  10.1016/j.physletb.2014.09.052} {\bibfield  {journal} {\bibinfo  {journal}
  {Phys. Lett.}\ }\textbf {\bibinfo {volume} {B738}},\ \bibinfo {pages} {305}
  (\bibinfo {year} {2014})},\ \Eprint {http://arxiv.org/abs/1403.4903}
  {arXiv:1403.4903 [hep-ph]} \BibitemShut {NoStop}%
%%CITATION = ARXIV:1403.4903;%%
\bibitem [{\citenamefont {Becattini}\ \emph {et~al.}(2017)\citenamefont
  {Becattini}, \citenamefont {Steinheimer}, \citenamefont {Stock},\ and\
  \citenamefont {Bleicher}}]{Becattini:2016xct}%
  \BibitemOpen
  \bibfield  {author} {\bibinfo {author} {\bibfnamefont {F.}~\bibnamefont
  {Becattini}}, \bibinfo {author} {\bibfnamefont {J.}~\bibnamefont
  {Steinheimer}}, \bibinfo {author} {\bibfnamefont {R.}~\bibnamefont {Stock}},
  \ and\ \bibinfo {author} {\bibfnamefont {M.}~\bibnamefont {Bleicher}},\
  }\href {\doibase 10.1016/j.physletb.2016.11.033} {\bibfield  {journal}
  {\bibinfo  {journal} {Phys. Lett.}\ }\textbf {\bibinfo {volume} {B764}},\
  \bibinfo {pages} {241} (\bibinfo {year} {2017})},\ \Eprint
  {http://arxiv.org/abs/1605.09694} {arXiv:1605.09694 [nucl-th]} \BibitemShut
  {NoStop}%
%%CITATION = ARXIV:1605.09694;%%
\bibitem [{\citenamefont {Vovchenko}\ \emph {et~al.}(2016)\citenamefont
  {Vovchenko}, \citenamefont {Begun},\ and\ \citenamefont
  {Gorenstein}}]{Vovchenko:2015idt}%
  \BibitemOpen
  \bibfield  {author} {\bibinfo {author} {\bibfnamefont {V.}~\bibnamefont
  {Vovchenko}}, \bibinfo {author} {\bibfnamefont {V.~V.}\ \bibnamefont
  {Begun}}, \ and\ \bibinfo {author} {\bibfnamefont {M.~I.}\ \bibnamefont
  {Gorenstein}},\ }\href {\doibase 10.1103/PhysRevC.93.064906} {\bibfield
  {journal} {\bibinfo  {journal} {Phys. Rev.}\ }\textbf {\bibinfo {volume}
  {C93}},\ \bibinfo {pages} {064906} (\bibinfo {year} {2016})},\ \Eprint
  {http://arxiv.org/abs/1512.08025} {arXiv:1512.08025 [nucl-th]} \BibitemShut
  {NoStop}%
%%CITATION = ARXIV:1512.08025;%%
\bibitem [{\citenamefont {Sagun}\ \emph {et~al.}(2018)\citenamefont {Sagun},
  \citenamefont {Bugaev}, \citenamefont {Ivanytskyi}, \citenamefont
  {Yakimenko}, \citenamefont {Nikonov}, \citenamefont {Taranenko},
  \citenamefont {Greiner}, \citenamefont {Blaschke},\ and\ \citenamefont
  {Zinovjev}}]{Sagun:2017eye}%
  \BibitemOpen
  \bibfield  {author} {\bibinfo {author} {\bibfnamefont {V.~V.}\ \bibnamefont
  {Sagun}}, \bibinfo {author} {\bibfnamefont {K.~A.}\ \bibnamefont {Bugaev}},
  \bibinfo {author} {\bibfnamefont {A.~I.}\ \bibnamefont {Ivanytskyi}},
  \bibinfo {author} {\bibfnamefont {I.~P.}\ \bibnamefont {Yakimenko}}, \bibinfo
  {author} {\bibfnamefont {E.~G.}\ \bibnamefont {Nikonov}}, \bibinfo {author}
  {\bibfnamefont {A.~V.}\ \bibnamefont {Taranenko}}, \bibinfo {author}
  {\bibfnamefont {C.}~\bibnamefont {Greiner}}, \bibinfo {author} {\bibfnamefont
  {D.~B.}\ \bibnamefont {Blaschke}}, \ and\ \bibinfo {author} {\bibfnamefont
  {G.~M.}\ \bibnamefont {Zinovjev}},\ }\href {\doibase
  10.1140/epja/i2018-12535-1} {\bibfield  {journal} {\bibinfo  {journal} {Eur.
  Phys. J.}\ }\textbf {\bibinfo {volume} {A54}},\ \bibinfo {pages} {100}
  (\bibinfo {year} {2018})},\ \Eprint {http://arxiv.org/abs/1703.00049}
  {arXiv:1703.00049 [hep-ph]} \BibitemShut {NoStop}%
%%CITATION = ARXIV:1703.00049;%%
\bibitem [{\citenamefont {Buballa}\ and\ \citenamefont
  {Carignano}(2015)}]{Buballa:2014tba}%
  \BibitemOpen
  \bibfield  {author} {\bibinfo {author} {\bibfnamefont {M.}~\bibnamefont
  {Buballa}}\ and\ \bibinfo {author} {\bibfnamefont {S.}~\bibnamefont
  {Carignano}},\ }\href {\doibase 10.1016/j.ppnp.2014.11.001} {\bibfield
  {journal} {\bibinfo  {journal} {Prog. Part. Nucl. Phys.}\ }\textbf {\bibinfo
  {volume} {81}},\ \bibinfo {pages} {39} (\bibinfo {year} {2015})},\ \Eprint
  {http://arxiv.org/abs/1406.1367} {arXiv:1406.1367 [hep-ph]} \BibitemShut
  {NoStop}%
%%CITATION = ARXIV:1406.1367;%%
\bibitem [{\citenamefont {Wygas}\ \emph {et~al.}(2018)\citenamefont {Wygas},
  \citenamefont {Oldengott}, \citenamefont {Bödeker},\ and\ \citenamefont
  {Schwarz}}]{Wygas:2018otj}%
  \BibitemOpen
  \bibfield  {author} {\bibinfo {author} {\bibfnamefont {M.~M.}\ \bibnamefont
  {Wygas}}, \bibinfo {author} {\bibfnamefont {I.~M.}\ \bibnamefont
  {Oldengott}}, \bibinfo {author} {\bibfnamefont {D.}~\bibnamefont {Bödeker}},
  \ and\ \bibinfo {author} {\bibfnamefont {D.~J.}\ \bibnamefont {Schwarz}},\
  }\href {\doibase 10.1103/PhysRevLett.121.201302} {\bibfield  {journal}
  {\bibinfo  {journal} {Phys. Rev. Lett.}\ }\textbf {\bibinfo {volume} {121}},\
  \bibinfo {pages} {201302} (\bibinfo {year} {2018})},\ \Eprint
  {http://arxiv.org/abs/1807.10815} {arXiv:1807.10815 [hep-ph]} \BibitemShut
  {NoStop}%
%%CITATION = ARXIV:1807.10815;%%
\end{thebibliography}%
\end{document}